\documentclass[12pt]{article}

\usepackage{a4,amsmath,amssymb,graphicx,color}
\usepackage[british]{babel}
\usepackage[hypertex]{hyperref}

\definecolor{mathblue}{rgb}{0.2472, 0.24, 0.6}
\definecolor{mathred}{rgb}{0.6, 0.24, 0.442893}
\definecolor{mathyellow}{rgb}{0.6, 0.547014, 0.24}
\definecolor{mathgreen}{rgb}{0.24, 0.6, 0.33692}

\newcommand{\di}{\genfrac{}{}{0pt}{}}
\def\under#1{\kern.4pt\underline{\kern-.4pt{}#1\kern-.4pt}\kern.4pt}

\makeatletter
\def\section{\@startsection{section}{1}{\z@}{-3.25ex plus -1ex minus
    -.2ex}{1.5ex plus .2ex}{\normalfont\large\bfseries}}
\def\subsection{\@startsection{subsection}{1}{\z@}{-3.25ex plus -1ex
    minus -.2ex}{1.5ex plus .2ex}{\normalfont\itshape}}

\renewenvironment{thebibliography}[1]
         {\section*{References}\frenchspacing\small
          \begin{list}{[\arabic{enumi}]}
         {\usecounter{enumi}\parsep=2pt\topsep 0pt
         \settowidth{\labelwidth}{[#1]}
         \leftmargin=\labelwidth\advance\leftmargin\labelsep
         \rightmargin=0pt\itemsep=0pt\sloppy}}{\end{list}}

\newcounter{calc}
\def\thecalc{\arabic{calc}}

\newenvironment{calc}
         {\begingroup
          \catcode`_=11 \catcode`^=11 \catcode`"=11 
         \begin{list}{\bf In[\thecalc]}
         {\@nmbrlisttrue\def\@listctr{calc}\parsep=2pt\topsep 1ex
         \settowidth{\labelwidth}{{\bf In[99]}}
         \leftmargin=\labelwidth\advance\leftmargin\labelsep
         \rightmargin=0pt\itemsep=1ex\ttfamily}}{\end{list}
         \endgroup}

\makeatother

\renewcommand{\title}[1]{\vspace{10mm}\noindent{\Large{\bf #1}}\vspace{8mm}}
\newcommand{\authors}[1]{\noindent{\large #1}\vspace{3mm}}
\newcommand{\address}[1]{{\itshape #1\vspace{2mm}}}

\newtheorem{Theorem}{Theorem}
\newtheorem{Proposition}[Theorem]{Proposition}

\newtheorem{Lemma}[Theorem]{Lemma}

\newtheorem{Conjecture}[Theorem]{Conjecture}
\allowdisplaybreaks[4]

\begin{document}

\begin{center}

\title{Solvable 4D noncommutative
QFT: \\[1mm] phase transitions and quest for reflection positivity}

\authors{Harald {\sc Grosse}$^1$ and Raimar {\sc Wulkenhaar}$^2$}

\address{$^{1}$\,Fakult\"at f\"ur Physik, Universit\"at Wien\\
Boltzmanngasse 5, A-1090 Wien, Austria}

\address{$^{2}$\,Mathematisches Institut der Westf\"alischen
  Wilhelms-Universit\"at\\
Einsteinstra\ss{}e 62, D-48149 M\"unster, Germany}

\footnotetext[1]{harald.grosse@univie.ac.at}
\footnotetext[2]{raimar@math.uni-muenster.de}

\vskip 1.5cm

\textbf{Abstract} \vskip 3mm
\begin{minipage}{14cm}%
\parindent 1.5em

\noindent
We provide further analytical and first numerical results on the
solvable $\lambda\phi^4_4$-NCQFT model.  We prove that for 
$\lambda<0$ the singular integral equation has a unique
solution, whereas for $\lambda>0$ there is considerable
freedom. Furthermore we provide integral formulae for partial
derivatives of the matrix 2-point function, which are the key to
investigate reflection positivity.

  The numerical implementation of these equations gives
  evidence for phase transitions. The derivative of the
  finite wavefunction renormalisation with respect to $\lambda$ is
  discontinuous at $\lambda_c \approx -0.39$. This leads to
  singularities in higher correlation functions for
  $\lambda<\lambda_c$. The phase $\lambda >0$ is not yet under
  control because of the freedom in the singular integral equation.

  Reflection positivity requires that the two-point function is
  Stieltjes.  Implementing Widder's criteria for Stieltjes functions
  we exclude reflection positivity outside the phase
  $[\lambda_c,0]$. For the phase $\lambda_c<\lambda \leq 0$ we show
  that refining the discrete approximation we satisfy Widder to higher
  and higher order. This is clear evidence, albeit no proof, of
  reflection positivity in that phase.
\end{minipage}

\end{center}

\section{Introduction}

The $\lambda\phi^4_4$-quantum field theory model \cite{Grosse:2004yu} on 
noncommutative Moyal space has surprising properties.
Although being the analogue of the ordinary $\lambda \phi^4_4$-model, it has
vanishing $\beta$-function, which was (first perturbatively and after
preliminary results in \cite{Grosse:2004by, Disertori:2006uy}) proved
by an ingenious combination of Ward identities related to a
$U(\infty)$-symmetry with Schwinger-Dyson equations
\cite{Disertori:2006nq}. This method was extended in
\cite{Grosse:2009pa} to obtain a closed equation for the 2-point
function of this model. 

In our previous work \cite{Grosse:2012uv} we have vastly extended 
the ideas of \cite{Grosse:2009pa} in two directions. We showed that 
Ward identity and reality
lead to an exact solution of the quartic matrix model
\begin{align}
\frac{1}{\text{volume}}\log \frac{\mathcal{Z}[E,J]}{\mathcal{Z}[E,0]}\;,\qquad
\mathcal{Z}[E,J]
= \int
\mathcal{D}[\Phi]\;
\exp(\mathrm{tr}(J\Phi{-}E\Phi^2{-}\tfrac{\lambda}{4} \Phi^4))
\label{QMM}
\end{align}
in terms of the solution of a non-linear equation. Here $E$ represents
an unbounded self\-adjoint positive operator with compact resolvent,
generalising the Laplacian, and $J$ is a test function operator used
to generate the correlation functions. Higher correlation functions
are given by purely algebraic recursion formulae in terms of the
eigenvalues of $E$ and the solution of the non-linear equation for the
2-point function. We proved that any renormalisable quartic matrix
model has vanishing $\beta$-function. The second extension achieved in
\cite{Grosse:2012uv} concerns the application to the noncommutative
$\lambda \phi^4_4$-model \cite{Grosse:2004yu} in the limit of extreme
noncommutativity $\theta\to \infty$. We observed that the
non-linear equation for the 2-point function can be split into a
linear singular integral equation of Carleman type \cite{Carleman,
  Tricomi} for the difference to the boundary and a resulting
fixed-point problem 
\begin{align}
G_{b0}=G_{0b}
& =  
\frac{1}{1+b}
\exp\Bigg(
{-} \lambda 
\int_0^b \!\!\! dt  \int_0^{\Lambda^2}  \!\!\!
\frac{dp}{(\lambda \pi p)^2 
+\big( t + \frac{1 +\lambda \pi p 
 \mathcal{H}_p^{\Lambda}[G_{\bullet 0}]}{G_{p0}}\big)^2} 
\Bigg)
\label{G0b-intro}
\end{align}
for the boundary 2-point function $G_{a0}$.
Here $\mathcal{H}_p^{\Lambda}$ denotes the finite Hilbert transform
over the interval ${]0,\Lambda^2[}$. 

In recent work \cite{Grosse:2013iva} we showed that the correlation
functions of \cite{Grosse:2012uv} lead to Schwinger functions for a
scalar field on $\mathbb{R}^4$ which satisfy the easy
Osterwalder-Schrader \cite{Osterwalder:1973dx, Osterwalder:1974tc}
axioms (OS0) growth conditions, (OS3) permutation symmetry and,
surprisingly for a highly noncommutative model, (OS1) Euclidean
invariance. We further proved that (OS2) reflection positivity of the
Schwinger 2-point function is equivalent to the requirement that the
diagonal matrix 2-point function is a Stieltjes function
\cite{Widder:1938??}. 

A simple perturbative argument shows that reflection positivity does
not hold for $\lambda>0$ \cite{Grosse:2013iva}. Looking closer at the
possibility of $\lambda<0$ we noticed that key formulae proved in
\cite{Grosse:2012uv} are only correct for $\lambda>0$. In sec.\
\ref{sec:winding} of this paper we carefully repeat this analysis for
either sign of $\lambda$. As by-product we clarify the freedom
resulting from the non-trivial solution of the homogeneous Carleman
equation \cite{Tricomi} which was left as an open problem in
\cite{Grosse:2012uv}. We prove the (lucky!) result that for $\lambda<0$
(which could possibly be reflection positive) the Carleman equation
has a unique solution, whereas for the less interesting 
case $\lambda>0$ (no reflection
positivity) there is considerable freedom. 

The fixed point equation (\ref{G0b-intro}) resulted from a symmetry
argument and not the true consistency equation for the boundary
two-point function $G_{a0}$. It was so far unclear whether
(\ref{G0b-intro}) admits false solutions which contradict the 
true consistency equation. In section~\ref{sec:master} we close this gap
and show that the true equation gives no further information.

In \cite{Grosse:2015fka} we prove, using the Schauder fixed point
theorem, that (\ref{G0b-intro}) has a solution (at least) for $-\frac{1}{6}\leq
\lambda\leq 0$ inside the region $\exp (\mathcal{K}_\lambda)$, with 
\begin{align}
\mathcal{K}_\lambda=\Big\{f\in \mathcal{C}^1(\mathbb{R}_+)\;:~ f(0)=0\;,
\quad -\frac{1-|\lambda|}{1+x}
\leq f'(x) \leq 
-\frac{1-\frac{|\lambda|}{1-2|\lambda|}}{1+x}  \Big\}\;.
\label{Klambda}
\end{align}
The much simpler case $\lambda>0$ was already treated
in \cite{Grosse:2012uv} under the (as we prove: false) assumption that
the non-trivial solution of the homogeneous Carleman
equation can be neglected. 

A first hint about reflection positivity can be obtained from a computer
simulation of the equations. Widder's criteria for Stieltjes functions
\cite{Widder:1938??} need derivatives of arbitrarily high order, which
is impossible for a discrete approximation of the equation. We
therefore derive in sec.~\ref{sec:intformula} an integral formula for arbitrary
partial derivatives of the 2-point function. 

In sec.~\ref{sec:computer} we present first results of a numerical
simulation of this model using \emph{Mathematica$^{TM}$}. The source
code is given in the appendix.  Starting point is the fixed point
equation  (\ref{G0b-intro}) for the boundary 2-point function. We view
$G_{0b}$ as a piecewise-linear function and (\ref{G0b-intro}) as
recursive definition of a sequence $\{G_{0b}^{i}\}_i$. We convince ourselves
that this sequence converges in Lipschitz norm. For given $\lambda$, a
sufficiently precise $G_{0b}^{i}$ is then used to compute
characterising data of the model. In this way we find clear evidence
for a phase transition at $\lambda_c \approx -0.39$ where the function
$\frac{\partial^2 G_{0b}(\lambda)}{\partial b \partial
  \lambda}\big|_{b=0}$ of $\lambda$ is
discontinuous. Within numerical error bounds we have\footnote{We 
  prove in the appendix of \cite{Grosse:2015fka} that $G_{0b}=1$ is an exact solution 
of (\ref{G0b-intro}) for
  any $\lambda<0$ and $\Lambda^2\to \infty$. This solution seems
 numerically unstable under small perturbations.}  $G_{0b} \equiv 1$ for $0\leq
b< b_\lambda$ and $\lambda<\lambda_c$, which would imply that higher
correlation functions do not exist for $\lambda<\lambda_c$. For
$\lambda>0$ we confirm an inconsistency due to neglecting the
freedom with the homogeneous Carleman equation. This leaves the region
$[\lambda_c,0]$ as the only interesting phase, and precisely here we
seem to have reflection positivity for the 2-point function. Of
course, a discrete approximation by piecewise-linear functions
cannot be Stieltjes. We show that the order where the Stieltjes
property fails increases significantly when the approximation is
refined; and this refinement slows down exactly at the same value
$\lambda_c\approx -0.39$. We view this as overwhelming support for the
conjecture that the boundary and diagonal 2-point functions $G_{0b}$
and $G_{aa}$, respectively, are Stieltjes functions. Together with
\cite{Grosse:2013iva} this would imply reflection positivity of the
Schwinger 2-point function.

\section{The 2-point function revisited}

\label{sec:2}

In \cite{Grosse:2012uv} we have studied the $\lambda\phi^4_4$-model on
noncommutative Moyal space in matrix representation. We showed that
the two-point function $G_{|\under{a}\under{b}|}$ satisfies a closed
non-linear equation in a scaling limit which simultaneously sends the
volume $V=(\frac{\theta}{4})^2$ and the size $\mathcal{N}$ of the
matrices to infinity with the ratio
$\frac{\mathcal{N}}{\sqrt{V}}=\mu^2 \Lambda^2(1+\mathcal{Y})$
fixed. In this limit, the 2-point function $G_{ab}$ depends on
`continuous matrix indices' $a,b\in [0,\Lambda^2]$ and satisfies a
non-linear integral equation $\mathcal{I}_a[G_{\bullet b}]=0$. It was
convenient to replace this equation by the coupled system
$\mathcal{I}_a[G_{\bullet b}]-\mathcal{I}_a[G_{\bullet 0}]=0$ and
$\mathcal{I}_a[G_{\bullet 0}]=0$. The difference equation admitted a
wavefunction renormalisation $Z\mapsto (1+\mathcal{Y})$ which reduced
the problem to a \emph{linear singular integral equation}
\cite{Carleman} for the difference $D_{ab} := a
\frac{G_{ab}-G_{a0}}{b}$. We treat this equation an its solution
$G_{ab}[G_{\bullet 0}]$ in sec.~\ref{sec:winding}. In
sec.~\ref{sec:master} we show that the boundary equation
$\mathcal{I}_a[G_{\bullet 0}]=0$ gives no other information than the
solution $G_{ab}[G_{\bullet 0}]$ plus symmetry $G_{ab}=G_{ba}$.

\subsection{Solution of the Carleman 
equation for any sign of $\lambda$}

\label{sec:winding}

As summarised above, the function $D_{ab} := a
\frac{G_{ab}-G_{a0}}{b}$ derived from the 2-point function $G_{ab}$ of
self-dual noncommutative $\lambda\phi^4_4$-theory \cite{Grosse:2004yu}
in the limit of continuous matrix indices $a,b\in [0,\Lambda^2]$
satisfies \cite{Grosse:2012uv} the Carleman singular integral equation
\cite{Carleman}
\begin{subequations}
\begin{align}
\Big(\frac{b}{a} + \frac{1+\lambda\pi a
\mathcal{H}_a^{\!\Lambda}[G_{\bullet 0}]}{a G_{a0}}
\Big) D_{ab}
-\lambda\pi \mathcal{H}_a^{\!\Lambda}[D_{\bullet b}]
&= -G_{a0}\;,
\label{D-Hilbert}
\\
\text{where}\qquad 
\mathcal{H}_a^{\!\Lambda}[f(\bullet)] &=\frac{1}{\pi}
\lim_{\epsilon\to 0}
\Big(\int_0^{a-\epsilon}+\int_{a+\epsilon}^{\Lambda^2}\Big) dp
\;\frac{f(p)}{p-a} 
\label{Hilbert}
\end{align}
\end{subequations}
denotes the \emph{finite Hilbert transform}
over the interval ${]0,\Lambda^2[}$.

The solution theory for such an equation over the interval ${]{-}1,1[}$ was
developed in Tricomi's book \cite{Tricomi} and, in much larger
generality, in \cite{Muskhelishvili}. Transforming the formulae 
given in \cite{Tricomi} for $x\in{]{-}1,1[}$ via 
$a=\frac{\Lambda^2}{2}(1+x)$ to $a\in
{]0,\Lambda^2[}$ we have
\begin{Proposition}[{\cite[\S 4.4]{Tricomi}}, $a=\frac{\Lambda^2}{2}(1+x)$]
\label{Prop:Carleman}
Let $h\in \mathcal{C}({]0,\Lambda^2[})$ and $f\in
L^q({]0,\Lambda^2[})$ for some $q>1$ (depending on $\vartheta$ defined
below).
Then the singular integral equation
\begin{align}
h(a) \varphi(a)-\lambda\pi \mathcal{H}_a^{\!\Lambda}[\varphi(\bullet)]=f(a)\;,
\qquad a\in {]0,\Lambda^2[}\;,
\label{Carleman}
\end{align}
has the solution
\begin{subequations}
\begin{align}
\varphi(a)&= \frac{e^{-\mathcal{H}_a^{\!\Lambda}[\pi-\vartheta]}
\sin (\vartheta(a)) }{\lambda \pi a} 
\nonumber
\\*
&\qquad\times \Big(
a f(a)  e^{\mathcal{H}_a^{\!\Lambda}[\pi-\vartheta]}\cos(\vartheta(a))
+ 
\mathcal{H}_a^{\!\Lambda}\big[e^{\mathcal{H}_\bullet[\pi-\vartheta]}
\bullet f(\bullet) \sin(\vartheta(\bullet)) \big]
+ C'\Big)
\label{Solution:Carleman-a}
\\
&\stackrel{*}{=} \frac{e^{\mathcal{H}_a^{\!\Lambda}[\vartheta]} 
\sin (\vartheta(a))}{\lambda \pi} \Big(
f(a) e^{-\mathcal{H}_a^{\!\Lambda}[\vartheta]} \cos(\vartheta(a))
+ 
\mathcal{H}_a^{\!\Lambda}\big[e^{-\mathcal{H}_\bullet[\vartheta]}
f(\bullet) \sin(\vartheta(\bullet)) \big]
+\frac{C }{\Lambda^2-a}\Big)\;,
\label{Solution:Carleman-b}
\end{align}
\end{subequations}
where $C,C'$ are arbitrary constants and the angle $\vartheta$ is
defined as $\displaystyle
\vartheta(a)=\di{\raisebox{-1.2ex}{\mbox{\normalsize$\arctan$}}}{
  \mbox{\scriptsize$[0,\pi]$}} \Big(\frac{\lambda\pi}{h(a)}\Big)
$. This angle obeys the identities \textup{\cite[\S
  4.4(28)]{Tricomi}}, \textup{\cite[\S 4.4(18)]{Tricomi}} and
\textup{\cite[\S 4.4(20)]{Tricomi}},
\begin{subequations}
\label{Tricomi}
\begin{align}
e^{-\mathcal{H}_a^{\!\Lambda}[\vartheta]}\cos(\vartheta(a))
+ \mathcal{H}_a^{\!\Lambda}\big[
e^{-\mathcal{H}_\bullet^{\!\Lambda}[\vartheta]}\sin(\vartheta(\bullet)\big]
&=1\;,  \label{Tricomi-28}
\\
e^{\mathcal{H}_a^{\!\Lambda}[\vartheta]}\cos (\vartheta(a))
-\mathcal{H}_a^{\!\Lambda}\big[e^{\mathcal{H}_\bullet^{\!\Lambda}[\vartheta]} 
\sin (\vartheta(\bullet))\big]
&= 1\;,
\label{Tricomi-18}
\\
\frac{e^{\mathcal{H}_a^{\!\Lambda}[\vartheta]}\cos (\vartheta(a))}{\Lambda^2-a}
-\mathcal{H}_a^{\!\Lambda}\Big[
\frac{e^{\mathcal{H}_\bullet^{\!\Lambda}[\vartheta]} \sin (\vartheta(\bullet))}{
\Lambda^2-\bullet}\Big]
&= 0\;.
\label{Tricomi-20}
\end{align}
\end{subequations}
\end{Proposition}
The relation $\stackrel{*}{=}$ between (\ref{Solution:Carleman-a}) and 
(\ref{Solution:Carleman-b}) follows from 
$e^{-\mathcal{H}_a^{\!\Lambda}[\pi]}=\frac{a}{\Lambda^2-a}$ and
consequently 
\[
\mathcal{H}^{\!\Lambda}_a[
e^{\mathcal{H}_\bullet^{\!\Lambda}[\pi]} \bullet F(\bullet)]
= \mathcal{H}^{\!\Lambda}_a[ ((\Lambda^2-a)-(\bullet-a))
F(\bullet)]
= a  e^{\mathcal{H}_a^{\!\Lambda}[\pi]}
\mathcal{H}^{\!\Lambda}_a[F(\bullet)]
- \frac{1}{\pi}\int_0^{\Lambda^2} \!\!\!dp \;F(p) \;.
\]
This means that if $p\mapsto 
F(p)=e^{-\mathcal{H}_p[\vartheta]} f(p) \sin(\vartheta(p))$ is integrable, 
(\ref{Solution:Carleman-a}) and 
(\ref{Solution:Carleman-b}) are equivalent with $C=C'- \frac{1}{\pi} 
\int_0^{\Lambda^2}\!\! dp 
\;F(p)$. The constants $C,C'$ are possibly restricted by normalisation
conditions which could prefer (\ref{Solution:Carleman-a}) or 
(\ref{Solution:Carleman-b}).

In \cite{Grosse:2012uv} we have studied the solution of
(\ref{D-Hilbert}) using (\ref{Solution:Carleman-b}) under the
assumption $C=0$. In the meantime we noticed that for $\lambda<0$ the
normalisation conditions do not permit the step from
(\ref{Solution:Carleman-a}) to (\ref{Solution:Carleman-b}). We
carefully repeat the solution of (\ref{D-Hilbert}) based on 
(\ref{Solution:Carleman-a}) and (\ref{Solution:Carleman-b}) where
$C,C'$ are taken into account:
\begin{subequations}
\begin{align}
D_{ab} &= - \frac{e^{-\mathcal{H}_a^{\!\Lambda}[\pi-\vartheta_b]}
\sin (\vartheta_b(a)) }{\lambda\pi a}  \Big(
a G_{a0}  e^{\mathcal{H}_a^{\!\Lambda}[\pi-\vartheta_b]} \cos(\vartheta_b(a))
\nonumber
\\*[-1ex]
&\hspace*{5cm} + 
\mathcal{H}_a^{\!\Lambda}\big[
e^{\mathcal{H}_\bullet^{\!\Lambda}[\pi-\vartheta_b]}
\bullet G_{\bullet 0} \sin(\vartheta_b(\bullet)) \big] 
-C'_{b,\lambda,\Lambda^2}\Big)
\label{solution-Dab-a}
\\
& \stackrel{*}{=} - \frac{e^{\mathcal{H}_a^{\!\Lambda}[\vartheta_b]}
\sin (\vartheta_b(a)) }{\lambda\pi}  \Big(
G_{a0}  e^{-\mathcal{H}_a^{\!\Lambda}[\vartheta_b]} \cos(\vartheta_b(a))
\nonumber
\\*[-1ex]
&\hspace*{5cm} + 
\mathcal{H}_a^{\!\Lambda}\big[
e^{-\mathcal{H}_\bullet^{\!\Lambda}[\vartheta_b]}
G_{\bullet 0} \sin(\vartheta_b(\bullet)) \big]
- \frac{\Lambda^2 C_{b,\lambda,\Lambda^2}}{\Lambda^2-a} \Big)\;,
\label{solution-Dab-b}
\\
\vartheta_b(a) &=\di{\raisebox{-1.2ex}{\mbox{\normalsize$\arctan$}}}{
\mbox{\scriptsize$[0,\pi]$}}
\Big(\frac{ \lambda\pi a G_{a0} }{
1 + b G_{a0}+\lambda \pi a \mathcal{H}_a^{\!\Lambda}
\big[G_{\bullet 0} \big] }\Big)\;.
\label{arctan-0}
\end{align}
\end{subequations}%
Since $b$ is merely a parameter for the function $h$ in the Carleman
equation, the constants $C,C'$ are actually functions
$C_{b,\lambda,\Lambda^2},C'_{b,\lambda,\Lambda^2}$ of $(b,\lambda,\Lambda^2)$. 
The starting point for the solution of (\ref{solution-Dab-a}) or 
(\ref{solution-Dab-b}) 
is the
observation \cite{Grosse:2012uv} that (\ref{arctan-0}) is, for $b=0$,
also a Carleman-type singular integral equation
\begin{align}
\lambda\pi \cot \vartheta_{0}(a) G_{a0}
- \lambda\pi \mathcal{H}_a^{\!\Lambda}[G_{a0}]
=\frac{1}{a}
\label{Galpha0-Carleman}
\end{align}
with solution
\begin{subequations}
\begin{align}
G_{a0} 
&=\frac{e^{-\mathcal{H}_a^{\!\Lambda}[\pi-\vartheta_0]} 
\sin (\vartheta_0(a))}{\lambda \pi a} \big(
e^{\mathcal{H}_a^{\!\Lambda}[\pi-\vartheta_0]}\cos(\vartheta_0(a))
+ 
\mathcal{H}_a^{\!\Lambda}\big[
e^{\mathcal{H}_\bullet^{\!\Lambda}[\pi-\vartheta_0]}
\sin(\vartheta_0(\bullet))\big]+C'_{\lambda,\Lambda^2}
\big)
\label{Ga0-pre-a}
\\
& \stackrel{*}{=} 
\frac{e^{\mathcal{H}_a^{\!\Lambda}[\vartheta_0]} 
\sin (\vartheta_0(a))}{\lambda \pi} \Big(
\frac{e^{-\mathcal{H}_a^{\!\Lambda}[\vartheta_0]}\cos(\vartheta_0(a))}{a}
+ 
\mathcal{H}_a^{\!\Lambda}\Big[
\frac{e^{-\mathcal{H}_\bullet^{\!\Lambda}[\vartheta_0]}
\sin(\vartheta_0(\bullet))}{\bullet}
\Big]
+\frac{\Lambda^2 \tilde{C}_{\lambda,\Lambda^2}}{\Lambda^2-a}
\Big)\;.
\label{Ga0-pre-b}
\end{align}
\end{subequations}
Writing $\sin(\vartheta_0(\bullet))=\sin(\pi{-}\vartheta_0(\bullet))$
and $\cos(\vartheta_0(\bullet))=-\cos(\pi{-}\vartheta_0(\bullet))$ 
in (\ref{Ga0-pre-a}) we can use (\ref{Tricomi-18}) to obtain 
$e^{\mathcal{H}_a^{\!\Lambda}[\pi-\vartheta_0]}\cos(\vartheta_0(a))
+ 
\mathcal{H}_a^{\!\Lambda}\big[
e^{\mathcal{H}_\bullet^{\!\Lambda}[\pi-\vartheta_0]}
\sin(\vartheta_0(\bullet))\big]=-1$. The normalisation $G_{a0}=1$ 
then forces 
$C'_{\lambda,\Lambda^2}-1= 
e^{\mathcal{H}_0^{\!\Lambda}[\pi-\vartheta_0]} \mathrm{sign}(\lambda)$
because 
$\lim_{a\to 0}
\frac{\sin (\vartheta_0(a))}{|\lambda| \pi a}=1$. 

In (\ref{Ga0-pre-b}) we use rational fraction expansion $\displaystyle
\mathcal{H}_a^{\!\Lambda}
\Big[\frac{f(\bullet)}{\bullet}\Big] = \frac{1}{a}\Big(
\mathcal{H}_a^{\!\Lambda} \big[f(\bullet)\big]
- \mathcal{H}_0^{\!\Lambda}\big[f(\bullet)\big]\Big)$ as in
\cite{Grosse:2012uv} to obtain
\begin{align*}
G_{a0} =
\frac{e^{\mathcal{H}_a^{\!\Lambda}[\vartheta_0]} 
\sin (\vartheta_0(a))}{\lambda \pi a} \Big(
&e^{-\mathcal{H}_a^{\!\Lambda}[\vartheta_0]}\cos(\vartheta_0(a))
+ 
\mathcal{H}_a^{\!\Lambda}\Big[
e^{-\mathcal{H}_\bullet^{\!\Lambda}[\vartheta_0]}
\sin(\vartheta_0(\bullet))
\Big]
\\*
&- \mathcal{H}_0^{\!\Lambda}\Big[
e^{-\mathcal{H}_\bullet^{\!\Lambda}[\vartheta_0]}
\sin(\vartheta_0(\bullet))
\Big]
+\frac{\Lambda^2 a \tilde{C}_{\lambda,\Lambda^2}}{\Lambda^2-a}
\Big)\;.
\end{align*}
From (\ref{Tricomi-28}) we have
$
\mathcal{H}_a^{\!\Lambda}\big[
e^{-\mathcal{H}_\bullet^{\!\Lambda}[\vartheta_0]}
\sin(\vartheta_0(\bullet))
\big]
- \mathcal{H}_0^{\!\Lambda}\big[
e^{-\mathcal{H}_\bullet^{\!\Lambda}[\vartheta_0]}
\sin(\vartheta_0(\bullet))
\big]
= e^{-\mathcal{H}_0^{\!\Lambda}[\vartheta_0]}
\cos(\vartheta_0(0))
- e^{-\mathcal{H}_a^{\!\Lambda}[\vartheta_0]}
\cos(\vartheta_0(a))$.
From (\ref{arctan-0}) one concludes 
\begin{align}
\lim_{p\to
  0}\vartheta_0(p)=\left\{ \begin{array}{cl}
    0 & \text{ for } \lambda \geq 0 \;,\\
    \pi & \text{ for } \lambda<0\;.
\end{array}\right.
\label{vartheta00}
\end{align}
This means
$\cos \vartheta_0(0)=\mathrm{sign}(\lambda)$ so that the two 
formulae (\ref{Ga0-pre-a}) and (\ref{Ga0-pre-b}) lead to
\begin{subequations}
\begin{align}
G_{a0}
&=\frac{e^{\mathcal{H}_0^{\Lambda}[\pi-\vartheta_0]
-\mathcal{H}_a^{\Lambda}[\pi-\vartheta_0]}
\sin (\vartheta_0(a))}{|\lambda| \pi a} 
\label{Gab-second-a}
\\
& \stackrel{*}{=}
\frac{e^{\mathcal{H}_a^{\Lambda}[\vartheta_0]
-\mathcal{H}_0^{\Lambda}[\vartheta_0]}
\sin (\vartheta_0(a))}{|\lambda| \pi a} \Big(
1+\frac{\Lambda^2 a C_{\lambda,\Lambda^2} }{\Lambda^2-a}
\Big) \;.
\label{Gab-second-b}
\end{align}
\end{subequations}
Both lines are \emph{formally equivalent, but they rely on the existence of
$\mathcal{H}_0^{\Lambda}[\pi-\vartheta_0]$ or
$\mathcal{H}_0^{\Lambda}[\vartheta_0]$.} For given $\lambda$, this turns 
out to be the case only for one of the equations.
The limit (\ref{vartheta00}) 
implies 
$e^{-\mathcal{H}_0^{\Lambda}[\vartheta_0]}=
\exp\big(-\frac{1}{\pi} \int_0^{\Lambda^2} \frac{dp}{p}\vartheta_0(p)\big)
\stackrel{\lambda<0}{\longrightarrow 0}$, which means that
(\ref{Gab-second-b}) reduces for $\lambda<0$ to (\ref{Gab-second-a}) 
after undoing the (incorrect) step from  $\tilde{C}_{\lambda,\Lambda^2}$ to
$C_{\lambda,\Lambda^2}$. Similarly, $\lim_{a\to 0}
e^{\mathcal{H}_a^{\Lambda}[\pi-\vartheta_0]}
\stackrel{\lambda>0}{=} \infty$, so that (\ref{Gab-second-a})
is only consistent with $\lambda<0$.  
These results can be summarised as follows:
\begin{Lemma}
\label{Lemma:Ga0}
\begin{subequations}
\begin{align}
G_{a0}  
&=\frac{e^{\mathcal{H}_0^{\!\Lambda}[\pi-\vartheta_0]
-\mathcal{H}_a^{\!\Lambda}[\pi-\vartheta_0]} 
\sin (\pi-\vartheta_0(a))}{|\lambda| \pi a} 
 && \quad \text{ for }\lambda < 0\;, \label{Ga0-a}
\\*[0.5ex]
&\stackrel{*}{=}
\frac{e^{\mathcal{H}_a^{\!\Lambda}[\vartheta_0]
-\mathcal{H}_0^{\!\Lambda}[\vartheta_0]} 
\sin (\vartheta_0(a))}{|\lambda| \pi a} \Big(
1+ \frac{\Lambda^2 a C_{\lambda,\Lambda^2}}{\Lambda^2-a}
\Big) && \quad \text{ for }\lambda > 0\;, 
\label{Ga0-b}
\end{align}
where $C_{\lambda,\Lambda^2}$ is an arbitrary constant. \hfill $\square$%
\end{subequations}
\end{Lemma}
Let us introduce the new angle function 
\begin{align}
\tau_b(a) :=\di{\raisebox{-1.2ex}{\mbox{\normalsize$\arctan$}}}{
\mbox{\scriptsize$[0,\pi]$}}
\Bigg(\dfrac{|\lambda| \pi a}{
b + \frac{1 +
\lambda \pi a \mathcal{H}_a^{\Lambda}[ G_{\bullet 0} ] }{G_{a 0}}}
\Bigg) = \left\{ 
\begin{array}{cl }\vartheta_b(a) & \text{ for } \lambda\geq 0 \;,\\
\pi-\vartheta_b(a) & \text{ for } \lambda<0 \;.
\end{array}\right.
\label{tau}
\end{align}
We have $\tau_b(0)=0$ independent of the sign of $\lambda$, and 
Lemma~\ref{Lemma:Ga0} can be written in the unified from
\begin{align}
G_{a0} &=
\dfrac{\sin(\tau_0(a))}{|\lambda| \pi a}
\mathrm{e}^{\mathrm{sign}(\lambda)(
\mathcal{H}_a^{\Lambda}[\tau_0]-\mathcal{H}_0^{\Lambda}[\tau_0])}
\cdot \left\{
\begin{array}{@{\!}cl@{}}
1
& \text{ for } \lambda <0 \;,
\\
\big(1{+}\frac{\Lambda^2 aC_{\lambda,\Lambda^2}  }{\Lambda^2{-}a}\big)
& \text{ for } \lambda > 0 \;.
\end{array}\right.
\label{Ga0}
\end{align}

In the next step we use the result of Lemma~\ref{Lemma:Ga0} to
explicitly compute $G_{ab}=G_{a0}+\frac{b}{a} D_{ab}$ with $D_{ab}$ given by 
(\ref{solution-Dab-a}) and (\ref{solution-Dab-b}), respectively.
A key is the addition theorem 
\begin{align}
\lambda \pi a \sin\big( \vartheta_d(a)-\vartheta_b(a)\big)
=  (b-d) \sin \vartheta_b(a)\sin \vartheta_d(a)
\label{add-thm}
\end{align}
obtained by insertion of (\ref{arctan-0}) into
$\cot \vartheta_b(a)-\cot \vartheta_d(a)$. 
For $\lambda<0$ we thus have in (\ref{solution-Dab-a})
\begin{subequations}
\begin{align}
G_{ab} &= G_{a0}- 
\frac{b \sin (\vartheta_b(a)) 
e^{-\mathcal{H}_a^{\!\Lambda}[\pi-\vartheta_b]}}{\lambda\pi a^2}  \Big(
e^{\mathcal{H}_a^{\!\Lambda}[\pi-\vartheta_b]
-\mathcal{H}_a^{\!\Lambda}[\pi-\vartheta_0]
+\mathcal{H}_0^{\!\Lambda}[\pi-\vartheta_0]} 
\frac{\cos(\vartheta_b(a))\sin(\vartheta_0(a))}{|\lambda|\pi }
\nonumber
\\
&\hspace*{2cm} + \frac{1}{|\lambda|\pi}
\mathcal{H}_a^{\!\Lambda}\big[
e^{\mathcal{H}_\bullet^{\!\Lambda}[\pi-\vartheta_b]
-\mathcal{H}_\bullet^{\!\Lambda}[\pi-\vartheta_0]
+\mathcal{H}_0^{\!\Lambda}[\pi-\vartheta_0]} 
\sin(\vartheta_0(\bullet))   
\sin(\vartheta_b(\bullet)) \big] -C'_{b,\lambda,\Lambda^2}\Big)
\nonumber
\\
&= \frac{e^{\mathcal{H}_0^{\!\Lambda}[\pi-\vartheta_0]
-\mathcal{H}_a^{\!\Lambda}[\pi-\vartheta_0]}}{|\lambda|\pi a}
\big( \sin \vartheta_0(a)-\cos \vartheta_b(a)
\sin (\vartheta_0(a)-\vartheta_b(a))\big)
\nonumber
\\
& - \frac{\sin (\vartheta_b(a)) 
e^{\mathcal{H}_0^{\!\Lambda}[\pi-\vartheta_0]-
\mathcal{H}_a^{\!\Lambda}[\pi-\vartheta_b]}}{|\lambda|\pi a^2} 
\mathcal{H}_a^{\!\Lambda}\big[
e^{\mathcal{H}_\bullet^{\!\Lambda}[\vartheta_0-\vartheta_b]}
(\bullet-a+a) \sin (\vartheta_0(\bullet)-\vartheta_b(\bullet))\big]
\nonumber
\\
&+ C'_{b,\lambda,\Lambda^2}\frac{b \sin (\vartheta_b(a)) 
e^{-\mathcal{H}_a^{\!\Lambda}[\pi-\vartheta_b]}}{\lambda\pi a^2} 
\nonumber
\\
&=\frac{\sin (\vartheta_b(a)) 
e^{\mathcal{H}_0^{\!\Lambda}[\pi-\vartheta_0]-
\mathcal{H}_a^{\!\Lambda}[\pi-\vartheta_b]}}{|\lambda|\pi a}
\Big(1 +  \frac{\tilde{C}'_{b,\lambda,\Lambda^2}}{a}\Big)\;,
\label{Gab-a-pre}
\end{align}
where 
$\displaystyle\tilde{C}'_{b,\lambda,\Lambda^2}:=b C'_{b,\lambda,\Lambda^2} \,
 \mathrm{sign}(\lambda)e^{-\mathcal{H}_0^{\!\Lambda}[\pi-\vartheta_0]}-
\frac{1}{\pi} \int_0^{\Lambda^2} \!\!\!dp\;
e^{\mathcal{H}_p^{\!\Lambda}[\vartheta_0-\vartheta_b]}
\sin (\vartheta_0(p)-\vartheta_b(p))$.
We have used (\ref{Tricomi-18}) and standard trigonometric addition
theorems to arrive at the last line of (\ref{Gab-a-pre}).
Existence of $\lim_{a\to 0} G_{ab}$ imposes 
$\tilde{C}'_{b,\lambda,\Lambda^2}=0$.

For $\lambda >0$ we combine (\ref{solution-Dab-b}) with 
(\ref{Ga0-b}) to obtain
\begin{align}
G_{ab} &= 
G_{a0}-\frac{b \sin (\vartheta_b(a)) 
e^{\mathcal{H}_a^{\!\Lambda}[\vartheta_b]}}{\lambda\pi a}
\Big( 
\frac{
e^{-\mathcal{H}_a^{\!\Lambda}[\vartheta_b]+\mathcal{H}_a^{\!\Lambda}[\vartheta_0]
-\mathcal{H}_0^{\!\Lambda}[\vartheta_0]}}{|\lambda|\pi a} 
\Big(1+ \frac{\Lambda^2 a C_{\lambda,\Lambda^2}}{\Lambda^2-a}
\Big)\sin \vartheta_0(a)\cos \vartheta_b(a)
\nonumber
\\
&+ 
\frac{e^{-\mathcal{H}_0^{\!\Lambda}[\vartheta_0]}}{|\lambda|\pi}
\mathcal{H}_a^{\!\Lambda}
\Big[
e^{\mathcal{H}_\bullet^{\!\Lambda}[\vartheta_0-\vartheta_b]}
\frac{\sin(\vartheta_0(\bullet)) \sin(\vartheta_b(\bullet))}{\bullet}
\Big(1+ \frac{\Lambda^2 C_{\lambda,\Lambda^2} \bullet}{
\Lambda^2-\bullet}\Big)
\Big]- \frac{\Lambda^2 C_{b,\lambda,\Lambda^2}}{\Lambda^2-a}\Big)
\nonumber
\\
&= \frac{
e^{\mathcal{H}_a^{\!\Lambda}[\vartheta_0]
-\mathcal{H}_0^{\!\Lambda}[\vartheta_0]}}{|\lambda|\pi a} 
\Big(1+ \frac{\Lambda^2 a C_{\lambda,\Lambda^2}}{\Lambda^2-a}
\Big)\Big( \sin \vartheta_0(a)-\cos \vartheta_b(a)
\sin (\vartheta_0(a){-}\vartheta_b(a))\Big)
\nonumber
\\
&-\frac{
e^{\mathcal{H}_a^{\!\Lambda}[\vartheta_b]-\mathcal{H}_0^{\!\Lambda}[\vartheta_0]}
\sin (\vartheta_b(a)) }{|\lambda|\pi a}
\mathcal{H}_a^{\!\Lambda}
\Big[
e^{\mathcal{H}_\bullet^{\!\Lambda}[\vartheta_0-\vartheta_b]}
\sin(\vartheta_0(\bullet){-}\vartheta_b(\bullet))
\Big(1+ \frac{\Lambda^2 C_{\lambda,\Lambda^2} (\bullet-\Lambda^2+\Lambda^2)}{
\Lambda^2-\bullet}\Big)
\Big]
\nonumber
\\
& +
\frac{b e^{\mathcal{H}_a^{\!\Lambda}[\vartheta_b]}
\sin (\vartheta_b(a)) }{\lambda\pi a} 
\frac{\Lambda^2 C_{b,\lambda,\Lambda^2}}{\Lambda^2-a}
\nonumber
\\
&= \frac{
e^{\mathcal{H}_a^{\!\Lambda}[\vartheta_b]-\mathcal{H}_0^{\!\Lambda}[\vartheta_0]}
\sin (\vartheta_b(a)) }{|\lambda|\pi a}
\Big(1-\Lambda^2 C_{\lambda,\Lambda^2}
- \frac{b \,\mathrm{sign}(\lambda) \Lambda^2 C_{b,\lambda,\Lambda^2} 
e^{\mathcal{H}_0^{\!\Lambda}[\vartheta_0]}}{\Lambda^2-a}
\Big)\;.
\label{Gab-b-pre}
\end{align}
\end{subequations}
To obtain the last line we have used both (\ref{Tricomi-18}) and 
(\ref{Tricomi-20}). The prefactor of 
$\big(1+ \frac{\Lambda^2 a C_{\lambda,\Lambda^2}}{\Lambda^2-a}
\big)$ vanishes by trigonometric addition theorems.
For $b=0$ the final formula must coincide with 
(\ref{Ga0-b}) which imposes
$C_{b,\lambda,\Lambda^2}= 
\mathrm{sign}(\lambda) 
e^{-\mathcal{H}_0^{\!\Lambda} [\vartheta_0]}
\Big(\frac{\Lambda^2 C_{\lambda,\Lambda^2}}{b}
+f_{\lambda,\Lambda}(b)\Big)$, where $f_{\lambda,\Lambda}$ is an
arbitrary function with $\lim_{b\to 0} b f_{\lambda,\Lambda}(b)=0$.

We can summarise (\ref{Gab-a-pre}) and 
(\ref{Gab-b-pre}) and the corresponding discussion of the limit $a\to
0$ in terms of the angle function $\tau_b(a)$ as follows:
\begin{Proposition}
\label{prop:Gab}
In terms of the function $\tau_b(a)$ of the boundary $2$-point
function $G_{a0}$, see (\ref{tau}), the full $2$-point function
is given by
\begin{align}
G_{ab} &=
\frac{e^{\mathrm{sign}(\lambda)(\mathcal{H}_a^{\!\Lambda}[\tau_b]
-\mathcal{H}_0^{\!\Lambda}[\tau_0])} 
\sin (\tau_b(a))}{|\lambda| \pi a} 
\cdot
\left\{\begin{array}{cl}
\Big(
1+ \frac{\Lambda^2 (a C_{\lambda,\Lambda^2}+b
  f_{\lambda,\Lambda^2}(b))}{
\Lambda^2-a}
\Big) \quad & \text{for } \lambda>0\;, \\
1 \quad & \text{for } \lambda<0\;,
\end{array}\right.
\label{Gab}
\end{align}
where $C_{\lambda,\Lambda^2}$ is an arbitrary constant and 
$f_{\lambda,\Lambda^2}$ an arbitrary function 
with $\lim_{b\to 0} b f_{\lambda,\Lambda}(b)=0$. \hfill $\square$%
\end{Proposition}
The limit $a\to 0$ of (\ref{Gab}) reads
\begin{align}
G_{0b} &=
\frac{e^{\mathrm{sign}(\lambda)(\mathcal{H}_0^{\!\Lambda}[\tau_b-\tau_0])}}{1+b} 
\cdot \left\{\begin{array}{cl}
\big(1+ bf_{\lambda,\Lambda^2}(b)\big) \quad & \text{for } \lambda>0\;, \\
1 \quad & \text{for } \lambda<0\;.
\end{array}\right.
\label{G0b}
\end{align}

Proposition \ref{prop:Gab} fills a gap in \cite{Grosse:2012uv}. We
knew that the freedom parametrised by constants $C,C'$ in the Carleman
solution in Proposition~\ref{Prop:Carleman} will influence the 2-point
function, but we ignored this possibility in \cite[Assumption
4.2]{Grosse:2012uv}.  Proposition~\ref{prop:Gab} tells us that this
Assumption is justified for $\lambda<0$ provided that the angle
function is suitably reflected $\vartheta_b(a)\mapsto \tau_b(a)$ for
$\lambda<0$ so that it vanishes at $a=0$. This vanishing at $0$ was
used in the perturbative expansion \cite[Appendix B]{Grosse:2012uv}
which agreed with a Feynman graph calculation. In terms of
$\tau_b(a)$, agreement with the Feynman graph expansion shows that
$C_{\lambda,\Lambda^2}$ and $f_{\lambda,\Lambda^2}$ are zero in
perturbation theory. If these happen to be not identically zero (as we
show by a numerical simulation), these must be flat functions of
$\lambda$, i.e.\ all derivatives of $C_{\lambda,\Lambda^2},
f_{\lambda,\Lambda^2}$ with respect to $\lambda$ vanish at
$\lambda=0$.  This suggest a phase transition of infinite order
$C_{\lambda,\Lambda^2},f_{\lambda,\Lambda^2} \left\{
  \begin{array}{ll} =0 & \text{ for } \lambda \leq 0 \;,\\
    \propto e^{-\frac{1}{\lambda}}
& \text{ for } \lambda > 0\;.
\end{array}\right.$

\subsection{Consisteny relations for the boundary function 
$G_{a0}$}

\label{sec:master}

Equation~(\ref{Gab}) gives the full two-point function $G_{ab}$ in terms
of the boundary $G_{a0}$. The boundary function should be obtained
from the equation symbolised by $\mathcal{I}_a[G_{\bullet 0}]$ in the
introduction to sec.~\ref{sec:2}. This equation is 
\cite[eqs.\ (4.33)+(4.17)]{Grosse:2012uv}:
\begin{align}               
a - \frac{1}{G_{a0}} + 1
&= -
\dfrac{\displaystyle\lambda 
\int_0^{\Lambda^2} \! q\,dq \;( G_{aq}-G_{0q})}{\displaystyle
1-\lambda \int_0^{\Lambda^2} dp \; G_{p0}}
-\lambda \int_0^{\Lambda^2} dp \; 
\frac{a-a\frac{G_{p0}}{G_{a0}}}{(p-a)}\;,
\end{align}
which we rewrite as 
\begin{align}               
&\Big((1+a)G_{a0} - 1
-\lambda \pi a \mathcal{H}^{\!\Lambda}_a[G_{\bullet 0}]
\Big)\Big(1-\lambda \int_0^{\Lambda^2} dp \; G_{p0}\Big)
\nonumber
\\
&= -
\lambda G_{a0}
\int_0^{\Lambda^2} \! q\,dq \;( G_{aq}-G_{0 q})
-\lambda \pi a G_{a0}\mathcal{H}^{\!\Lambda}_a[1]
\Big(1-\lambda \int_0^{\Lambda^2} dp \; G_{p0}\Big)\;.
\label{calT}
\end{align}
We show that this consistency condition gives no other information 
than symmetry $G_{ab}=G_{ba}$.

\begin{Lemma}
The solution (\ref{Gab}) implies
\begin{subequations}
\label{HGab}
\begin{align}
\lambda\pi \cot \vartheta_b(a) \cdot G_{ab} 
-\lambda \pi \mathcal{H}^{\!\Lambda}_a[G_{\bullet b}]
&=\frac{(1+b)G_{0b}}{a}
\qquad \text{ or}
\label{HGab-first}
\\
\lambda\pi a \big(G_{a0}\mathcal{H}^{\!\Lambda}_a[G_{\bullet b}]
-G_{ab}\mathcal{H}^{\!\Lambda}_a[G_{\bullet 0}]\big)
&= b(G_{ab}-G_{0b}) G_{a0}
+(G_{ab}-G_{a0} G_{0b})
\label{HGab-expl}
\end{align}
\end{subequations}
and
\begin{align}
1+\lambda \int_0^{\infty} dp\;(G_{pb}-G_{p0})=(1+b)G_{0b}\;.
\label{conj-b}
\end{align}
\end{Lemma}
\emph{Proof.}
Let $\Theta$ be the step function $\Theta(\lambda)=1$ for $\lambda>0$
and $\Theta(\lambda)=0$ for $\lambda<0$.
The Hilbert transform of (\ref{Gab}) reads with (\ref{Tricomi})
and rational fraction expansion
\begin{align}
\mathcal{H}^{\!\Lambda}_a[G_{\bullet b}]
&= \Theta(\lambda)
\frac{\cos \tau_b(a) \Lambda^2C_{\lambda,\Lambda^2}}{
|\lambda|\pi (\Lambda^2-a)} \mathrm{e}^{\mathrm{sign}(\lambda)
(\mathcal{H}_a^{\Lambda}[\tau_b]-\mathcal{H}_0^{\Lambda}[\tau_0])}
\nonumber
\\
& + \frac{\mathrm{e}^{-\mathrm{sign}(\lambda)
\mathcal{H}_0^{\Lambda}[\tau_0]}}{|\lambda|\pi a}
\Big(
\mathcal{H}^{\!\Lambda}_a\big[
\sin(\tau_b(\bullet))\mathrm{e}^{\mathrm{sign}(\lambda)
\mathcal{H}_\bullet^{\Lambda}[\tau_b]}
(1+\Theta(\lambda)\tfrac{\Lambda^2 bf_{\lambda,\Lambda^2}(b)}{
\Lambda^2-\bullet} )\big] 
\nonumber
\\
&\qquad\qquad - 
\mathcal{H}^{\!\Lambda}_0\big[
\sin(\tau_b(\bullet))
\mathrm{e}^{\mathrm{sign}(\lambda)
\mathcal{H}_\bullet^{\Lambda}[\tau_b]}
(1+\Theta(\lambda)\tfrac{\Lambda^2 bf_{\lambda,\Lambda^2}(b)}{
\Lambda^2-\bullet} )\big] 
\nonumber
\\
&=  \Theta(\lambda)
\frac{\cos \tau_b(a) \Lambda^2C_{\lambda,\Lambda^2}}{
|\lambda|\pi (\Lambda^2-a)} \mathrm{e}^{\mathrm{sign}(\lambda)
(\mathcal{H}_a^{\Lambda}[\tau_b]-\mathcal{H}_0^{\Lambda}[\tau_0])}
\nonumber
\\
&+ \frac{\mathrm{e}^{-\mathrm{sign}(\lambda)
\mathcal{H}_0^{\Lambda}[\tau_0]}}{|\lambda| \pi a}
\Big(\mathrm{sign}(\lambda)\cos (\tau_b(a))
\mathrm{e}^{\mathrm{sign}(\lambda)
\mathcal{H}_a^{\Lambda}[\tau_b]}
(1+\Theta(\lambda)\tfrac{\Lambda^2 bf_{\lambda,\Lambda^2}(b)}{
\Lambda^2-a} )
\nonumber
\\
& \qquad\qquad
-\mathrm{sign}(\lambda)
\mathrm{e}^{\mathrm{sign}(\lambda)
\mathcal{H}_0^{\Lambda}[\tau_b]}
(1+\Theta(\lambda)bf_{\lambda,\Lambda^2}(b))
\nonumber
\\
&= \frac{\cos \vartheta_b(a)}{|\lambda|\pi a}
\mathrm{e}^{\mathrm{sign}(\lambda)
(\mathcal{H}_a^{\Lambda}[\tau_b]
- \mathcal{H}_0^{\Lambda}[\tau_0])}
\Big(1+\Theta(\lambda) \frac{\Lambda^2(C_{\lambda,\Lambda^2} a
+bf_{\lambda,\Lambda^2}(b))}{\Lambda^2-a} \Big)
\nonumber
\\
&-\frac{
\mathrm{e}^{\mathrm{sign}(\lambda)
(\mathcal{H}_0^{\Lambda}[\tau_b]-
\mathcal{H}_0^{\Lambda}[\tau_0])}
(1+\Theta(\lambda) bf_{\lambda,\Lambda^2}(b) )
}{\lambda\pi a}
\nonumber
\\
& = \cot \vartheta_b(a) \cdot G_{ab} 
-\frac{(1+b)G_{0b}}{\lambda\pi a}\;,
\end{align}
which can be rearranged to (\ref{HGab-first}). 
We have used $\cos\tau_b(a)=\mathrm{sign}(\lambda) 
\cos\vartheta_b(a)$. Equation 
(\ref{HGab-expl}) then results from (\ref{arctan-0}). 
Integration of (\ref{Gab}) over $a=p$ gives with (\ref{Tricomi})
\begin{align}
&\int_0^{\Lambda^2} \!\!\! dp \;G_{pb}
\nonumber
\\
&= \mathcal{H}^{\!\Lambda}_0 \Big[ 
\frac{\sin(\tau_b(\bullet))}{|\lambda|}
\mathrm{e}^{\mathrm{sign}(\lambda)(
\mathcal{H}_\bullet^{\Lambda}[\tau_b]-
\mathcal{H}_0^{\Lambda}[\tau_0])}
\Big(1+\Theta(\lambda)\Big(-C_{\lambda,\Lambda^2} \Lambda^2 + 
\frac{\Lambda^2(C_{\lambda,\Lambda^2} \Lambda^2{+} 
bf_{\lambda,\Lambda^2}(b))}{\Lambda^2-\bullet}\Big)\Big)\Big]
\nonumber
\\*
& =
\frac{\cos(\tau_b(0))}{\lambda}
\mathrm{e}^{\mathrm{sign}(\lambda)(
\mathcal{H}_0^{\Lambda}[\tau_b]-
\mathcal{H}_0^{\Lambda}[\tau_0])}
\big(1 + \Theta(\lambda) bf_{\lambda,\Lambda^2}(b)\big)
-\frac{(1-\Theta(\lambda) C_{\lambda,\Lambda^2} \Lambda^2 )}{\lambda}
\mathrm{e}^{-\mathrm{sign}(\lambda)
\mathcal{H}_0^{\Lambda}[\tau_0]}
\nonumber
\\*
&= \frac{(1+b)G_{0b}}{\lambda} 
- \frac{(1-\Theta(\lambda) C_{\lambda,\Lambda^2}\Lambda^2)}{
\lambda}\mathrm{e}^{-\mathrm{sign}(\lambda)
\mathcal{H}_0^{\Lambda}[\tau_0]}\;.
\label{Int-Gpb}
\end{align}
Subtraction of the same equation at $b=0$ gives the assertion
(\ref{conj-b}). 
\hfill $\square$%

\bigskip

As by-product we obtain, setting $b=0$ in (\ref{Int-Gpb}), 
for the wavefunction renormalisation 
$Z(1+\mathcal{Y})$ given by \cite[eq.\ (4.17)]{Grosse:2012uv} the formula
\begin{align}
\frac{1}{Z(1+\mathcal{Y})} = 1-\lambda \int_0^{\Lambda^2} \!\!dp
\;G_{p0} = \left\{
\begin{array}{cl} 
e^{\mathcal{H}_0^{\Lambda}[\tau_0]} & \quad\text{for } \lambda<0 \;,\\
(1-C_{\lambda,\Lambda^2}\Lambda^2) 
e^{-\mathcal{H}_0^{\Lambda}[\tau_0]} & \quad\text{for } \lambda>0\;.
\end{array}\right.
\label{Z}
\end{align}
This shows, in contrast to the interpretation in \cite{Grosse:2012uv},
that \emph{$Z$ is positive for $\lambda<0$ (assuming $1+\mathcal{Y}>0$) but 
negative for $\lambda>0$ (assuming
$C_{\lambda,\Lambda^2}=\mathcal{O}(1)$)}. 
We conclude that the `wrong sign' $\lambda<0$ is the good phase, and
$\lambda>0$ is the bad phase. This conclusion will repeatedly be
confirmed throughout this paper.

We return to (\ref{calT}). 
Integrating (\ref{HGab-expl}) over $b=q$ and multiplying by
$\lambda$ gives
\begin{align}
&-\lambda G_{a0} \int_0^{\Lambda^2} \!\! dq\; q(G_{aq}-G_{0q}) 
+\lambda\pi a G_{a0}\mathcal{H}^{\!\Lambda}_a[
\lambda {\textstyle\int_0^{\Lambda^2}} dq\;G_{\bullet q}]
\nonumber
\\
&=(1+\lambda\pi a \mathcal{H}_a^{\!\Lambda}[G_{\bullet 0}] ) 
\lambda \int_0^{\Lambda^2} \!\! dq\;G_{aq}
-\lambda G_{a0} \int_0^{\Lambda^2} \!\! dq\;G_{0q}\;.
\label{HGab-int}
\end{align}
Let us define a function $f$ by 
\begin{align}
\lambda \int_0^{\Lambda^2} dq\;G_{aq}=:
f(a)+(1+a)G_{a0}-\Big(1-\lambda \int_0^{\Lambda^2} dq \;G_{0q}\Big)
\;. \label{conj-a}
\end{align}
For $f(a)=0$ this is (\ref{conj-b}) for exchanged indices 
$G_{ab}\mapsto G_{ba}$. For the moment we keep $f(a)$ arbitrary in order 
to derive $f(a)=0$ from (\ref{calT}).
Equation (\ref{HGab-int}) reads
\begin{align}
&-\lambda G_{a0} \int_0^{\Lambda^2}\!\! dq\; q(G_{aq}-G_{0q}) 
-\Big(1-\lambda \int_0^{\Lambda^2} \!\! dq \;G_{0q}\Big)
\lambda\pi a G_{a0}\mathcal{H}^{\!\Lambda}_a[1]
\nonumber
\\
&=
-\lambda\pi a G_{a0}\mathcal{H}^{\!\Lambda}_a
\big[
f(\bullet)+(1+\bullet)G_{\bullet 0}\big]
-\lambda G_{a0} \int_0^{\Lambda^2} \!\! dq\;G_{0q}
\nonumber
\\
&+(1+\lambda\pi a  \mathcal{H}_a^{\!\Lambda}[G_{\bullet 0}] ) 
\Big(f(a)+(1+a)G_{0a}-\Big(1-\lambda \int_0^{\Lambda^2} \!\!dq \;G_{0q}\Big)\Big)
\;.
\end{align}
This provides an alternative formula for the rhs of (\ref{calT}). 
Rewriting $
\mathcal{H}^{\!\Lambda}_a[\bullet G_{\bullet 0}]
=a\mathcal{H}^{\!\Lambda}_a[G_{\bullet 0}]
+\frac{1}{\pi} \int_0^{\Lambda^2} dp \; G_{p0}$ 
and using the symmetry $G_{0p}=G_{p0}$
we arrive at
\begin{align}               
&\Big((1+a)G_{a0} - 1
-\lambda \pi a \mathcal{H}^{\!\Lambda}_a[G_{\bullet 0}]
\Big)\Big(1-\lambda \int_0^{\Lambda^2} \!\! dp \; G_{p0}\Big)
\nonumber
\\
&= 
-\lambda\pi a G_{a0}\mathcal{H}^{\!\Lambda}_a[f(\bullet)]
-\lambda\pi a(1+a) G_{a0}\mathcal{H}^{\!\Lambda}_a\big[G_{\bullet 0}\big]
-\lambda (1+a) G_{a0}\int_0^{\Lambda^2} \!\! dp \;G_{p0}
\nonumber
\\
&+(1+\lambda\pi a  \mathcal{H}_a^{\!\Lambda}[G_{\bullet 0}] ) 
\Big(f(a)+(1+a)G_{0a}-\Big(1-\lambda \int_0^{\infty} \!\! dp \;G_{p0}\Big)\Big)
\;,
\end{align}
which reduces to 
\begin{align}
\lambda\pi \cot \vartheta_0(a)
\cdot f(a)
-\lambda\pi \mathcal{H}^{\!\Lambda}_a[f(\bullet)]=0\;.
\label{Carleman-hom}
\end{align}
In other words, the consistency equation (\ref{calT}) reduces to a homogeneous 
Carleman equation for $f(a)$. Symmetry $G_{ab}=G_{ba}$ implies $f(a)=0$, and 
(\ref{Carleman-hom}) is automatically fulfilled. Conversely,
we proved in sec.~\ref{sec:winding} that for $\lambda<0$ the homogeneous 
Carleman equation only has the trivial solution. For $\lambda>0$ equation
(\ref{Carleman-hom}) could have solutions
\[
f(a)= \tilde{C} \frac{\sin(\tau_b(a))}{|\lambda|\pi (\Lambda^2-a)}
\mathrm{e}^{\mathrm{sign}(\lambda)(
\mathcal{H}_a^{\Lambda}[\tau_b]-\mathcal{H}_0^{\Lambda}[\tau_0])}
\]
which contradict symmetry in case of $\tilde{C}\neq 0$.
In summary, the consistency equation (\ref{calT}) contains no other 
information than (\ref{Gab}) plus the symmetry requirement 
$G_{ab}=G_{ba}$.

Always for $\lambda <0$ and for $\lambda >0$ in a region (if existent)
where $f_{\lambda,\Lambda^2}=0$ we can use the symmetry requirement
$G_{ab}=G_{ba}$, in particular $G_{0b}=G_{b0}$, to turn (\ref{G0b})
into the fixed point problem 
\begin{align}
f_{\lambda,\Lambda^2}=0\quad \Rightarrow\quad G_{b0}=G_{0b}
& =  
\frac{1}{1{+}b}
\exp\Bigg(\!
{-} \lambda \!
\int_0^b \!\! dt \! \int_0^{\Lambda^2}  \!\!\!\!\!
\frac{dp}{(\lambda \pi p)^2 
+\big( t {+} \frac{1 {+}\lambda \pi p 
 \mathcal{H}_p^{\Lambda}[G_{\bullet 0}]}{G_{p0}}\big)^2} 
\Bigg)\;.
\label{master}
\end{align}
This formula holds independently of the sign of $\lambda$, and in this
way we rigorously confirm \cite[eq.\ (4.37)]{Grosse:2012uv} for
$\lambda<0$. As shown in \cite{Grosse:2015fka}, (\ref{master}) has for
$-\frac{1}{6}\leq \lambda\leq 0$ a solution in $\exp
\mathcal{K}_\lambda$, with $\mathcal{K}_\lambda$ given in (\ref{Klambda}).
For $\lambda>0$ but $f_{\lambda,\Lambda^2}=0$, we can still use (\ref{master}) 
to define $G_{0b}$, and the symmetry 
condition $G_{0a}=G_{a0}$ is actually an equation for the 
constant $C_{\lambda,\Lambda^2}$:
\begin{align}
\frac{e^{-\mathcal{H}_0^{\!\Lambda}[\tau_0-\tau_a]}}{1+a} 
&=
\frac{e^{\mathcal{H}_a^{\!\Lambda}[\tau_0]
-\mathcal{H}_0^{\!\Lambda}[\tau_0]} 
\sin (\tau_0(a))}{\lambda \pi a} \Big(
1+ \frac{\Lambda^2 a C_{\lambda,\Lambda^2}}{
\Lambda^2-a}
\Big)
\nonumber
\\
\Rightarrow \qquad 
C_{\lambda,\Lambda^2}
&= \Big(1-\frac{a}{\Lambda^2}\Big)
\frac{e^{\mathcal{H}_0^{\!\Lambda}[\tau_a]-\mathcal{H}^{\!\Lambda}_a
    [\tau_0]}
\sqrt{\big(\frac{\lambda\pi a}{1+a}\big)^2+ \big(\frac{1+\lambda\pi a 
\mathcal{H}^{\!\Lambda}_a[G_{0\bullet}]}{(1+a)G_{0a}} \big)^2}-1}{a}\;.
\label{ClL}
\end{align}
If this is not a constant function of $a$, then the assumption
$f_{\lambda,\Lambda^2}=0$ was wrong.

\section{Integral formulae for the derivative}

\label{sec:intformula}

\subsection{Stieltjes functions}

In \cite{Grosse:2013iva} we have identified a limit in which the
matrix correlation functions constructed in \cite{Grosse:2012uv}
converge to (connected) Schwinger functions in position space:
\begin{align}
&\mathcal{S}_c(\mu x_1,\dots,\mu x_N)
\nonumber
\\*
&= \frac{1}{64\pi^2}\!\!\!\!\!
\sum_{\di{N_1{+}\dots{+}N_B=N}{N_\beta\,\mathrm{even}}}
\sum_{\sigma \in \mathrm{S}_N} \!\!
\bigg(\prod_{\beta=1}^B \!\frac{4^{N_\beta}}{N_\beta}\!
\int_{\mathbb{R}^4} \frac{dp_\beta}{4\pi^2\mu^4}
\;e^{\mathrm{i}\big\langle
\frac{p_\beta}{\mu},\sum_{i=1}^{N_\beta} ({-}1)^{i{-}1} \mu
x_{\sigma(N_1{+}\dots{+}N_{\beta-1}+i)}\big\rangle} \bigg)
\nonumber
\\*[-0.5ex]
&\qquad\quad \times
{\mbox{\Large$G$}}_{\!\!{\underbrace{\tfrac{\|p_1\|^2}{
2\mu^2(1+\mathcal{Y})},\cdots,
\tfrac{\|p_1\|^2}{2\mu^2(1+\mathcal{Y})}}_{N_1}}\big| \dots \big|
{\underbrace{\tfrac{\|p_B\|^2}{2\mu^2(1+\mathcal{Y})},\cdots,
\tfrac{\|p_B\|^2}{2\mu^2(1+\mathcal{Y})}}_{N_B}}}\;.
\label{Schwinger-final}
\end{align}
Here, $\mu$ defines the mass scale so that (\ref{Schwinger-final})
only involves densities, and
$1+\mathcal{Y}:=-\frac{dG_{0b}}{db}\big|_{b=0}$ 
is the finite wavefunction renormalisation. These Schwinger
functions satisfy the Osterwalder-Schrader \cite{Osterwalder:1974tc} axioms
\emph{(OS3) permutation symmetry} for trivial reasons but also 
\emph{(OS1) Euclidean invariance}, which is highly surprising for a
field theory on noncommutative Moyal space. The axiom 
\emph{(OS4) clustering} is not satisfied, but also not strictly required.

In this section we prove integral equations for the partial
derivatives $\frac{\partial^{n+\ell}G_{ab}}{\partial a^n\partial^\ell
  b}$ of the matrix 2-point function (\ref{Gab}) assuming
$C_{\lambda,\Lambda^2}=f_{\lambda,\Lambda^2}=0$ (which is the case for
$\lambda<0$). On one hand this establishes
explicit factorial growth 
$\big|\frac{\partial^{n+\ell}G_{ab}}{\partial a^n\partial^\ell
  b}\big| \leq C_{n\ell} n!\ell!$. Bounds on $C_{n\ell}$ are left for
future work, but already at this point a bound of the type 
$\big|\frac{\partial^{n+\ell}G_{ab}}{\partial a^n\partial^\ell
  b}\big| \leq C (n!\ell!)^\alpha$ is plausible. Together with the
recursion formulae for higher correlation functions 
\cite{Grosse:2012uv}, such bounds would be enough to prove
the axiom \emph{(OS0) growth conditions}. 

In this paper we focus on another application of integral formulae for 
$\frac{\partial^{n+\ell}G_{ab}}{\partial a^n\partial^\ell b}$.
We have shown in \cite{Grosse:2013iva} that 
the Schwinger 2-point function satisfies the axiom
\emph{(OS2) reflection
positivity} iff the diagonal matrix 2-point
function $a\mapsto G_{aa}$ is a Stieltjes function, i.e.\ 
\begin{align}
G_{aa}=\int_0^\infty  \frac{d\rho(t)}{a+t}
\label{Gaa-Stieltjes}
\end{align}
for some positive non-decreasing function $\rho$. This is essentially
a consequence of the K\"all\'en-Lehmann spectral representation. 

Stieltjes functions form an important subclass of the class
$\mathcal{C}$ of \emph{completely
monotonic functions}. We refer to \cite{Berg:2008??} for an overview
about completely monotonic functions and their relations to other
classes of functions. The class $\mathcal{C}$ characterises
the positive definite functions on $\mathbb{R}_+$, i.e.\ 
for any $x_1,\dots,x_n \geq 0$ the
matrix $a_{ij}=f(x_i+x_j)$, with $f\in \mathcal{C}$, 
is positive (semi-)definite. A function $f:\mathbb{R}_+\to \mathbb{R}$
is positive definite, bounded and continuous if and only if 
it is the Laplace transform of a positive finite measure, 
$\displaystyle f(x)=\int_0^\infty e^{-xt} d\mu(t)$. This
representation provides a unique analytic continuation of such
functions to the half space $\mathrm{Re}(z) > 0$. Remarkably, such
analyticity is a consequence of the purely real conditions 
$(-1)^n f^{(n)}(x)\geq 0$ for all $n\in \mathbb{N}$ and $x>0$. 

The Stieltjes integral (\ref{Gaa-Stieltjes}) provides a unique
analytic continuation of a Stieltjes function to the cut plane 
$\mathbb{C} \setminus {]{-}\infty,0[}$.  Remarkably again, 
this analyticity can be tested by purely real conditions identified by 
Widder \cite{Widder:1938??}: A smooth non-negative function $f$ on
$\mathbb{R}_+$ is Stieltjes iff $L_{n,t}[f(\bullet)]\geq 0$ for all
$n\in \mathbb{N}$ and $t\in \mathbb{R}_+$, where
$L_{0,t}[f(\bullet)]=f(t)$,
$L_{1,t}[f(\bullet)]=\frac{d}{dt}\big(tf(t)\big)$
and  
\begin{align}
L_{n,t}[f(\bullet)]
:= \frac{(-t)^{n-1}}{n!(n-2)!} \frac{d^{2n-1}}{dt^{2n-1}}\big( 
t^n f(t)\big)\;,\qquad n\geq 2\;.
\label{Widder}
\end{align}
If Widder's criterion is satisfied, the sequence
$\{L_{n,t}[f(\bullet)]\}$ converges for $n\to \infty$
in distributional sense and 
almost everywhere to the measure function of the Stieltjes transform, 
\begin{align}
f(x)=\int_0^\infty \frac{\rho'(t) \,dt}{t+x}\;,\qquad 
\int_0^T\;\rho'(t) \,dt= \lim_{n\to \infty} 
\int_0^T dt\;L_{n,t}[f(\bullet)] \quad
\text{ a.e.}\;.
\end{align}

\subsection{Derivatives of the 2-point function}

In \cite{Grosse:2014nza} we have given first results on
$L_{n,t}[G_{\bullet\bullet}]$ based on numerically obtained
interpolations and $n\leq 4$. For larger $n$ this method becomes too
noisy so that integral formulae for derivatives of $G_{ab}$ 
become indispensable. 

We start from (\ref{Gab}) which we write for
$C_{\lambda,\Lambda^2}=f_{\lambda,\Lambda^2}(b)=0$, which is always the
case for $\lambda<0$, as 
\begin{align}
\log G_{ab}&= 
-\mathrm{sign}(\lambda)\mathcal{H}_0^{\!\Lambda}[\tau_0]
+\mathrm{sign}(\lambda) \mathcal{H}_a^{\!\Lambda}[\tau_b]
-\frac{1}{2} \log \Big((\lambda\pi a)^2 + \Big(b+\frac{1+\lambda\pi a 
\mathcal{H}_a^{\!\Lambda}[G_{\bullet 0}]}{G_{a0}}\Big)^2\Big)
\nonumber
\\*
&=-\mathrm{sign}(\lambda)\mathcal{H}_0^{\!\Lambda}[\tau_0]
-\log(|\lambda|\pi a) 
+\mathrm{sign}(\lambda) \mathcal{H}_a^{\!\Lambda}[A(\cot
\tau_b(\bullet)) ]+L(\cot\tau_b(a))\;,
\label{logGab}
\\
A(x)&:=\mathrm{arccot}(x)=\frac{\pi}{2} -\frac{1}{2\mathrm{i}} 
\Big(\log (1+\mathrm{i}x)-\log(1-\mathrm{i} x)\Big)\;,\qquad 
\nonumber
\\*
L(x)&:=-\frac{1}{2}\log(1+x^2)= -\frac{1}{2}\Big(
\log (1+\mathrm{i}x) +\log(1-\mathrm{i} x)\Big)\;.
\nonumber
\end{align}
Here (\ref{tau}) has been used. For these functions $A,L$ one has
\begin{Lemma}
The functions $A,L$ introduced in (\ref{logGab}) have the 
following $(k\geq 1)$-fold derivatives:
\begin{align}
A^{(k)}(\cot t)= (-1)^k (k{-}1)! \sin(k t) \sin^k t\;,\qquad
L^{(k)}(\cot t)= (-1)^k (k{-}1)! \cos(k t) \sin^k t\;.
\end{align}
\label{Lemma:dAdL}
\end{Lemma}
\emph{Proof.}
An elementary calculation yields
\begin{align*}
\frac{d^k A(x)}{dx^k}
&= \frac{(k-1)!(-1)^{k}}{(1+x^2)^k}
\sum_{j=0}^{[\frac{k-1}{2}]} \binom{k}{2j+1} (-1)^j x^{k-2j-1}\;,
\\
\frac{d^k L(x)}{dx^k}
&= \frac{(k-1)!(-1)^{k}}{(1+x^2)^k}
\sum_{j=0}^{[\frac{k}{2}]} \binom{k}{2j} (-1)^j x^{k-2j}\;.
\end{align*}
The assertion
follows from the identities \cite{Gradsteyn:1994??}[\S 1.331.1+3],
\begin{align}
\frac{\sin (kt)}{ \sin^k t} =\sum_{j=1}^{[\frac{k-1}{2}]} 
\binom{k}{2j+1} (-1)^j \cot^{k-2j-1} t\;, \qquad
\frac{\cos (kt)}{ \sin^k t} =\sum_{j=0}^{[\frac{k}{2}]} 
\binom{k}{2j} (-1)^j \cot^{k-2j} t\;.
\tag*{\mbox{$\square$}}
\end{align}

The derivative $G^{(n)}_{ab}:=\frac{d^n}{dt^n} G_{a+t,b+t}\big|_{t=0}$
will be traced back to 
\begin{align}
(\log G_{ab})^{(n)}:=\frac{d^n}{dt^n} (\log G_{a+t,b+t})\Big|_{t=0}
= \sum_{\ell=0}^n \binom{n}{\ell} \frac{\partial^n (\log G_{ab})}{
\partial a^{n-\ell} \partial b^{\ell}}\;.
\label{DG-diag}
\end{align}
This is achieved by Fa\`a di Bruno's
formula, i.e.\ the higher order analogue of the chain rule:
\begin{align}
\frac{d^n}{dx^n} (g\circ f)(x)= \sum_{k=0}^n g^{(k)}(f(x)) 
Y_{n,k}(f'(x),f''(x),\dots,f^{(n-k+1)}(x))\;,
\label{FaaDiBruno}
\end{align}
where the $Y_{n,k}$ (also denoted $B_{n,k}$) are the Bell polynomials 
\cite{Bell}
\begin{align}
Y_{n,k}(x_1,\dots,x_{n-k+1})= \!\!\!\sum_{\di{j_1+2j_2+
    \dots=n}{j_1+j_2+\dots =k}}
\frac{n!}{j_1!j_2!\cdots j_n!}
\Big(\frac{x_1}{1!}\Big)^{j_1} 
\Big(\frac{x_2}{2!}\Big)^{j_2} \cdots 
\Big(\frac{x_{n-k+1}}{(n{-}k{+}1)!}\Big)^{j_{n-k+1}}\;.
\label{Bell}
\end{align}
For $n\geq 1$ the summation over $k$ actually starts with $k=1$
because $Y_{n,0}=0$ for $n\geq 1$. In many cases it will be useful to
include in (\ref{FaaDiBruno}) the case $n=0$ via the convention
$Y_{0,0}=1$.

In a first step we have
\begin{align}
G^{(h)}_{ab} = G_{ab} \sum_{k=0}^h Y_{h,k} 
\big(\{(\log G_{ab})^{(n)}\}_{n=1}^{h-k+1}\big)\;.
\label{DGab}
\end{align}
Since $\frac{\partial \cot \tau_b(a)}{\partial b}= \frac{1}{|\lambda|\pi a}$, 
Lemma~\ref{Lemma:dAdL} and (\ref{FaaDiBruno}) yield for
(\ref{logGab}) 
\begin{subequations}
\begin{align}
\frac{\partial^\ell \log G_{ab}}{\partial b^\ell}\Big|_{\ell\geq 1}
&= 
\mathrm{sign}(\lambda) \,
\mathcal{H}_a^{\!\Lambda}\Big[\frac{A^{(\ell)}(\cot
  \tau_b(\bullet))}{(|\lambda|\pi \bullet)^\ell}\Big]
+ \frac{L^{(\ell)}(\cot  \tau_b(a))}{(|\lambda|\pi a)^\ell}
\label{dlogGabdb-ell}
\\
&= 
(-1)^{\ell} (\ell{-}1)! \,\mathrm{sign}(\lambda) \,
\mathcal{H}_a^{\!\Lambda}\Big[
\sin\big( \ell \tau_b(\bullet)\big)
\Big(\frac{\sin \tau_b(\bullet)}{|\lambda|\pi \bullet}\Big)^\ell
\Big]
\nonumber
\\
& + (-1)^{\ell} (\ell{-}1)! 
\cos\big( \ell \tau_b(a)\big)
\Big(\frac{\sin \tau_b(a)}{|\lambda|\pi a}\Big)^\ell\;.
\label{dlogGab-b}
\end{align}
\end{subequations}
For $a=0$ we obtain with $\lim_{a\to 0} \tau_b(a)=0$ and 
$\lim_{a\to 0} \frac{\sin \tau_b(a)}{|\lambda|\pi a}=\frac{1}{1+b}$ 
the equation
\begin{align}
(\log G_{0b})^{(\ell)}:=
\frac{d^\ell \log G_{0b}}{db^\ell} 
&=  \frac{(-1)^{\ell}(\ell{-}1)!} {(1+b)^\ell}
+ (-1)^{\ell} \ell! \lambda 
\int_0^{\Lambda^2} \!\!\! dp \;
\frac{\sin\big( \ell \tau_b(p)\big)}{\ell |\lambda|\pi p}
\Big(\frac{\sin \tau_b(p)}{|\lambda|\pi p}\Big)^{\ell}\;.
\label{dlogG0b-b}
\end{align}

The asymptotic similarity of the functions $G_{aa}$ and $G_{0a}$ (see
Section~\ref{sec:computer}) lets
us conjecture that $a\mapsto G_{aa}$ is Stieltjes iff $b\mapsto G_{0b}$
is Stieltjes. Stieltjes functions are logarithmically completely
monotonic \cite{Berg:2008??}, i.e.\
\begin{align}
f(x) =\int_0^\infty \frac{d\rho(t)}{x+t} \qquad \Rightarrow \quad
(-1)^n\frac{d^n}{dx^n} (\log f(x)) \geq 0\;.
\end{align}
Therefore, necessary for $b\mapsto G_{0b}$ being a Stieltjes function
is $(-1)^\ell (\log G_{0b})^{(\ell)}\big|_{b=0}\geq 0$ or
$\mathcal{Y}_\ell \geq -1$ for all $\ell$, where
\begin{align}
\mathcal{Y}_\ell
:= \lambda
\int_0^{\Lambda^2} \!\!\! dp \;
\frac{\sin\big( \ell \tau_0(p)\big)}{|\lambda|\pi p}
\Big(\frac{\sin \tau_0(p)}{|\lambda|\pi p}\Big)^{\ell}
= \mathrm{sign}(\lambda) \mathcal{H}^{\!\Lambda}_0\Big[
\sin( \ell \tau_0(\bullet))
\Big(\frac{\sin \tau_0(\bullet)}{|\lambda|\pi
  \bullet}\Big)^{\ell}\Big]\;.
\label{calY-ell}
\end{align}
In particular, $\mathcal{Y}_1=: \mathcal{Y}$ is the finite
wavefunction renormalisation \cite[eq.\ (4.30)]{Grosse:2012uv}.

Differentiation of the Hilbert transform (\ref{Hilbert}) leads
after integration by parts to
\begin{subequations}
\begin{align}
\frac{d}{da} \mathcal{H}^{\!\Lambda}_a[f(\bullet)]
&=\frac{1}{\pi} \lim_{\epsilon\to 0}
\Big(
\frac{f(a-\epsilon)}{a-\epsilon-a}-
\frac{f(a+\epsilon)}{a+\epsilon-a}
+ \Big(\int_0^{a-\epsilon} +\int_{a+\epsilon}^{\Lambda^2}\Big) 
\frac{f(p)dp}{(p-a)^2} \Big)
\nonumber
\\
&= 
- \frac{f(0)}{\pi a}
- \frac{f(\Lambda^2)}{\pi(\Lambda^2-a)}
+ \mathcal{H}_a^{\!\Lambda}[f'(\bullet)]
\label{Diff-Hilbert-1}
\\
&= - \frac{f(\Lambda^2)}{\pi a}+ 
\frac{1}{\pi a} \int_0^{\Lambda^2} \!\! dp\; f'(p) \frac{p-a}{p-a}
- \frac{f(\Lambda^2)}{\pi(\Lambda^2-a)}
+ \frac{1}{a} \mathcal{H}_a^{\!\Lambda}[a f'(\bullet)]
\nonumber
\\
&= - \frac{f(\Lambda^2)}{\pi a(1-\frac{a}{\Lambda^2})}
+ \frac{1}{a} \mathcal{H}_a^{\!\Lambda}[\bullet f'(\bullet)]\;.
\label{Diff-Hilbert-2}
\end{align}
\end{subequations}
The two equations (\ref{Diff-Hilbert-1}) and 
(\ref{Diff-Hilbert-2}) are equivalent, but in the numerical simulation  
one of them is preferred. We can only use (\ref{Diff-Hilbert-1}) for
higher derivatives as long as $f^{(k)}(0)$ exists. Such existence
could rely on cancellations which are numerically not guaranteed; 
we would prefer (\ref{Diff-Hilbert-2}) in these cases.
The same method as employed in (\ref{Diff-Hilbert-2}),
$\displaystyle 
\mathcal{H}_a^{\!\Lambda}[(\bullet^n f^{(n)}(\bullet))']
= \frac{1}{a}\mathcal{H}_a^{\!\Lambda}[((a-\bullet)+\bullet)
(\bullet^n f^{(n)}(\bullet))']
$,
leads to the following generalisation of (\ref{Diff-Hilbert-2}):
\begin{align*}
&\frac{d}{da}\Big(\frac{1}{a^n} \mathcal{H}^{\!\Lambda}_a[\bullet^n 
f^{(n)}(\bullet)] \Big) 
= - 
\frac{(\Lambda^2)^nf^{(n)}(\Lambda^2) }{\pi a^{n+1}(1-\frac{a}{\Lambda^2})}
+ \frac{1}{a^{n+1}}\mathcal{H}_a^{\!\Lambda}[\bullet^{n+1} f^{(n+1)}(\bullet)]\;.
\end{align*}
Homogeneous contribution of $f^{(k)}(\Lambda^2)$ are then collected to
\begin{subequations}
\label{Diff-Hilbert-Fn}
\begin{align}
\frac{d^n}{da^n}\mathcal{H}_a^{\!\Lambda}[f(\bullet)]
&=
\frac{1}{a^n}
\mathcal{H}_a^{\!\Lambda}[\bullet^n  f^{(n)}(\bullet)]
+ \frac{(-1)^{n}(n-1)!}{\pi a^n}
\sum_{k=0}^{n-1} 
\frac{(-\Lambda^2)^kf^{(k)}(\Lambda^2)}{k!} F_{n,k}^\Lambda(a)\;,
\label{Diff-Hilbert-2n}
\\
F_{n,k}^\Lambda(a) &:= 
\frac{1}{(1-\frac{a}{\Lambda^2})} 
\sum_{p=0}^{n-k-1} \frac{\binom{n-k-1}{p}}{\binom{n-1}{p}}
\Big(\frac{-a}{\Lambda^2-a}\Big)^p\;.
\label{Diff-Hilbert-F}
\end{align}
Alternatively, we can use (\ref{Diff-Hilbert-1}) for the first
derivative and  (\ref{Diff-Hilbert-2n}) for higher derivatives:
\begin{align}
\frac{d^n}{da^n}\mathcal{H}_a^{\!\Lambda}[f(\bullet)]
&= \frac{(-1)^{n}(n-1)!}{\pi a^n} f(0)-
\frac{(n-1)!}{\pi (\Lambda^2-a)^n} f(\Lambda^2)
+
\frac{1}{a^{n-1}}
\mathcal{H}_a^{\!\Lambda}[\bullet^{n-1}  f^{(n)}(\bullet)]
\nonumber
\\*
& 
+ \frac{(-1)^{n}(n{-}2)!}{\pi a^{n-1} \Lambda^2}
\sum_{k=1}^{n-1} k
\frac{(-\Lambda^2)^{k}f^{(k)}(\Lambda^2)}{k!} F_{n-1,k-1}^\Lambda(a)\;.
\label{Diff-Hilbert-1+2n}
\end{align}
\end{subequations}
This version is particularly useful if $f(0)=0$.

Next we provide an equation for derivatives of
$\cot \tau_b(a)$ with respect to $a$. In a first step we need
\begin{align}
\Gamma_{s}(a)
:= \frac{(-1)^s a^s G_{a0}}{s!} \frac{d^s}{da^s} \frac{1}{G_{a0}}
& = \sum_{l=0}^s \frac{(-1)^l l!}{s!} Y_{s,l}\Big( \Big\{\frac{(-a)^j 
G^{(j)}_{a0}}{G_{a0}}\Big\}_{j=1}^{s-l+1}\Big)
\nonumber
\\*
& = 
\sum_{l=0}^s \frac{(-1)^{l}}{s!} Y_{s,l}
\Big( \big\{(-a)^j (\log G_{a0})^{(j)}\}_{j=1}^{s-l+1}\Big)\;.
\label{Gamma-sa}
\end{align}
Using (\ref{Diff-Hilbert-2n}) we obtain for 
the derivatives of $\cot \tau_b(a)$ the formula
\begin{align}
C_b^n(a) &:= (-1)^{n}|\lambda|\pi a^{n+1}
\frac{d^n \cot \tau_b(a)}{da^n}
=
(-1)^{n}a^{n+1}
\frac{d^n}{da^n} \Big(\frac{b}{a}
+ \frac{1}{G_{a0}} 
\Big(\frac{1}{a} 
+ \lambda\pi \mathcal{H}^{\!\Lambda}_a[G_{\bullet 0}]\Big)\Big)
\nonumber
\\
&= n! \bigg\{ b 
+ \frac{1+\lambda\pi a \mathcal{H}^{\!\Lambda}_a[G_{\bullet 0}]}{G_{a0}} 
\Gamma_n(a)
\nonumber
\\
&
\qquad + \sum_{k=1}^n 
\frac{1 
+\frac{\lambda a}{k}\sum_{l=0}^{k-1}
\frac{(-\Lambda^2)^l G^{(l)}_{0\Lambda^2}}{l!} F_{k,l}^\Lambda(a)
+ \lambda\pi a
\mathcal{H}^{\!\Lambda}_a\big[\frac{(-\bullet)^k}{k!} 
G^{(k)}_{\bullet 0}\big] }{G_{a0}} \Gamma_{n-k}(a) 
\bigg\}\;.
\label{Cbna}
\end{align}
We prefer here (\ref{Diff-Hilbert-2n}) to 
(\ref{Diff-Hilbert-1+2n}) because the latter leads to Hilbert
transforms $\lambda\pi a^2
\mathcal{H}^{\!\Lambda}_a\big[\frac{(-\bullet)^{k-1}}{k!}
G^{(k)}_{\bullet 0}\big]$ instead of $\lambda\pi a
\mathcal{H}^{\!\Lambda}_a\big[\frac{(-\bullet)^{k}}{k!}
G^{(k)}_{\bullet 0}\big]$. 
Although both results must agree, the increased exponent 
$\lambda\pi a^2$ compared with $\lambda\pi a$ might lead in numerical
simulations to larger errors\footnote{
At this point a remark on the limit $\Lambda^2\to \infty$ is in
order. Of course $\lim_{\Lambda^2\to \infty} \cot \tau_b(a)$ is
expected to exist, and its derivatives should reproduce the
limit $\Lambda^2\to \infty$ of (\ref{Cbna}) as a whole. But there is no
reason to assume that all individual terms in  (\ref{Cbna}) converge
for $\Lambda^2\to \infty$.}.

According to (\ref{dlogGabdb-ell}) and (\ref{Diff-Hilbert-Fn}),
$a$-derivatives of $\frac{\partial^\ell \log G_{ab}}{\partial b^\ell}$
involve the functions
\begin{subequations}
\label{Anl}
\begin{align}
&A^{(n,\ell)}(a,b):=
\frac{(-a)^n}{n!} \frac{(-b)^\ell}{\ell!} 
\frac{\partial^{n+\ell} A[\cot \tau_b(a)]}{\partial a^n\;
\partial^\ell b} 
 =
\frac{(-a)^n}{n!} \frac{(-b)^\ell}{\ell!} 
\frac{\partial^n}{\partial a^n} \Big(\frac{A^{(\ell)}(\cot
  \tau_b(a))}{(|\lambda|\pi a)^\ell}\Big)
\nonumber
\\
&=\frac{(-a)^n}{n!} \frac{(-b)^\ell}{\ell!} \sum_{m=0}^{n} \binom{n}{m} 
\frac{\partial^m A^{(\ell)} [\cot \tau_b(a)]}{\partial a^m}
\frac{(-1)^{n-m} (\ell{+}n{-}m{-}1)!}{(\ell-1)! 
(|\lambda|\pi)^\ell a^{\ell+n-m}}
\nonumber
\\
& = 
\sum_{m=0}^{n} \binom{n{-}m{+}\ell{-}1}{\ell-1} 
\sum_{k=0}^m 
\frac{(-a)^m}{m!\ell!} A^{(\ell+k)} [\cot \tau_b(a)]
Y_{m,k}\Big( \Big\{\frac{(-1)^\kappa C_b^\kappa(a)}{|\lambda|\pi
  a^{\kappa+1}}
\Big\}_{\kappa=1}^{m-k+1}\Big)
\cdot 
\Big(\frac{-b}{|\lambda|\pi a}\Big)^\ell
\nonumber
\\
& = 
\sum_{m=0}^{n} \sum_{k=0}^m 
\binom{n{-}m{+}\ell{-}1}{\ell-1} 
\frac{(-1)^{k}(\ell+k)!}{m!\ell!} \frac{\sin((\ell{+}k)
  \tau_b(a))}{(\ell+k)}
\nonumber
\\*
&\qquad\qquad\qquad\qquad \times 
\Big(\frac{b \sin \tau_b(a)}{|\lambda|\pi a}\Big)^\ell
Y_{m,k}\Big( \Big\{\frac{C_b^\kappa(a)\sin \tau_b(a)}{|\lambda|\pi  a}
\Big\}_{\kappa=1}^{m-k+1}\Big)\;.
\label{Anl-1}
\end{align}
In the third line Fa\`a di Bruno and the definition (\ref{Cbna}) have
been used. To obtain the last equation we have reinserted
Lemma~\ref{Lemma:dAdL} and used the homogeneity properties of the Bell
polynomials.  For $m\geq 1$ the second sum actually restricts to
$k\geq 1$.  Formula (\ref{Anl-1}) extends to the case
$(n{=}0,\;\ell{\geq}1)$ where it reproduces $\frac{(-b)^\ell}{\ell!}$
times the function under the Hilbert transform in (\ref{dlogGab-b}).
For $(n{\geq}1,\;\ell{=}0)$ only the terms with $m=n$ survive (which
forces $k\geq 1$), and $(n{=}0,\;\ell{=}0)$ is easily included:
\begin{align}
\label{Anl-0}
A^{(n,0)}(a,b)&:=
\frac{(-a)^n}{n!} 
\frac{\partial^{n} A[\cot \tau_b(a)]}{\partial a^n} 
\nonumber
\\
&=\left\{\begin{array}{ll}
\tau_b(a) & \text{for } n=0\;, \\[2ex]
\displaystyle \sum_{k=1}^n
\frac{(-1)^{k} k!}{n!} \frac{\sin(k \tau_b(a))}{k}
Y_{n,k}\Big( \Big\{\frac{C_b^\kappa(a)\sin \tau_b(a)}{|\lambda|\pi  a}
\Big\}_{\kappa=1}^{n-k+1}\Big)
& \text{for } n\geq 1\;.
\end{array}\right.
\end{align}
\end{subequations}

In complete analogy one finds for $\ell\geq 1$ and any $n$
\begin{subequations}
\label{Lnl}
\begin{align}
&L^{(n,\ell)}(a,b):=
\frac{(-a)^n}{n!} \frac{(-b)^\ell}{\ell!} 
\frac{\partial^{n+\ell}}{\partial a^n\;
\partial^\ell b} \Big( L[\cot \tau_b(a)] -\log(|\lambda|\pi a)\Big)
\nonumber
\\
& = \sum_{m=0}^{n} \sum_{k=0}^m 
\binom{n{-}m{+}\ell{-}1}{\ell-1} 
\frac{(-1)^{k}(\ell+k)!}{m!\ell!} \frac{\cos((\ell{+}k)
  \tau_b(a))}{(\ell+k)}
\nonumber
\\
&\qquad\qquad\qquad\qquad \times 
\Big(\frac{b \sin \tau_b(a)}{|\lambda|\pi a}\Big)^\ell
Y_{m,k}\Big( \Big\{\frac{C_b^\kappa(a)\sin \tau_b(a)}{|\lambda|\pi  a}
\Big\}_{\kappa=1}^{m-k+1}\Big)\;,
\end{align}
whereas for $\ell=0$, $n\geq 1$ one has
\begin{align}
L^{(n,0)}(a,b)&:=
\frac{(-a)^n}{n!} 
\frac{\partial^{n}}{\partial a^n} 
\Big( L[\cot \tau_b(a)] -\log(|\lambda|\pi a)\Big)
\nonumber
\\*
& = \frac{1}{n}+ \sum_{k=1}^n 
\frac{(-1)^{k} k!}{n!} \frac{\cos( k \tau_b(a))}{k}
Y_{n,k}\Big( \Big\{\frac{C_b^\kappa(a)\sin \tau_b(a)}{|\lambda|\pi  a}
\Big\}_{\kappa=1}^{n-k+1}\Big)\;.
\end{align}
\end{subequations}

These formulae are inserted into the $a$-derivatives of 
(\ref{dlogGabdb-ell}), however with a smaller cut-off
$\Lambda\mapsto \tilde{\Lambda}<\Lambda$. The reason is
the singularity of $C_b^n(a)$ at $a=\Lambda^2$, which excludes the
values at $\Lambda^2$ in (\ref{Diff-Hilbert-1+2n}). 
Since we have 
$A^{(n,\ell)}(0,b)=0$, we can use 
(\ref{Diff-Hilbert-1}) for $n=1$ and (\ref{Diff-Hilbert-1+2n}) for
 $n\geq 2$ to obtain:
\begin{align}
\frac{(-a)^n}{n!}& \frac{(-b)^\ell}{\ell!} 
\frac{\partial^{n+\ell} \log G_{ab}}{\partial a^n \partial b^\ell}
\Big|_{n\geq 1}
\label{dlogGab-ab}
\\
&=
\lambda\pi
a\,\mathcal{H}^{\tilde{\Lambda}}_a\Big[\frac{A^{(n,\ell)}(\bullet,b)}{
|\lambda|\pi \bullet} \Big] +L^{(n,\ell)}(a,b)
- \frac{\mathrm{sign}(\lambda)}{n \pi} 
\Big(\frac{-a}{\tilde{\Lambda}^2-a}\Big)^n
A^{(0,\ell)}(\tilde{\Lambda}^2,b)
\nonumber
\\* 
& \qquad\quad+ \left\{ \begin{array}{cl}
\displaystyle 
\frac{\mathrm{sign}(\lambda)}{\pi n(n-1)}\frac{a}{\tilde{\Lambda}^2}
\sum_{k=1}^{n-1} k A^{(k,\ell)}(\tilde{\Lambda}^2,b)\,
F_{n-1,k-1}^{\tilde{\Lambda}}(a)
& \text{ for } n\geq 2\;,
\\
0 & \text{ for } n=1\;.
\end{array}\right.
\nonumber
\end{align}
We prefer here (\ref{Diff-Hilbert-1+2n}) to (\ref{Diff-Hilbert-2n}) in
order to make explicit that (\ref{dlogGab-ab}) vanishes for $a=0$. We
only rely on cancellations giving $L^{(n,\ell)}(0,b)=0$ (we prove this
in Appendix~\ref{app:limit}, together with the related computation of
$\lim_{a\to 0} \frac{A^{(n,\ell)}(a,b)}{|\lambda|\pi a}$) but not on
cancellations under the Hilbert transform which numerically are not
guaranteed.  We thus conclude that Widder's operators
$L_{n,t}[G_{\bullet\bullet}]$ defined by (\ref{Widder}) vanish at
$t=0$.  Finally we remark that the case $\ell=0$ of (\ref{dlogGab-ab})
should be the symmetric partner to (\ref{dlogGab-b}). This would imply
that $\frac{(-a)^n}{n!} \frac{(-b)^\ell}{\ell!}
\frac{\partial^{n+\ell} \log G_{ab}}{\partial a^n \partial b^\ell}$
vanishes to $n^\mathrm{th}$ order in $a=0$. The proof relies on subtle
cancellations which cannot be expected for the numerical result.

We have thus established:
\begin{Proposition}
\label{Prop:Stieltjes}
For natural numbers $s\geq 2, h\geq 1$ one has
\begin{subequations}
\begin{align}
L_{s,a}[G_{\bullet\bullet}]
&= \sum_{l=0}^s \frac{(-1)^{s-l}(2s-1)!}{
(2s-l-1)!(s-2)!(s-l)!l!} \cdot 
(-a)^{2s-l-1} G_{aa}^{(2s-l-1)}\;,
\label{Widder-K}
\\
(-a)^h G_{aa}^{(h)}
&= G_{aa} \sum_{k=0}^h Y_{h,k}\big(
\{(-a)^n (\log G_{aa})^{(n)}\}_{n=1}^{h-k+1}\big)\;.
\label{aGaa-n}
\end{align}
The occurring functions $(-a)^n (\log G_{aa})^{(n)}$, for $n\geq 1$, are given in 
terms of $A^{(n,\ell)},L^{(n,\ell)}$ defined in 
(\ref{Anl})+(\ref{Lnl}), which rely on
$\tau_b(a)$ defined in (\ref{tau}),
$C^\kappa_b(a)$ defined in (\ref{Cbna}), 
$\Gamma_l(a)$ defined in (\ref{Gamma-sa}) and
$F^\Lambda_{n,k}(a)$ defined in (\ref{Diff-Hilbert-F}), by
\begin{align}
&(-a)^n (\log G_{aa})^{(n)}
\nonumber
\\
&= n! \bigg\{\mathrm{sign}(\lambda)\,
\mathcal{H}_a^{\!\tilde{\Lambda}}\big[ A^{(0,n)}(\bullet,a)\big]
+ L^{(0,n)}(a,a)
\nonumber
\\
& + \sum_{\ell=0}^{n-1}\Big\{ 
\lambda\pi a
\mathcal{H}^{\tilde{\Lambda}}_a\Big[\frac{A^{(n-\ell,\ell)}(\bullet,a)}{|\lambda|\pi
  \bullet} \Big]
+L^{(n-\ell,\ell)}(a,a)
- \frac{\mathrm{sign}(\lambda)\;A^{(0,\ell)}(\tilde{\Lambda}^2,a)}{
\pi (n{-}\ell)} \Big(\frac{-a}{\tilde{\Lambda}^2-a}\Big)^{n-\ell}\Big\}
\nonumber
\\
& \qquad 
+ \sum_{\ell=0}^{n-2} \sum_{k=1}^{n-\ell-1} 
\frac{\mathrm{sign}(\lambda)}{\pi (n{-}\ell)(n{-}\ell{-}1)} 
\Big(\frac{a}{\tilde{\Lambda}^2}\Big)\,
 k A^{(k,\ell)}(\tilde{\Lambda}^2,a)\,
F_{n-\ell-1,k-1}^{\tilde{\Lambda}}(a)
\bigg\}.
\label{dlogGaa-a}
\end{align}
\end{subequations}
In this way, $L_{s,a}[G_{\bullet\bullet}]$ is eventually expressed in
terms of $G_{a0}$ and its derivatives, which under use of
$G_{a0}=G_{0a}$ are given by (\ref{master}) and its 
derivatives (\ref{dlogG0b-b}).
\end{Proposition}
\emph{Proof.} (\ref{Widder-K}) is an obvious rewriting of
(\ref{Widder}), and (\ref{aGaa-n}) is Fa\`a di Bruno applied to
$G_{aa}=\exp(\log(G_{aa}))$ together with homogeneity properties of the
Bell polynomials (\ref{Bell}). 
Finally, (\ref{dlogGaa-a}) follows from 
(\ref{DG-diag}) taken at $b\mapsto a$ together with (\ref{dlogGab-ab}) 
and insertion of (\ref{Anl}) and (\ref{Lnl}). 
\hspace*{\fill}$\square$%

\section{Numerical results}

\label{sec:computer}

We use the computer algebra system \emph{Mathematica$^{TM}$} for a numerical
approximation of the two-point function. We need no sophisticated
tools of \emph{Mathematica$^{TM}$}; everything boils down to standard
manipulations of arrays and basic mathematical functions 
$\exp,\log,\sqrt{},\arctan,\sin,\cos,+,-,\times,\div$. We provide in
Appendix~\ref{sec:implementation} the source code, together with
additional explanations, of our
implementation so that the reader can check and adapt our calculation
or reimplement it in other computer languages. We do not claim that
our code is optimal, and we do not provide any error handling.

In this section we summarise the main results of this numerical
implementation:
\begin{enumerate} \itemsep 0pt \parskip 0.5ex
\item The fixed point equation~(\ref{master}) satisfies (numerically) 
the assumptions of the Banach fixed point theorem for any $\lambda\in
\mathbb{R}$. We can thus compute $G_{0b}$ and hence, in principle, all
correlation functions and Widder's operators
$L_{n,t}[G_{\bullet\bullet}]$ with sufficient precision.

\item For $\lambda>0$, neglecting $C_{\lambda,\Lambda^2}$ and 
$f_{\lambda,\Lambda^2}$ in (\ref{Gab}) is \emph{not justified}. At the moment
we see no possibility to improve this situation for $\lambda>0$.

\item For $\lambda<0$ everything is consistent within small numerical
  error bounds. The symmetry $G_{ab}=G_{ba}$ is confirmed. 
 The derivative $\frac{d\mathcal{Y}}{d\lambda}$ of the finite wavefunction
  renormalisation $\mathcal{Y}=\mathcal{Y}_1$ is discontinuous 
at some critical (negative)
  coupling constant $\lambda_c \approx -0.39$, which we interpret as a
  phase transition at $\lambda_c$. Whereas the phase $\lambda_c<\lambda\leq 0$
  has good properties, we have $\mathcal{Y}=-1$ within small error
  bounds for $\lambda<\lambda_c$. This implies that higher
  correlation functions loose their meaning in the phase $\lambda <
  \lambda_c$.

\item The Stieltjes property of the diagonal 2-point function,
  equivalent to reflection positivity of the Schwinger 2-point
  function \cite{Grosse:2013iva}, is
  excluded \emph{outside} the window $[\lambda_c,\,0]$. We have good
  reasons to assume that the Stieltjes property holds
for $\lambda_c <  \lambda \leq 0$, but this needs further verification.

\end{enumerate}

\subsection{Convergence of the 
iteration for $\lambda=\frac{1}{\pi}$}

The fixed point equation (\ref{master}) defines an operator $\tilde{T}:=\exp\circ T \circ \log $ via 
$G_{0b}=(\tilde{T}G)(b)$. We have shown in \cite{Grosse:2015fka}
that $T:\mathcal{K}_\lambda\to \mathcal{K}_\lambda$ satisfies for
$-\frac{1}{6}\leq \lambda\leq 0$ the assumptions of the Schauder fixed point
theorem. 
In addition, $(\tilde{T}1)(b)=1$ for any $\lambda<0$ 
\cite[Appendix A]{Grosse:2015fka}.
For $\lambda>0$ we have already seen in 
\cite{Grosse:2012uv} that $\tilde{T}$ has a fixed point 
by the Schauder fixed point theorem. 

Here we test numerically the conjecture that $\tilde{T}$ also
satisfies the assumptions of the Banach fixed point theorem. This
would imply that starting from an arbitrary initial function $G^{0}
\in \mathcal{K}$ in a closed subset of a Banach space, the iteration
$G^{i+1}:=\tilde{T}G^{i}$ converges to the fixed point solution
$G=\tilde{T}G=\lim_{i\to \infty} G^{i}$.  It would be most natural to take the
Banach space $\mathcal{C}_0^1(\mathbb{R}^+)$ of differentiable
functions vanishing at $\infty$ as in \cite{Grosse:2012uv}. A good
numerical substitute is the Banach space
$\mathcal{C}^{0,1}([0,\Lambda^2])$ of Lipschitz continuous functions
on $[0,\Lambda^2]$. We approximate $G_{0b}$ by a piecewise linear
function (which is Lipschitz) determined by its corner values 
at $L+1$ sample points $0=b_1<b_2\dots < b_L < b_{L+1}=\Lambda^2$. 
After initialisation of these sample points in\footnote{{\bf In[\dots]}
  refers to the implementation in Appendix~\ref{sec:implementation}.}  
{\tt\bf In[\ref{co}]} we compute the corner values of $G_{0b}$ by an
iteration {\tt\bf In[\ref{For}]}. 

\begin{figure}[h!]
\begin{picture}(150,78)
  \put(0,0){\includegraphics[width=10cm,
bb=0 0 206 156]{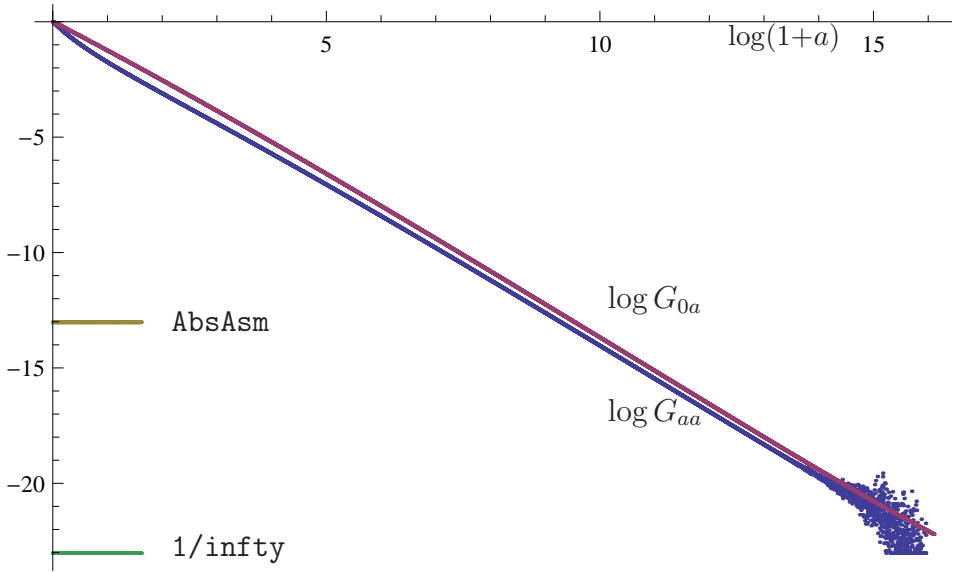}}
  \put(96,70){\mbox{\small$\log (1{+}a)$}}
  \put(80,35){\mbox{$\log G_{0a}$}}
  \put(80,20){\mbox{$\log G_{aa}$}}
  \put(22,32){\mbox{\tt AbsAsm}}
  \put(22,2){\mbox{\tt 1/infty}}
\end{picture}
\caption{Plot of the logarithm of the boundary and diagonal
  2-point functions $G_{0a}$ (red, upper curve) and 
$G_{aa}$ (blue, lower curve) over $\log(1+a)$, 
for $\lambda=\frac{1}{\pi}$ in numerical approximation  
{\tt i=20}, {\tt co=10\^{}7},
{\tt len=10\^{}4}. The noise in $G_{aa}$ appears near {\tt 1/infty}, 
hence much beyond the asymmetry {\tt AbsAsm}. 
\label{fig:G}}
\end{figure}

In table~\ref{tab:la=1/pi} we study for $\lambda=\frac{1}{\pi}$ the
dependence {\tt\bf In[\ref{co}]} of $G_{0b}$ and $G_{aa}$ on the
parameters $\Lambda^2=\,${\tt co}, $L=\,${\tt len} and 
{\tt infty} of 
the \emph{Mathematica}$^{TM}$-implementation.
\begin{table}[t]
$\begin{array}{|@{\,}c@{\,}|@{\,}c@{\,}|@{\,}c@{\,}|@{\,}c@{\,}|@{\,}c@{\,}|@{\,}c@{\,}|@{\,}c@{\,}|@{\,}c@{\,}|@{\,}c@{\,}|}
\hline
\multicolumn{9}{|l|}{\text{Approximation $G^{\tt i}$ for $G_{0b}$ at
    $\lambda=\frac{1}{\pi}$}, {\tt i=20} \text{ iterations of $G_{0b}^0=1$} 
\rule[-3mm]{0pt}{9mm}}
\\ \hline
{\tt co} & {\tt len} & \infty & G^{\tt i}_{0,100} & 
G^{\tt i}_{0,{\tt co}} & \|G^{\tt i}{-}G^{\text{\tt i-1}}\| & {\tt
  AbsAsm} & 
A+Bx & C+Dx \\ \hline
10^2 & 10^3 & 10^8 & 0.0027095 & 2.7{\times} 10^{-3} & 
1.9{\times} 10^{-11} &
2.7{\times} 10^{-3} & 0.030{-}1.290x & {-}0.344 {-} 1.351x 
\\ \hline
10^3 & 10^3 & 10^8 & 0.0023612 & 1.1{\times}10^{-4} & 
1.7{\times} 10^{-9} &
1.4{\times} 10^{-4} & 0.085{-}1.330x & {-}0.365 {-} 1.341x 
\\ \hline
10^4 & 10^3 & 10^8 & 0.0023225 & 4.2{\times}10^{-6} & 
6.6{\times} 10^{-9} &
4.6{\times} 10^{-5} & 0.148{-}1.358x & {-}0.340 {-} 1.353x 
\\ \hline
10^5 & 10^3 & 10^8 & 0.0023180 & 1.5{\times}10^{-7} & 
1.0{\times} 10^{-8} &
7.1{\times} 10^{-5} & 0.211{-}1.379x & {-}0.293 {-} 1.368x 
\\
10^5 & 10^3 & {\bf 10^9} & 0.0023180 & 1.5{\times}10^{-7} & 
1.0{\times} 10^{-8} &
7.1{\times} 10^{-5} & 0.211{-}1.379x & {-}0.293 {-} 1.368x 
\\ \hline
10^6 & 10^3 & 10^9 & 0.0023174 & 5.5{\times}10^{-9} & 
1.2{\times} 10^{-8} &
9.7{\times} 10^{-5} & 0.272{-}1.395x & {-}0.239 {-} 1.382x 
\\
10^6 & 10^4 & 10^9 & 0.0023177 & 5.5{\times}10^{-9} & 
1.2{\times} 10^{-8} &
\mbox{\boldmath$1.7{\!\times\!} 10^{-6}$} & 0.272{-}1.395x 
& {-}0.226 {-} 1.384x 
\\ \hline
10^7 & 10^3 & 10^{10} & 0.0023173 & 2.0{\times}10^{-10} & 
1.3{\times} 10^{-8} &
1.3{\times} 10^{-4} & 0.318{-}1.405x & {-}0.196 {-} 1.391x 
\\
10^7 & {\bf 10^4} & 10^{10} & 0.0023176 & 2.2{\times}10^{-10} & 
1.3{\times} 10^{-8} &
\mbox{\boldmath$2.2{\!\times\!} 10^{-6}$} & 
0.304{-}1.402x & {-}0.168 {-} 1.396x 
\\ \hline
10^8 &   10^3 & 10^{12} & 0.0023172 & 1.2{\times}10^{-11} & 
1.3{\times} 10^{-8} & 1.6\times 10^{-4} & 
0.277{-}1.398x & {-}0.172 {-} 1.455x 
\\ 
10^8 & {\bf 10^4} & 10^{12} & 0.0023172 & 1.2{\times}10^{-11} & 
1.4{\times} 10^{-8} &
\mbox{\boldmath$2.8{\!\times\!} 10^{-6}$} & 
0.244{-}1.392x & {-}0.769 {-} 1.557x 
\\ \hline
\end{array}$
\caption{The 2-point function $G_{0b}$ at $\lambda=\frac{1}{\pi}$ for various
  cut-offs {\tt co} and resolutions {\tt len}.\label{tab:la=1/pi}}
\end{table}
Typical numerical results for 
{\tt len=10\^{}4}, {\tt co=10\^{}7} are visualised in fig.~\ref{fig:G}.
We observe:
\begin{itemize}\itemsep 0pt \parskip 0.5ex

\item The convergence of the iteration $G^{\tt i}\mapsto G^{\tt i+1}$ 
in Lipschitz norm 
$\|f\|={\displaystyle\sup_{0\leq a\leq \Lambda^2}|f(a)|}+
{\displaystyle\sup_{0\leq a < b \leq \Lambda^2}}
\big|\frac{f(a)-f(b)}{a-b}\big|$ is
remarkably good for any cutoff {\tt co}=$\Lambda^2$. Each iteration step
reduces the norm error by a factor bigger than 2. 

\item If the cut-off {\tt co}=$\Lambda^2$ is too small, then the
  absolute asymmetry $\displaystyle{\tt AbsAsm}= \sup_{0\leq a < b
    < \Lambda^2} |G_{ab}-G_{ba}|$ is governed by boundary effects
  at $b \approx \Lambda^2$. For larger cut-off the largest asymmetry
  is located where $a,b\approx 0$ because $G_{ab}$ is largest there.
  The asymmetry is reduced if more sample points (increased {\tt
    len}=$L$, see {\tt len=10\^{}4} versus {\tt len=10\^{}3} at {\tt
    co=10\^{}7}) are included. The computation time of our algorithm
  grows at least quadratically in {\tt len}.
  The relative asymmetry turns out large near the
  boundary. We address this problem in sec.~\ref{sec:la>0}.

\item The numerical choice of {\tt infty} for
  $\infty=\lim_{\epsilon\to 0} \log \frac{a_2-a_1}{\epsilon}$ has
  no influence. For {\tt co}=$10^5$, {\tt len}=$10^3$ the sup-norm
  difference $\sup_a |G_{0a}^{{\tt infty}=10^8}-
G_{0a}^{{\tt infty}=10^9}|$ is of the order $10^{-14}$. We
nevertheless keep {\tt 1/infty} of the same order as $G_{0,{\tt co}}$.

\item The cut-off 2-point function $G_{0a}(\Lambda^2)$ is pointwise
  convergent for $\Lambda\to \infty$.

\item Previous numerical simulations in the preprint arXiv:1205.0465v1 of
\cite{Grosse:2012uv} had suggested an asymptotic behaviour 
$G_{0a}\propto \frac{1}{(1+a)^{1+\lambda}}$, i.e.\ 
$\log  G_{0a}\propto -(1+\lambda)\log (1+a)$ is linear. The
corresponding fits of 
$\log G_{0,\exp (x)-1}$ to a line $A+Bx$ and of 
$\log G_{\exp (x)-1,\exp(x)-1}$ to a line $C+Dx$ are also indicated in
table~\ref{tab:la=1/pi}. We thus confirm that the slope actually
decreases with the cut-off without any hint of convergence.

\item As shown in fig.~\ref{fig:G} there is clear evidence that
  $G_{0a}$ and $G_{aa}$ have the same asymptotic behaviour for $a\to
  \infty$. The previously conjectured asymptotics holds without any
  doubt in form of inequalities
\[
0 < G_{0a} < \frac{C}{(1+a)^{1+\lambda}}\;,\qquad 
0 < G_{aa} < \frac{C'}{(1+a)^{1+\lambda}}\;.
\]
This is enough to state that the function $a\mapsto G_{aa}$ is for
$\lambda>0$ not a Stieltjes function. The results of
\cite{Grosse:2013iva} then imply that the corresponding Euclidean quantum
field theory does not have an analytic continuation to a Wightman
theory for $\lambda>0$. 

\end{itemize}

\subsection{Varying $\lambda > 0$: Inconsistency}
\label{sec:la>0}

Next we study the dependence of $G_{0b}$ and $G_{aa}$ on $\lambda \geq
0$.  The convergence rate of the iteration {\tt
  i}$\to${\tt i+1} is highly sensitive to $\lambda$.  To have
comparable results we run the iteration until $G^{\tt i}$ and $G^{\tt
  i-1}$ differ by $<4\times 10^{-8}$ in supremum norm. This is
achieved by replacing in the {\tt For[ ]}-loop of {\tt\bf
  In[\ref{For}]} the termination condition {\tt If[i>=20, Break[]];}
by
\begin{calc}
\item[]
If[SupNorm[gs[i], gs[i-1]] < 4*10^(-8), Break[]];
\end{calc}

\begin{figure}[h!]
\begin{picture}(150,50)
  \put(0,0){\includegraphics[width=7.8cm,
bb=0 0 288 181]{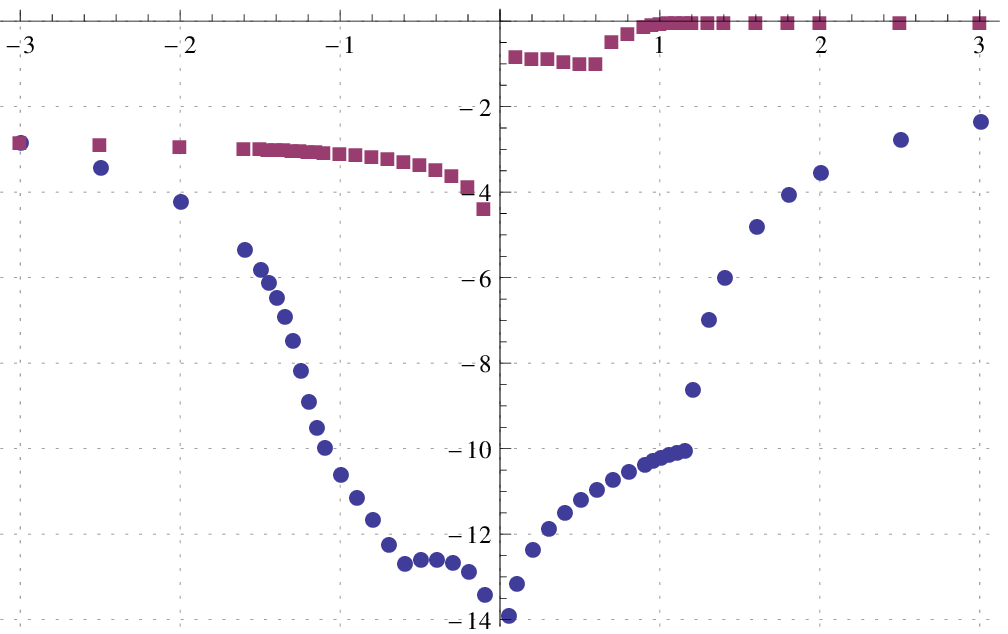}}
  \put(69,50){\mbox{\scriptsize$\lambda\pi$}}
  \put(3,5){\mbox{\textcolor{mathred}{{\tiny$\blacksquare$} 
             \small{$\log({\tt RelAsm})$}}}}
  \put(3,2){\mbox{\small\textcolor{mathblue}{\textbullet\ $\log({\tt AbsAsm})$}}}
  \put(152,0){\mbox{\scriptsize$\lambda\pi$}}
  \put(140,8){\mbox{\small\textcolor{mathred}{\textbullet\ {\tt RelAsm}}}}
  \put(82,0){\includegraphics[width=7.8cm,
bb=0 0 288 189]{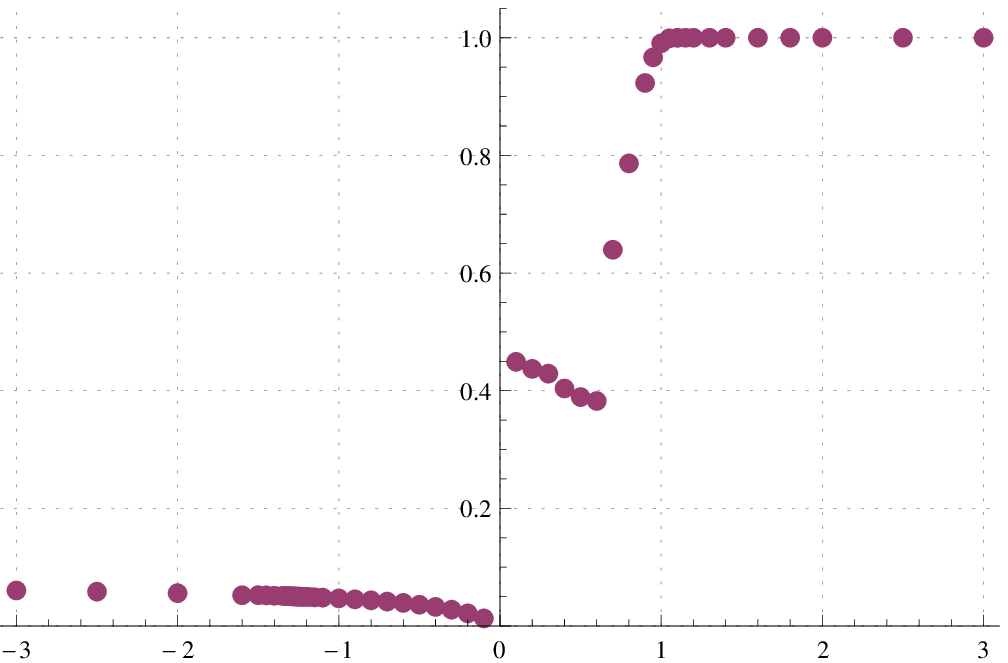}}
\end{picture}
\caption{Left: Plot of the logarithms of absolute (blue dots) and relative
  (red squares) asymmetries as function of $\lambda$. 
\newline 
Right:  Plot of the relative asymmetry as function of $\lambda$.
\label{fig:asymm}}
\end{figure}

\noindent
The results are given in table~\ref{tab1}.
\begin{table}[t]
$\begin{array}{|@{\,}l@{\,}|@{\,}c@{\,}|@{\,}c@{\,}|@{\,}c@{\,}|@{\,}c@{\,}|@{\,}r@{\,}|@{\,}r@{\,}|@{\,}c@{\,}|@{\,}c@{\,}|}
\hline
\multicolumn{9}{|l|}{\text{Approximation $G^{\tt i}$ as
function of  $\lambda$, with $\sup_b|G^{\tt i}_{0b}
-G^{\tt i-1}_{0b}|<4{\times}10^{-8}$
\rule{0pt}{6mm}}}
\\
\multicolumn{9}{|l|}{\text{{\tt co=10\^{}7}, {\tt len=2000}, 
{\tt infty=10\^{}(10)}}
\rule[-3mm]{0pt}{8mm}}
\\ \hline
\;\lambda\pi  & {\tt i} & 
G^{\tt i}_{{\tt co},0} & {\tt AbsAsm} & \texttt{RelAsm} & 
A+Bx~~~ & C+Dx~~~ & \mathcal{Y} & \lambda_{\mathit{eff}}\pi
\\ \hline
3.0 &  7  & 
1.0{\times} 10^{-11} & 
1.0{\times} 10^{-1} & 1.0000 & 0.433{-}1.588x 
& 1.47{-} 2.595 x & 0.340 & 3.117
\\ \hline
2.5 &  7 & 
1.1{\times} 10^{-11} & 6.5{\times} 10^{-2} & 1.0000 & 0.446{-}1.587x 
& 1.87{-} 2.571 x & 0.334 & 2.613
\\ \hline
2.0 &  7 & 
1.1{\times} 10^{-11} & 
3.0{\times} 10^{-2} & 1.0000 & 0.473{-}1.583x 
& 2.35{-} 2.520 x & 0.322 & 2.107
\\ \hline
1.8 &  7 & 
1.2{\times} 10^{-11} & 1.8{\times} 10^{-2} & 1.0000 & 0.492{-}1.580x 
& 2.57{-} 2.482 x & 0.314 & 1.904
\\ \hline
1.6 &  8 & 
1.3{\times} 10^{-11} & 8.6{\times} 10^{-3} & 1.0000 & 0.518{-}1.575x 
& 2.76{-} 2.420 x & 0.303 & 1.700
\\ \hline 
1.4 &  9 & 
1.5{\times} 10^{-11} & 2.6{\times} 10^{-3} & 1.0000 & 0.556{-}1.566x 
& 2.85{-} 2.304 x & 0.288 & 1.494
\\ \hline 
1.3 &  9 & 
1.7{\times} 10^{-11} & 9.6{\times} 10^{-4} & 1.0000 & 0.580{-}1.558x 
& 2.75{-} 2.198 x & 0.279 & 1.389
\\ \hline 
1.2 &  11 & 
2.1{\times} 10^{-11} & 1.9{\times} 10^{-4} & 1.0000 & 0.603{-}1.544x 
& 2.34{-} 2.017 x & 0.267 & 1.282
\\ \hline 
1.15 &  11 & 
2.5{\times} 10^{-11} & 4.6{\times} 10^{-5} & 1.0000 & 0.608{-}1.531x 
& 1.86{-} 1.868 x & 0.261 & 1.228
\\ \hline 
1.1 &  12 & 
3.3{\times} 10^{-11} & 4.3{\times} 10^{-5} & 0.9999 & 0.594{-}1.511x 
& 0.99{-} 1.649 x
& 0.254 & 1.171
\\ \hline 
1.05 &  14 & 
8.5{\times} 10^{-11} & 4.1{\times} 10^{-5} & 0.9990 & 0.474{-}1.462x 
& 0.006 {-} 1.443x & 0.246 & 1.113
\\ \hline 
1.0 &  18 & 
2.1{\times} 10^{-10} & 3.8{\times} 10^{-5} & 0.9906 &  0.313{-}1.404x 
& {-}0.188 {-} 1.392x & 0.238 & 1.055
\\ \hline
0.95 &  25 & 
3.7{\times} 10^{-10} & 3.5{\times} 10^{-5} & 0.9667  & 0.235{-}1.365x 
& {-}0.274 {-} 1.360x & 0.229 & 0.997
\\ \hline
0.9 &  36 & 
5.7{\times} 10^{-10} & 3.3{\times} 10^{-5} & 0.9231 & 0.189{-}1.335x 
& {-}0.329 {-} 1.333x & 0.220 & 0.940
\\ \hline
0.8 &  58 & 
1.2{\times} 10^{-9} & 2.8{\times} 10^{-5} & 0.7864 & 0.124{-}1.284x 
& {-}0.403 {-} 1.288x & 0.202 & 0.828
\\ \hline
0.7 &  57 & 
2.4{\times} 10^{-9} & 2.3{\times} 10^{-5} & 0.6396 & 0.081{-}1.240x & 
{-}0.455 {-} 1.247x & 0.182 & 0.719
\\ \hline
0.6 &  42 & 
4.4{\times} 10^{-9} & 1.8{\times} 10^{-5} & 0.3825 & 0.053{-}1.200x & 
{-}0.493 {-} 1.209x & 0.160 & 0.612
\\ \hline
0.5 &  29 & 
7.8{\times} 10^{-9} & 1.4{\times} 10^{-5} & 0.3892 & 
0.032{-} 1.164 x & {-}0.521 {-} 1.174 x & 0.137 & 0.507
\\ \hline
0.4 &  20 & 
1.3{\times} 10^{-8} & 1.1{\times} 10^{-5} & 0.4038 & 
0.017{-} 1.129 x & {-}0.541 {-} 1.140 x & 0.113 & 0.404
\\ \hline
0.3 &  13 & 
2.2{\times} 10^{-8} & 7.3{\times} 10^{-6} & 0.4240 & 
0.007{-} 1.095 x & {-}0.554 {-} 1.107 x & 0.087 & 0.303
\\ \hline 
0.2 &  10 & 
3.7{\times} 10^{-8} & 4.5{\times} 10^{-6} & 0.4371 & 
0.001{-} 1.063 x & {-}0.560 {-} 1.075 x & 0.060 & 0.200
\\ \hline 
0.1 &  8 & 
6.1{\times} 10^{-8} & 2.0{\times} 10^{-6} & 0.4491 & 
{-}0.001{-} 1.031 x & {-}0.562 {-} 1.043 x & 0.031 & 0.100
\\ \hline 
0.05 &  7 & 
7.8{\times} 10^{-8} & 9.5{\times} 10^{-7} & 0.4546 & 
{-}0.001{-} 1.016 x & {-}0.561 {-} 1.028 x & 0.016 & 0.050
\\ \hline 
0.0 &  2 & 
1.0{\times} 10^{-7} & 0 & 0 & 0.000{-} 1.000 x & 
{-}0.558 {-} 1.012 x & 0 & 0
\\ \hline 
\end{array}$
\caption{Fixed point solution $G_{0a}$ of (\ref{master}) for 
$\lambda \geq 0$ \label{tab1}}
\end{table}
We notice a sudden increase of the absolute asymmetry 
$\displaystyle{\tt AbsAsm}= \sup_{0\leq a < b
    \leq \Lambda^2} |G_{ab}-G_{ba}|$
for $\lambda >
\frac{1.1}{\pi}$ which goes hand in hand with qualitative change of the
function $a\mapsto G_{aa}$: Whereas for $0\leq \lambda <
\frac{1.05}{\pi}$ the slopes $B,C$ are comparable, we find for $\lambda >
\frac{1.1}{\pi}$ that $D$ grows much faster than $B$ and stabilises at
$D\approx B+1$ for $\lambda> \frac{2}{\pi}$. A look at the relative
asymmetry
$\displaystyle{\tt RelAsm}= \sup_{0\leq a < b
    < \Lambda^2} \tfrac{|G_{ab}-G_{ba}|}{G_{ab}+G_{ba}}$
shows, however, that the true transition already occurs
at\footnote{Interestingly, this value coincides with slowest
  convergence of the iteration. The phase transition 
at $\lambda_c \approx -\frac{1.24}{\pi}$ identified in the next
subsection also coincides with slowest convergence rate for $\lambda<0$.
A possible origin is a change of sign of $Z$ for finite $\Lambda$, see (\ref{Z}).
}
$\lambda \approx \frac{0.7}{\pi}$. We show on the left of 
fig.~\ref{fig:asymm}
the absolute and relative asymmetry in a logarithmic scale and on the
right of fig.~\ref{fig:asymm} the  
relative asymmetry in a linear scale, including the results
for $\lambda<0$ obtained in sec.~\ref{sec:la<0}. The plot of the
relative asymmetry identifies three clearly different regions in $\lambda$: For
$\lambda<0$ the function $G_{ab}$ is symmetric up to small discretisation errors
of a few percent. For $0<\lambda < \frac{0.7}{\pi}$ we have some
$40\%$ asymmetry, whereas for larger $\lambda$ the asymmetry strongly
increases to nearly $100\%$ for $\lambda> \frac{1}{\pi}$.

\begin{figure}[b]
\begin{picture}(150,50)
  \put(0,0){\includegraphics[width=7.8cm,
bb=0 0 288 173]{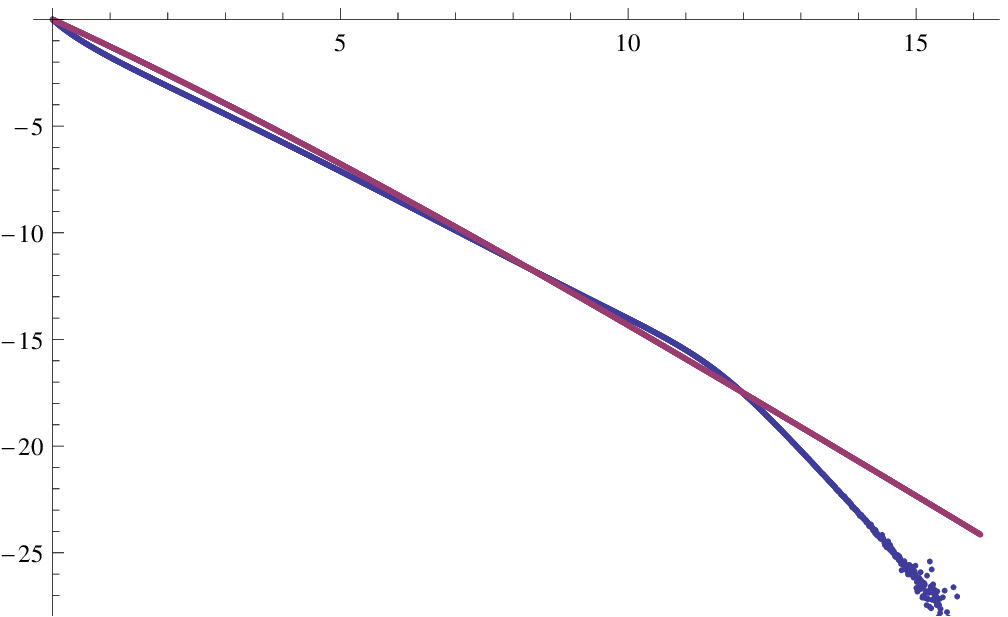}}
  \put(57,44){\mbox{\scriptsize$\log(1{+}a)$}}
  \put(20,15){\fbox{\mbox{$\lambda=\frac{1.1}{\pi}$}}}
  \put(100,15){\fbox{\mbox{$\lambda=\frac{1.0}{\pi}$}}}
  \put(138,44){\mbox{\scriptsize$\log(1{+}a)$}}
  \put(57,5){\mbox{\small$\log G_{aa}$}}
  \put(142,8){\mbox{\small$\log G_{aa}$}}
  \put(65,15){\mbox{\small$\log G_{a0}$}}
  \put(147,17){\mbox{\small$\log G_{a0}$}}
  \put(82,0){\includegraphics[width=7.8cm,
bb=0 0 288 173]{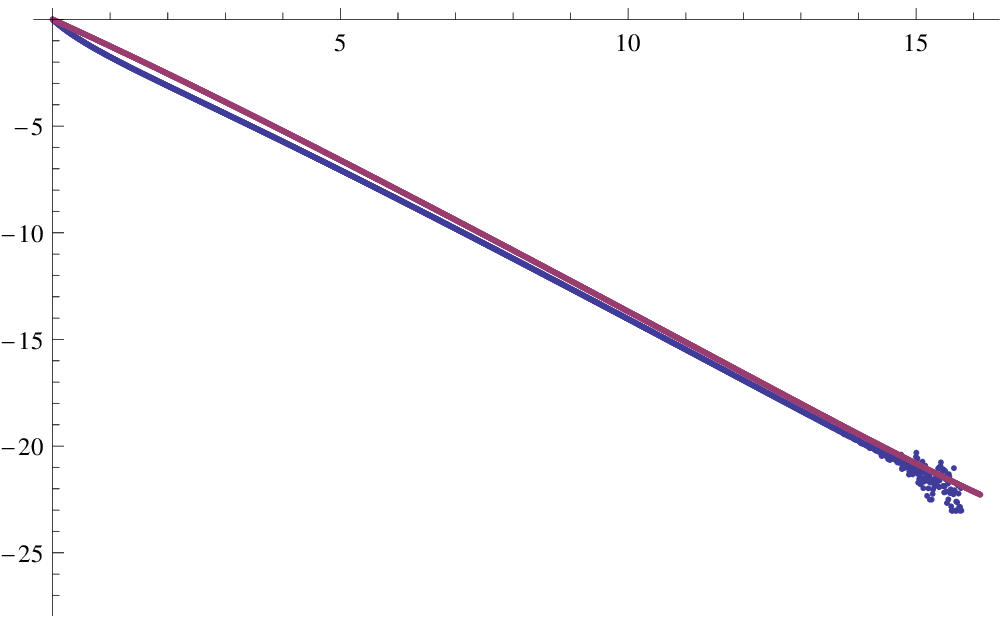}}
\end{picture}
\caption{Comparison of the functions $\log G_{a0}$ (red) and $\log
  G_{aa}$ (blue)  over $\log(1+a)$ for $\lambda=\frac{1.1}{\pi}$
  (left) with  $\lambda=\frac{1.0}{\pi}$
  (right). The buckle in $\log G_{aa}$ for $\lambda=\frac{1.1}{\pi}$
  moves to smaller $a$ with increasing $\lambda$. At larger cut-off
  $\Lambda^2=10^8$ there is also a buckle for $\lambda=\frac{1.0}{\pi}$.
\label{fig:buckle}}
\end{figure}

The region of $\lambda> \frac{1}{\pi}$ where the relative asymmetry
nearly reaches $100\%$ also shows a qualitative change of the
function $a\mapsto G_{aa}$. As shown in fig.~\ref{fig:buckle},
there is a critical $a$ where the slope of $\log(1+a)\mapsto \log
G_{aa}$ suddenly decreases by $1$. This critical $a$ gets larger for
smaller $\lambda$. It is near the cut-off \texttt{co=}$10^7$ 
for $\lambda=\frac{1}{\pi}$ and
moves into $[0,10^7]$ for larger $\lambda$. This observation
lets us conjecture that \emph{the entire region
$\lambda>0$  shows a critical value of $a$
where the slope  of $\log(1+a)\mapsto \log
G_{aa}$ decreases by $1$}. We tend to think that as $\lambda$ increases
from $0$ to $\frac{1.0}{\pi}$, the decrease of slope becomes more and more
visible and hence induces the jump of the relative asymmetry to nearly
100\% in the region $\frac{0.6}{\pi} < \lambda < \frac{1.0}{\pi}$.

\begin{figure}[t]
\begin{picture}(150,47)
  \put(0,0){\includegraphics[width=7.8cm,
bb=0 0 288 162]{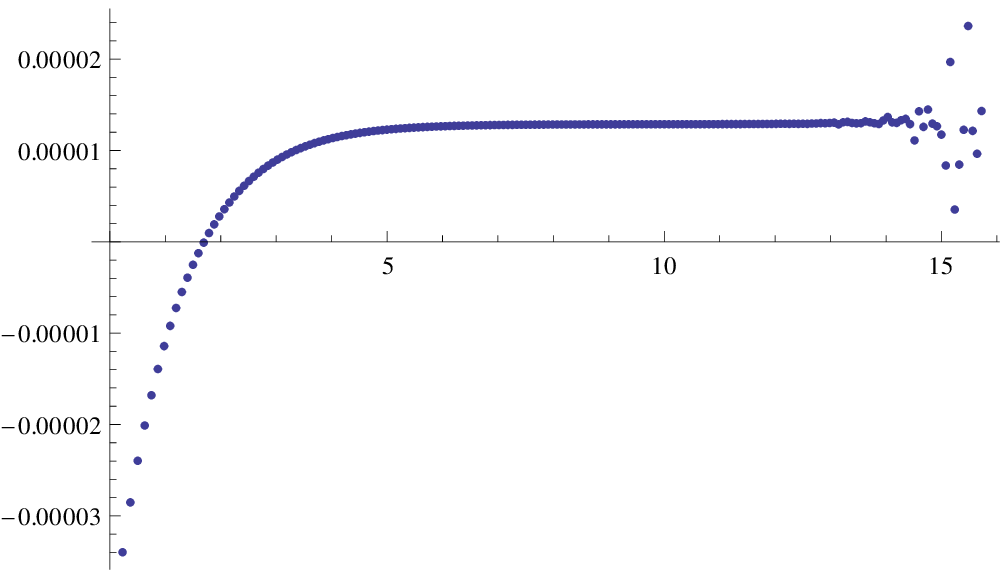}}
  \put(60,23){\mbox{\scriptsize$\log(1{+}a)$}}
  \put(20,15){\mbox{$C_{\lambda,\Lambda^2}$ for
      $\lambda=\frac{1.1}{\pi}$}}
  \put(100,15){\mbox{$C_{\lambda,\Lambda^2}$ for $\lambda=\frac{2.0}{\pi}$}}
  \put(140,0){\mbox{\scriptsize$\log(1{+}a)$}}
  \put(82,0){\includegraphics[width=7.8cm,
bb=0 0 288 178]{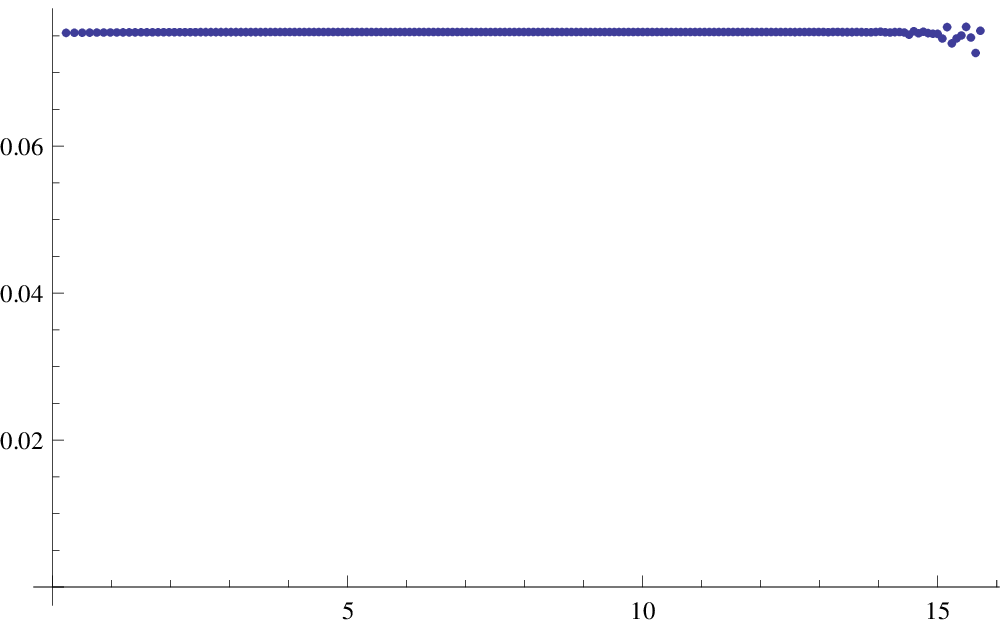}}
\end{picture}
\caption{The  parameter $C_{\lambda,\Lambda}$ plotted as
numerical function of $a$. The result should be constant within
numerical errors.
\label{fig:ClL}}
\end{figure}

There remains this background relative asymmetry of $\approx 40\%$
which is not explained by the decrease of slope. We trace this back to
$f_{\lambda,\Lambda^2}\neq 0$ in (\ref{Gab}) by the following
investigation: The decreased slope by $1$ is naturally interpreted as
missing factor $1+\frac{C_{\lambda,\Lambda^2} \Lambda^2
  a}{\Lambda^2-a}$.  We derive in (\ref{ClL}) the formula for
$C_{\lambda,\Lambda^2}$ under the assumption
$f_{\lambda,\Lambda^2}(b)=0$. This formula is implemented in {\tt\bf
  In[\ref{calc:ClL}]}.  Necessary for validity of
$f_{\lambda,\Lambda^2}(b)=0$ is that {\tt ClL[lis\_, hilb\_, xi\_,
  a\_]} is (up to numerical errors) independent of $a$.  Typical
results are shown in fig.~\ref{fig:ClL}.
We notice that the function is approximately constant in a middle
region $\exp(5)\approx 150 \leq a \leq \exp(14) \approx 1.2\times
10^6$. For larger $a$ the noise is too large, whereas for small $a$
together with smaller $\lambda$ there is a clear discrepancy. We see
this as indication that \emph{also the assumption
  $f_{\lambda,\Lambda^2}=0$ is not justified for $\lambda>0$}.

\subsection{Varying $\lambda \leq 0$: Consistency and 
evidence for phase transition}
\label{sec:la<0}

\begin{table}[h!] 
$\begin{array}{|c|c|c|c|c|c|c|c|}
\hline
\multicolumn{8}{|l|}{\text{Approximation $G^{\tt i}$ as
function of  $\lambda$, with $\sup_b|G^{\tt i}_{0b}
-G^{\tt i-1}_{0b}|<4{\times}10^{-8}$
\rule{0pt}{6mm}}}
\\
\multicolumn{8}{|l|}{\text{{\tt co=10\^{}7}, {\tt len=2000}, 
{\tt infty=10\^{}(10)}}
\rule[-3mm]{0pt}{8mm}}
\\ \hline
{\tt \lambda\pi}  & {\tt i} & G^{\tt i}_{0,100} &
G^{\tt i}_{0,{\tt co}} & {\tt AbsAsm} & {\texttt{RelAsm}} & 
\mathcal{Y} & \lambda_{\mathit{eff}}\pi
\\ \hline
\phantom{-}0.00 &  2 & 0.00990 & 1.0{\times} 10^{-7} & 0 & 0 & 
0 & 0
\\ \hline 
-0.10 &  14 & 0.01149 & 1.6{\times} 10^{-7} & 
1.6{\times}10^{-6} & 0.0129 & 
-0.0329 & -0.100
\\ \hline 
-0.20 &  20 & 0.01339 & 
2.7{\times} 10^{-7} & 2.6{\times}10^{-6} &  0.0215 & 
-0.0680 & -0.201
\\ \hline
-0.30 &  25 & 0.01569 & 
4.4{\times} 10^{-7} & 3.3{\times}10^{-6} &0.0278 & 
-0.1059 & -0.302
\\ \hline
-0.40 &  32 & 0.01854 & 
7.4{\times} 10^{-7} & 3.5{\times} 10^{-6} & 0.0325 & 
-0.1468 & -0.405
\\ \hline
-0.50 &  40 & 0.02213 & 
1.3{\times} 10^{-6} & 3.5{\times} 10^{-6} & 0.0362 & 
-0.1915 & -0.511
\\ \hline
-0.60 &  50 & 0.02678 & 2.2{\times} 10^{-6} & 
3.2{\times}10^{-6} & 0.0392 & 
-0.2409 & -0.621
\\ \hline
-0.70 &  64 & 0.03305 & 
3.9{\times} 10^{-6} & 5.1{\times}10^{-6} & 0.0416 & 
-0.2969 & -0.737
\\ \hline
-0.80 &  83 & 0.04199 & 
7.4{\times} 10^{-6} & 9.0{\times}10^{-6} &0.0437 & 
-0.3620 & -0.864
\\ \hline
-0.90 &  113 & 0.05582 & 
1.5{\times} 10^{-5} & 1.5{\times}10^{-5} &0.0454 & 
-0.4416 & -1.009
\\ \hline
-1.00 &  162 & 0.08016 & 
3.4{\times} 10^{-5} & 2.6{\times}10^{-5} & 0.0470 & 
-0.5459 & -1.196
\\ \hline
-1.05 &  200 & 0.10071 & 
5.5{\times} 10^{-5} & 3.4{\times} 10^{-5} & 0.0476 & 
-0.6132 & -1.321
\\ \hline
-1.10 &  251 & 0.13275 & 
9.4{\times} 10^{-5} & 4.9{\times}10^{-5} & 0.0483 & 
-0.6947 & -1.490
\\ \hline
-1.15 &  320 & 0.18679 & 1.7{\times} 10^{-4} & 
7.7{\times}10^{-5} & 0.0489 & 
-0.7912 & -1.737
\\ \hline
-1.18 &  366 & 0.23790 & 
2.7{\times} 10^{-4} & 1.1{\times}10^{-4} & 0.0493 & 
-0.8526 & -1.943
\\ \hline
-1.20 &  396 & 0.28352 & 
3.6{\times} 10^{-4} & 1.4{\times} 10^{-4}  & 0.0495 & 
-0.8913 & -2.106
\\ \hline
-1.22 &  420 & 0.34038 & 
4.9{\times} 10^{-4} & 1.9{\times} 10^{-4}  & 0.0497 & 
-0.9253  & -2.279
\\ \hline
-1.23 &  428 & 0.37323 & 
5.8{\times} 10^{-4} & 2.2{\times} 10^{-4}  & 0.0498 & 
-0.9397  & -2.361
\\ \hline
-1.24 &  433 & 0.40886 & 
6.8{\times} 10^{-4} & 2.5{\times} 10^{-4}  & 0.0499 & 
-0.9521  & -2.434
\\ \hline
-1.25 &  434 & 0.44695 & 
7.9{\times} 10^{-4} & 2.9{\times} 10^{-4}  & 0.0500 & 
-0.9626  & -2.487
\\ \hline
-1.26 &  432 & 0.48697 & 
9.3{\times} 10^{-4} & 3.4{\times} 10^{-4}  & 0.0501 & 
-0.9711  & -2.506
\\ \hline
-1.28 &  419 & 0.57004 & 
1.3{\times} 10^{-3} & 4.5{\times} 10^{-4}  & 0.0503 & 
-0.9831  & -2.341
\\ \hline
-1.30 &  399 & 0.65194 & 
1.7{\times} 10^{-3} & 5.9{\times} 10^{-4}  & 0.0506 & 
-0.9902 & -1.480
\\ \hline
-1.32 &  375 & 0.72699 & 
2.2{\times} 10^{-3} & 8.1{\times} 10^{-4}  & 0.0508 & 
-0.9943 & 1.477
\\ \hline
-1.35 &  340 & 0.81897 & 
3.1{\times} 10^{-3} & 1.0{\times} 10^{-3}  & 0.0511 & 
-0.9974 & 21.37
\\ \hline
-1.40 &  288 & 0.91426 & 
5.2{\times} 10^{-3} & 1.6{\times} 10^{-3}  & 0.0515 &
-0.9995 & 614.6
\\ \hline
-1.45 &  248 & 0.95914 & 
7.9{\times} 10^{-3} & 2.3{\times} 10^{-3}  & 0.0520 &
-1.0002 & 7812
\\ \hline
-1.50 &  217 & 0.97952 & 
1.1{\times} 10^{-2} & 3.1{\times} 10^{-3}  & 0.0524 & 
-1.0005 & 956.5
\\ \hline
-1.60 &  173 & 0.99421 & 
2.0{\times} 10^{-2} & 5.0{\times} 10^{-3}  & 0.0532 & 
-1.0008 & 399.4
\\ \hline
-2.00 &  93 & 1.00193 & 
7.1{\times} 10^{-2} & 1.5{\times} 10^{-2}  & 0.0559 & 
-1.0017 & 162.2
\\ \hline
-2.50 &  59 & 1.00383 & 
1.5{\times} 10^{-1} & 3.4{\times} 10^{-2}  & 0.0584 & 
-1.0030 & 88.61
\\ \hline
-3.00 &  44 & 1.00561 & 
2.3{\times} 10^{-1} & 6.1{\times} 10^{-2}  & 0.0604 & 
-1.0052 & 98.29
\\ \hline
\end{array}$
\caption{The fixed point solution $G_{0b}$ of (\ref{master}) for
  $\lambda \leq 0$ for $L=2000$ sample points. \label{tab:la<0}}
\end{table}

\begin{figure}[h!]
\begin{picture}(110,110)
  \put(0,0){\includegraphics[width=11cm,bb=0 0 260 260]{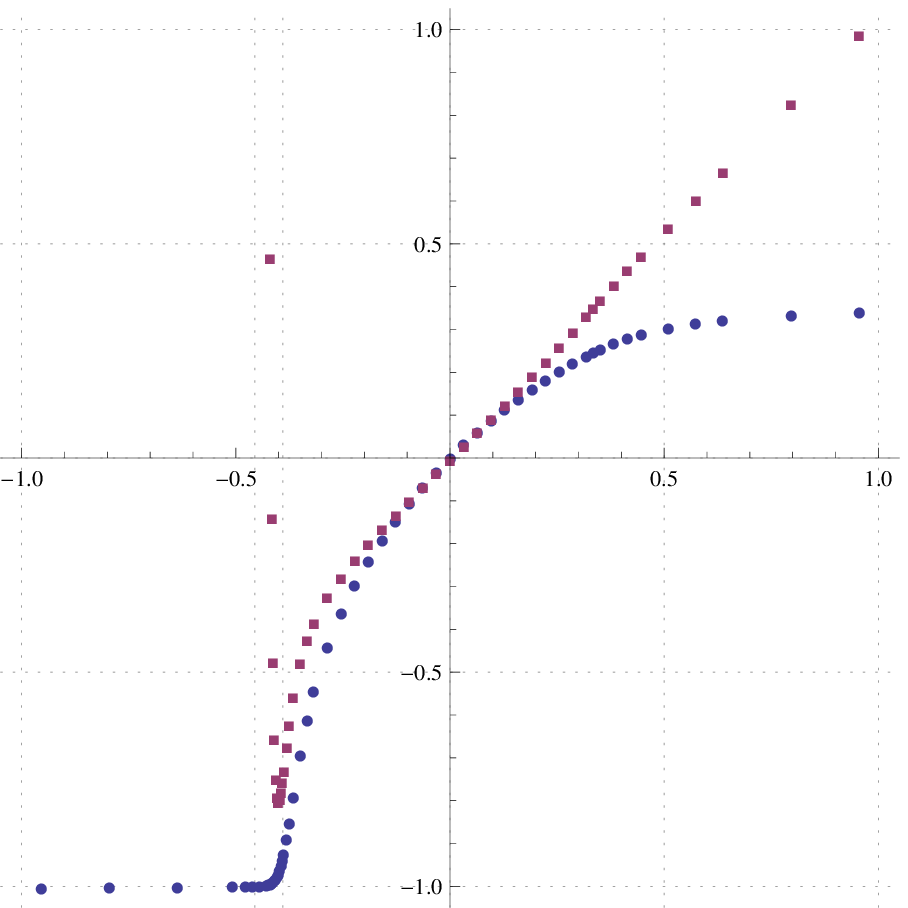}}
  \put(100,51){\mbox{\small$\lambda$}}
  \put(27,57){\mbox{\scriptsize$\lambda_0$}}
  \put(34,57){\mbox{\scriptsize$\lambda_c$}}
  \put(92,10){\mbox{\textcolor{mathblue}{
  \textbullet{\small$\;\mathcal{Y}$}}}}
  \put(92,5){\mbox{\textcolor{mathred}{{\tiny$\blacksquare$}
             {\small$\lambda_{\mathit{eff}}$}}}}
\end{picture}
\caption{Plot of the finite wavefunction renormalisation $\mathcal{Y}$
  (blue dots) and of the effective coupling constant
  $\lambda_{\mathit{eff}}$ (red squares) as function of $\lambda$. The
  points $\lambda_0 \approx -\frac{1.432}{\pi}$ where
  $\mathcal{Y}(\lambda_0)=-1$ and $\lambda_c\approx
  -0.39$ where $\mathcal{Y}'(\lambda_c)$ is discontinuous
  are indicated as two of the dashed grid lines.
\label{fig:calY}}
\end{figure}

Next we study the dependence of $G_{0b}$ and $G_{aa}$ on $\lambda \leq
0$. As before we start with $G^0_{0b}=1$ and run the iteration until
$|G^{\tt i}-G^{\tt i-1}|_\infty <4\times 10^{-8}$. The results are
listed in table~\ref{tab:la<0} for $L=2000$ sample points.  We partly
use these results in \cite{Grosse:2015fka} to check that $G^{\tt i}$
falls into the expected region $\exp (\mathcal{K}_\lambda)$ given in
(\ref{Klambda}) \emph{although we start from $G^{\tt 0}=1$ which
  should be an exact solution for $\Lambda^2\to \infty$ \cite[Appendix
  A]{Grosse:2015fka}}. We therefore conclude that at finite $\Lambda$
the difference between
$G^1_{0b}=\frac{1}{1+\frac{b}{1+|\lambda|\Lambda^2}}$ and $G^0_{0b}=1$
is enough to drive the iteration away from $1$ and into another fixed
point solution $G^\infty_{0b}\neq 1$. We find a monotonic
convergence\footnote{For $\lambda>0$ there was always an alternating
  convergence $G^{\tt i} < G^{\tt i+2}< G^{\tt i+3} < G^{\tt i+1}$
  (for {\tt i} either even or odd.} of $\{G^{\tt i}\}$ for {\tt i}
sufficiently large. Together with the boundedness of $\{G^{\tt i}\}$
proved in \cite{Grosse:2015fka}, monotonicity (if rigorously proved)
is enough for uniqueness of the fixed point soulution.

We notice that the relative asymmetry is
roughly constant at a few percent and thus much smaller than for
$\lambda>0$ (see fig.~\ref{fig:asymm}). This is a clear signal that
the sector $\lambda>0$ is affected by the undetermined quantities
$C_{\lambda,\Lambda^2}$ and $f_{\lambda,\Lambda^2}(b)$ of
(\ref{Gab}) whereas the sector $\lambda<0$ is completely determined.

\begin{figure}[h!]
\begin{picture}(150,50)
  \put(0,0){\includegraphics[width=7.8cm,bb=0 0 288 175]{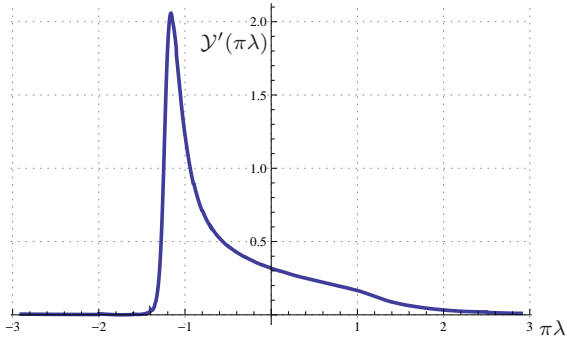}}
  \put(71,0){\mbox{\scriptsize$\pi\lambda$}}
  \put(26,38){\mbox{\scriptsize$\mathcal{Y}'(\pi\lambda)$}}
\end{picture}
\caption{Plot of $\mathcal{Y}'(\lambda)$ as function of $\pi\lambda$.
\label{fig:DcalY}}
\end{figure}

\begin{figure}[t]
\begin{picture}(150,50)
  \put(0,0){\includegraphics[width=7.8cm,bb=0 0 288 175]{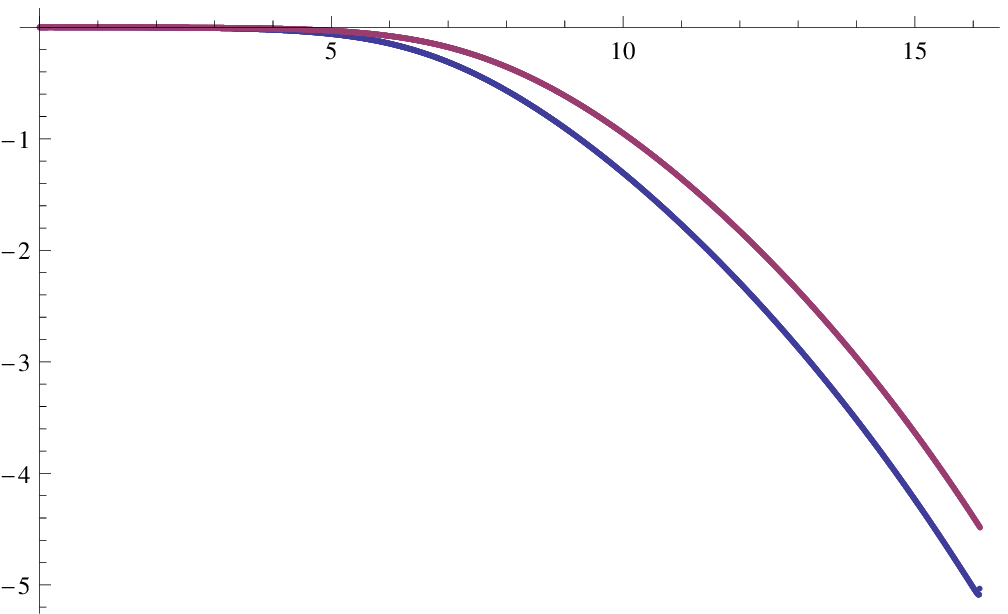}}
  \put(60,47){\mbox{\scriptsize$\log(1{+}a)$}}
  \put(57,32){\mbox{\textcolor{mathred}{\small$\log G_{0a}$}}}
  \put(50,18){\mbox{\textcolor{mathblue}{\small$\log G_{aa}$}}}
  \put(10,10){\fbox{\mbox{\small$\lambda=\frac{1.5}{\pi}$}}}
  \put(82,0){\includegraphics[width=7.8cm,bb=0 0 288 171]{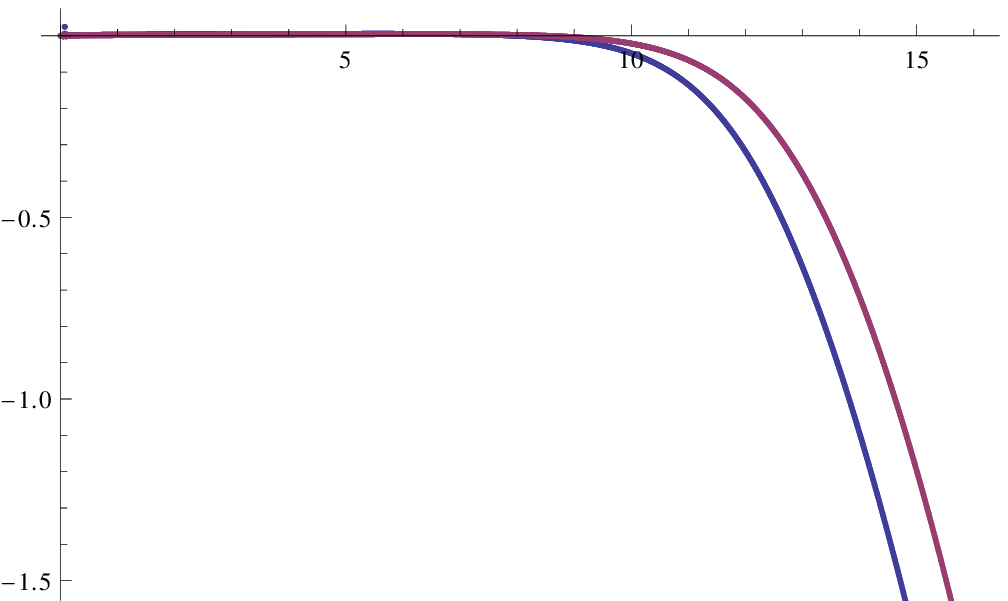}}
  \put(142,47){\mbox{\scriptsize$\log(1{+}a)$}}
  \put(146,34){\mbox{\textcolor{mathred}{\small$\log G_{0a}$}}}
  \put(132,20){\mbox{\textcolor{mathblue}{\small$\log G_{aa}$}}}
  \put(92,10){\fbox{\mbox{\small$\lambda=\frac{2.5}{\pi}$}}}
\end{picture}
\caption{Plot of $\log G_{0a}$ (red) and 
$\log G_{aa}$ (blue) as function of $\log(1+a)$ for $\lambda<\lambda_c$. 
\label{fig:Y=-1}}
\end{figure}

The most striking observation is the behaviour of the finite
wavefunction renormalisation $\mathcal{Y}=\mathcal{Y}_1$ 
which determines $\frac{dG_{0b}}{db}\big|_{b=0}=-(1+\mathcal{Y})$. 
As shown in fig.~\ref{fig:calY} (which also includes $\lambda>0$),
\emph{$\mathcal{Y}(\lambda)$ undergoes a second order phase transition at
$\lambda_c \approx 0.39$} where
$\mathcal{Y}'(\lambda)$ is discontinuous. To be precise, there is no
discontinuity in $\mathcal{Y}'(\lambda)$ at finite cut-off
$\Lambda^2={\tt co}$, only a large jump. Fig.~\ref{fig:DcalY}
shows $\mathcal{Y}'(\lambda)$ as function of $\pi\lambda$, where
$\mathcal{Y}(\pi\lambda)$ is obtained by cubic interpolation of
tables~\ref{tab1} and \ref{tab:la<0}.  The maximum
$\mathcal{Y}'(\pi\lambda_{\max})=2.058$ is attained at
$\pi\lambda_{\max}=-1.163$, i.e.\ $\lambda_{\max}=-0.370$, and the half value 
$\mathcal{Y}'(\pi\lambda_h)=
\frac{1}{2}\mathcal{Y}'(\pi\lambda_{\max})
\big|_{\lambda_c<\lambda_{\max}}$ at $\pi\lambda_h=-1.246$. Combined with
results on the Stieltjes property we assign the value 
$\lambda_c \approx -0.39$, i.e.\ $\pi\lambda_c \approx -1.225$ as
critical coupling constant.
Another possibility would be the point
$\mathcal{Y}(\lambda_0)=-1$ for which we find $\lambda_0 \approx
-\frac{1.432}{\pi} \approx -0.455$. At $\lambda_0$ all higher
correlation functions, in particular $\lambda_{\mathit{eff}}$ become
singular\footnote{Note that $\lambda_0$ is far beyond the pole
  $\lambda_B$ of Borel resummation of planar graphs. This pole is
  given as $\hat{\lambda}=-\frac{1}{12}$ in the literature, but for
  the normalisation $-\frac{\hat{\lambda}}{4!}\phi^4$ of the quartic
  interaction whereas we worked with $-\frac{\lambda}{4}\phi^4=
  -\frac{6\lambda}{4!}\phi^4$.  This means that the Borel pole would
  be at $6\lambda_B=-\frac{1}{12}$, i.e.\ $\lambda_B=-\frac{1}{72}
  =-0.013888\dots$. We cannot identify any particular behaviour at
  $\lambda_B$.}. It came as surprise to us that the iteration is still
convergent for $\lambda<\lambda_0$. Even more surprising is that
$\mathcal{Y}(\lambda)\approx \mathcal{Y}(\lambda_0)=-1$ stays roughly
constant for $\lambda<\lambda_0$. In fact, within the reliability
bound given by the relative asymmetry of $5\%$ we can regard
$\mathcal{Y}(\lambda)=-1$ for all $\lambda \leq \lambda_c$. This is
also supported by fig.~\ref{fig:Y=-1}
which shows plots of both
$G_{0a},G_{aa}$ over $a$ for $\lambda>\lambda_c$. We notice that both
$G_{0a},G_{aa}$ and in fact also $G_{ab}$ all equal $1$ within the
reliability bound of $5\%$ for $0\leq a,b \leq A(\lambda)$. 
In other words, the exact solution $G_{0b}=1$ of 
\cite[Appendix A]{Grosse:2015fka} 
becomes stable for $b\in [0,A(\lambda)]$. The end point
$A(\lambda)=\sup\{b\;:G_{0b}=1\}$ serves as an order
parameter: We have $A(\lambda)=0$ in the phase $\lambda_c < \lambda \leq 0$
and $A(\lambda)>0$ for $\lambda < \lambda_c$.

For some values of $\lambda$ we have computed $G_{0b}$ at finer
resolution, see table \ref{tab:la<0-L}. 
\begin{table}[h!] \vspace*{-2mm}
$\begin{array}{|c|c|c|c|c|c|c|c|c|}
\hline
\multicolumn{9}{|l|}{\text{Approximation $G^{\tt i}$ as
function of  $\lambda$ and $L$, with $\sup_b|G^{\tt i}_{0b}
-G^{\tt i-1}_{0b}|<4{\times}10^{-8}$
\rule{0pt}{6mm}}}
\\
\multicolumn{9}{|l|}{\text{{\tt co=10\^{}7}, 
{\tt infty=10\^{}(10)}}
\rule[-3mm]{0pt}{8mm}}
\\ \hline
{\tt \lambda\pi} & {\tt len}  & {\tt i} & G^{\tt i}_{0,100} &
G^{\tt i}_{0,{\tt co}} & {\tt AbsAsm} & {\texttt{RelAsm}} & 
\mathcal{Y} & \lambda_{\mathit{eff}}\pi
\\ \hline
-0.80 & 2000&  83 & 0.04199 & 
7.4{\times} 10^{-6} & 9.0{\times}10^{-6} &0.0437 & 
-0.3620 & -0.864
\\
-0.80 & 10000&  3 & 0.04199 & 
7.4{\times} 10^{-6} & 5.9{\times}10^{-7} &0.0385 & 
-0.3620 & -0.864
\\
-0.80 & 40000&  2 & 0.04199 & 
7.4{\times} 10^{-6} & 5.4{\times}10^{-7} &0.0350 & 
-0.3620 & -0.864
\\ \hline
-0.90 & 2000& 113 & 0.05582 & 
1.5{\times} 10^{-5} & 1.5{\times}10^{-5} &0.0454 & 
-0.4416 & -1.009
\\
-0.90 & 10000& 13 & 0.05582 & 
1.5{\times} 10^{-5} & 1.2{\times}10^{-6} &0.0399 & 
-0.4415 & -1.009
\\
-0.90 & 40000& 5 & 0.05582 & 
1.5{\times} 10^{-5} & 1.1{\times}10^{-6} &0.0361 & 
-0.4415 & -1.009
\\ \hline
-1.00 & 2000&  162 & 0.08016 & 
3.4{\times} 10^{-5} & 2.6{\times}10^{-5} & 0.0470 & 
-0.5459 & -1.196
\\
-1.00 & 10000&  35 & 0.08015 & 
3.4{\times} 10^{-5} & 2.9{\times}10^{-6} & 0.0411 & 
-0.5458 & -1.196
\\
-1.00 & 40000&  12 & 0.08015 & 
3.4{\times} 10^{-5} & 2.6{\times}10^{-6} & 0.0370 & 
-0.5458 & -1.196
\\ \hline
-1.05 & 2000&  200 & 0.10071 & 
5.5{\times} 10^{-5} & 3.4{\times} 10^{-5} & 0.0476 & 
-0.6132 & -1.321
\\
-1.05 & 10000&  52 & 0.10069 & 
5.5{\times} 10^{-5} & 4.8{\times} 10^{-6} & 0.0416 & 
-0.6131 & -1.321
\\
-1.05 & 40000&  20 & 0.10069 & 
5.5{\times} 10^{-5} & 4.3{\times} 10^{-6} & 0.0374 & 
-0.6130 & -1.321
\\ \hline
-1.10 & 2000&  251 & 0.13275 & 
9.4{\times} 10^{-5} & 4.9{\times}10^{-5} & 0.0483 & 
-0.6947 & -1.490
\\
-1.10 & 10000&  78 & 0.13270 & 
9.4{\times} 10^{-5} & 8.2{\times}10^{-6} & 0.0421 & 
-0.6945 & -1.490
\\
-1.10 & 40000&  34 & 0.13270 & 
9.4{\times} 10^{-5} & 7.4{\times}10^{-6} & 0.0378 & 
-0.6945 & -1.490
\\ \hline
-1.15 &2000&   320 & 0.18679 & 1.7{\times} 10^{-4} & 
7.7{\times}10^{-5} & 0.0489 & 
-0.7912 & -1.737
\\
-1.15 &10000&  117 & 0.18665 & 1.7{\times} 10^{-4} & 
1.5{\times}10^{-5} & 0.0426 & 
-0.7909 & -1.737
\\
-1.15 &40000&  56 & 0.18665 & 1.7{\times} 10^{-4} & 
1.4{\times}10^{-5} & 0.0382 & 
-0.7909 & -1.737
\\ \hline
-1.22 & 2000&  420 & 0.34038 & 
4.9{\times} 10^{-4} & 1.9{\times} 10^{-4}  & 0.0497 & 
-0.9253  & -2.279
\\
-1.22 & 10000&  192 & 0.33979 & 
4.9{\times} 10^{-4} & 4.4{\times} 10^{-5}  & 0.0432 & 
-0.9248  & -2.287
\\
-1.22 & 40000&  103 & 0.33976 & 
4.9{\times} 10^{-4} & 3.9{\times} 10^{-5}  & 0.0387 & 
-0.9248  & -2.288
\\ \hline
\end{array}$
\caption{The fixed point solution $G_{0b}$ of (\ref{master}) for
  $\lambda \leq 0$ for various resolutions {\tt len}$=L$. Finer
  resolutions provide a significant decrease of asymmetry.
The iteration for {\tt len=2000} starts with $G^0_{0b}=1$, whereas
for {\tt len=10000} and {\tt len=40000} we start with $G^0_{0b}$ given 
by an interpolation of the previous $G^{\tt i}_{0b}$ at  
{\tt len=2000} and {\tt len=10000}, respectively. The values for {\tt len=40000} 
had a total computation time of 4 months!
\label{tab:la<0-L}}
\end{table}
We notice a considerable
improvement of the asymmetry\footnote{$G_{\Lambda^2,b}$ differs significantly from
$G_{b,\Lambda^2}$ so that we measure the relative asymmetry only for 
$a,b<\Lambda^2$.}, whereas $G_{0b}$, $\mathcal{Y}$ and 
$\lambda_{\mathit{eff}}$ are nearly independent of $L$.

\subsection{The Stieltjes property of the 2-point function}

Necessary for reflection positivity is that $a\mapsto G_{aa}$ is a
Stieltjes function \cite{Grosse:2013iva}, which by
\cite{Widder:1938??} is equivalent to $L_{n,t}[G_{\bullet\bullet}]\geq
0$ for all $n,t$. These functions (\ref{Widder}) are implemented 
in {\bf In[\ref{WidderI}]} using an
interpolation method. We have shown results for several values of 
$\lambda<0$ in \cite[Fig.~3]{Grosse:2014nza}. 
These interpolation results allowed to exclude the Stieltjes property
for $\lambda<-\frac{1.25}{\pi}\approx -0.398$. 
For $\lambda>0$ we already have
$L_{1,t}[G_{\bullet\bullet}]<0$ for sufficiently large $t$. Hence
there remained a window $\lambda \in [-0.398,0]$ where the 
interpolation results were not conclusive. In this paper we
investigate the remaining window by means of the formulae of
Proposition~\ref{Prop:Stieltjes}. 

It turns out that our numerical results based on the 
\emph{Mathematica}$^{TM}$ implementation in 
Appendix~\ref{sec:implementation} are
affected by systematic discretisation errors. 
These errors are unavoidable. The Stieltjes property or the (weaker
but better accessible)
complete monotonicity property encode a strong form of analyticity in 
the cut plane $\mathbb{C}\setminus {]{-}\infty,0]}$ or the half space 
$\mathrm{Re}(z)>0$, respectively. A
piecewise-linear approximation cannot share such properties. But because we test
the decisive properties by integral formulae, we expect that at finer
resolution $L$ and larger cut-off $\Lambda^2$ we recover more and more
the true behaviour of the solution. Our numerical results confirm this
expectation and thus provide strong support, albeit no proof, of the 
following
\begin{Conjecture}
The diagonal matrix 2-point function $G_{aa}$ of the $\lambda \phi^4_4$-model 
on noncommutative Moyal space in the
limit of infinite noncommutativity is a Stieltjes function for 
$\lambda_c<\lambda \leq 0$.
\label{conj-Stieltjes}
\end{Conjecture}

\begin{figure}[t]
\begin{picture}(150,43)
  \put(0,0){\includegraphics[width=7.8cm,bb=0 0 288 175]{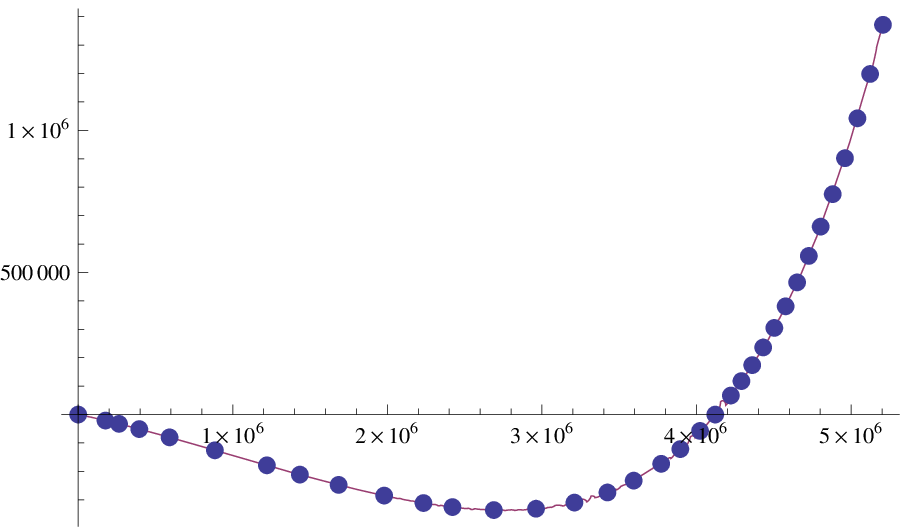}}
  \put(82,0){\includegraphics[width=7.8cm,bb=0 0 288 171]{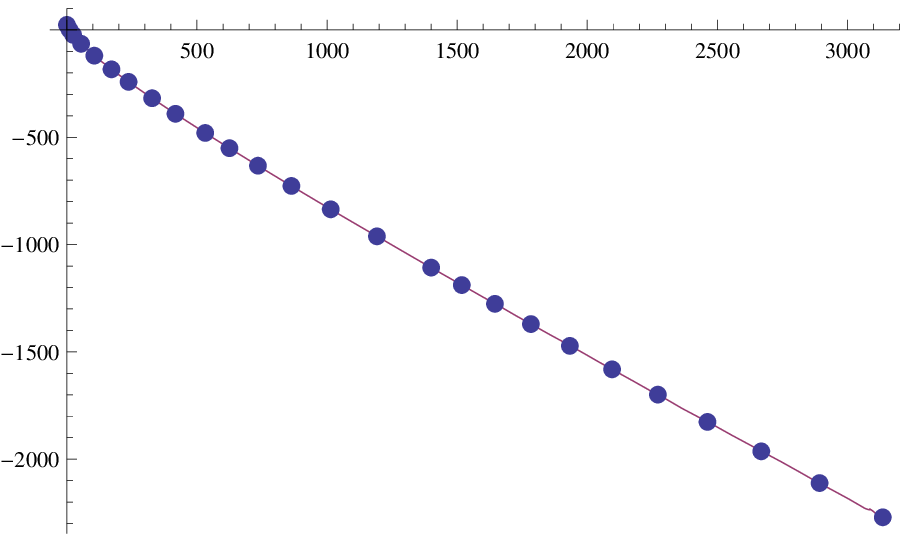}}
\end{picture}
\caption{Comparison of interpolation (solid line) and integral 
formulae (dots) for 
$C^2_0(a)$ (left) and 
$C^4_0(a)$ (right) at $\lambda\pi=-1.25$. The integral formulae are
based on $\Lambda^2=10^7$ and $L=2000$ sample points.
\label{fig:dcottau}}
\end{figure}

\begin{figure}[t]
\begin{picture}(150,43)
  \put(0,0){\includegraphics[width=7.8cm,bb=0 0 288 175]{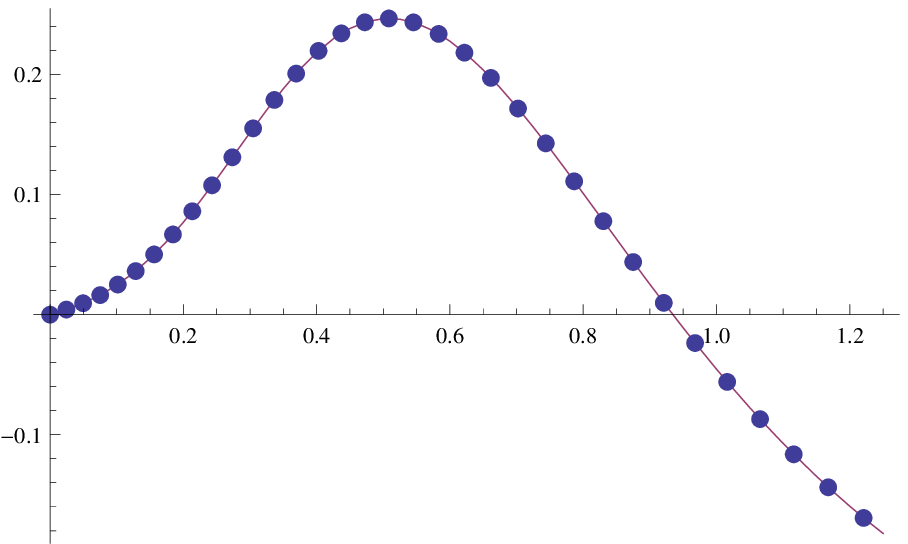}}
  \put(82,0){\includegraphics[width=7.8cm,bb=0 0 288 171]{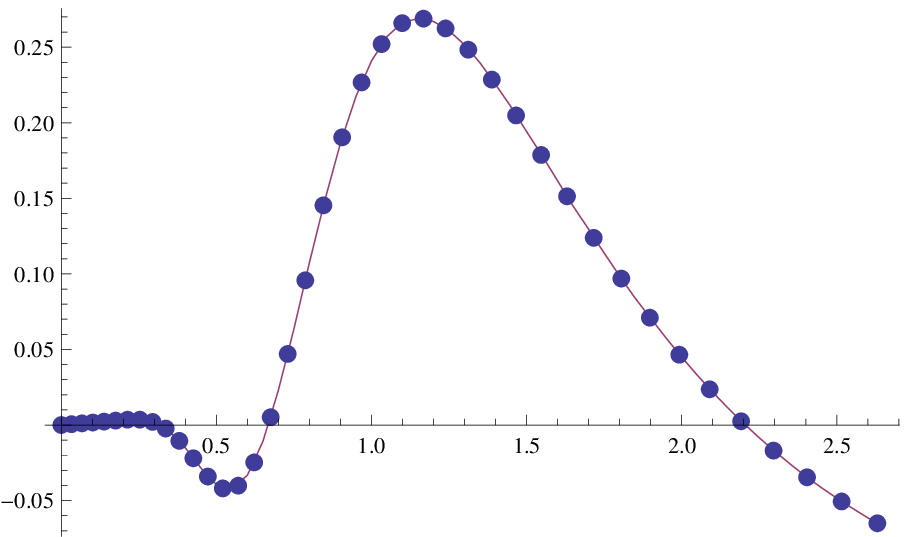}}
\end{picture}
\caption{Comparison of interpolation (solid line) and integral 
formulae (dots) for 
$\frac{(-a)^2}{2!} \frac{d^2}{da^2}
(\mathrm{sign}(\lambda)
\mathcal{H}_a^{\tilde{\Lambda}}[\tau_0(\bullet)])$ 
(left) and 
$\frac{(-a)^5}{5!} \frac{d^5}{da^5}
(\mathrm{sign}(\lambda)
\mathcal{H}_a^{\tilde{\Lambda}}[\tau_0(\bullet)])$ 
(right) at $\lambda\pi=-1.10$,  $\Lambda^2=10^7$ and 
 $\tilde{\Lambda}^2=(10^7+1)^{\frac{3}{5}}-1$.
The integral formulae are based on $L=2000$ sample points.
\label{fig:DHtau}}
\end{figure}

We first provide some checks for the correctness of 
Proposition~\ref{Prop:Stieltjes} and 
its implementation. We show in fig.~\ref{fig:dcottau} 
for the function $C_0^n(a)$ defined in 
(\ref{Cbna}) a comparison between the numerical differentiation of 
$\cot \tau_0(a)$ and the integral formula.
In fig.~\ref{fig:DHtau} we compare for the function
$\frac{(-a)^n}{n!} \frac{d^n}{da^n}
(\mathrm{sign}(\lambda)
\mathcal{H}_a^{\tilde{\Lambda}}[\tau_0(\bullet)])$ 
the numerical differentiation with the integral formula (which is 
(\ref{dlogGab-ab})  without the $L^{(n,\ell)}(a,b)$-term taken at
$\ell=0$ and $b=0$, and with insertion of (\ref{Anl})).
For small $a$ this function should be independent of the cut-off
$\tilde{\Lambda}$. We confirm in table~\ref{tab:dHtau}
\begin{table}[h]
$\begin{array}{|c||r|r|r|r|r|} \hline
K & (n,\ell){=}(6,1) & (n,\ell){=}(5,0) & (n,\ell){=}(0,5) & 
(n,\ell){=}(1,0)&(n,\ell){=}(0,1) \\ \hline
50  & 2.40622\phantom{000}  & -6.21325\phantom{0000} & 0.0000129836
& -0.556715\phantom{0} & -0.0065814\\
80  & 0.00182617&-0.00479778\phantom{0}& 0.0000162935&-0.0837978&-0.0548360\\
100 & 0.00145747&-0.00021191\phantom{0}& 0.0000170968&-0.0149419&-0.0714913\\
150 & 0.00143641&  0.000222779 & 0.0000171231 &0.0426437&-0.0931625\\
200 & 0.00143644&  0.000229322 & 0.0000171114 &0.0592362&-0.102702\phantom{0}\\
400 & 0.00143644&  0.000229763 & 0.0000171104 &0.0724140&-0.113218\phantom{0}\\
800 & 0.00143643&  0.000229765 & 0.0000171104 &0.0743314&-0.115140\phantom{0}\\
1200& 0.00143643&  0.000229765 & 0.0000171104 &0.0744010&-0.115211\phantom{0}\\
1800& 0.00143643&  0.000229765 & 0.0000171104 &0.0744037&-0.115214\phantom{0}\\
1900& 0.00143643&  0.000229764 & 0.0000171104 &0.0744037&-0.115214\phantom{0}\\
1990& 0.00143643&  0.000229636 & 0.0000171104 &0.0744037&-0.115214\phantom{0}\\
2000& 0.00143643&  0.00474417\phantom{0} & 0.0000171104&0.0744037
&-0.115214\phantom{0}\\ \hline
\end{array}$
\caption{$A^{(n,\ell)}(x_{40},x_{20})$ for $\lambda\pi=-1.10$ as
  function of the secondary cutoff $\tilde{\Lambda}^2=x_{K}$. The
  primary cut-off is $x_{L+1}=10^7$ with $L=2000$.
\label{tab:dHtau}}
\end{table}
that, as long as $\tilde{\Lambda}^2 \gg a$ and up to boundary artifacts for 
$\tilde{\Lambda}^2=\Lambda^2$, the function $A^{(n,\ell)}$ is indeed independent
of $\tilde{\Lambda}^2$. We use this independence in order to take 
a comparably low value $K=1200$, corresponding to 
$1+\tilde{\Lambda}^2 = (1+\tilde{\Lambda}^2)^{\frac{3}{5}}$, for our
simulation in order to save computing time.

\begin{figure}[t]
\begin{picture}(150,95)
  \put(0,50){\includegraphics[width=8.4cm,bb=0 0 288
    175]{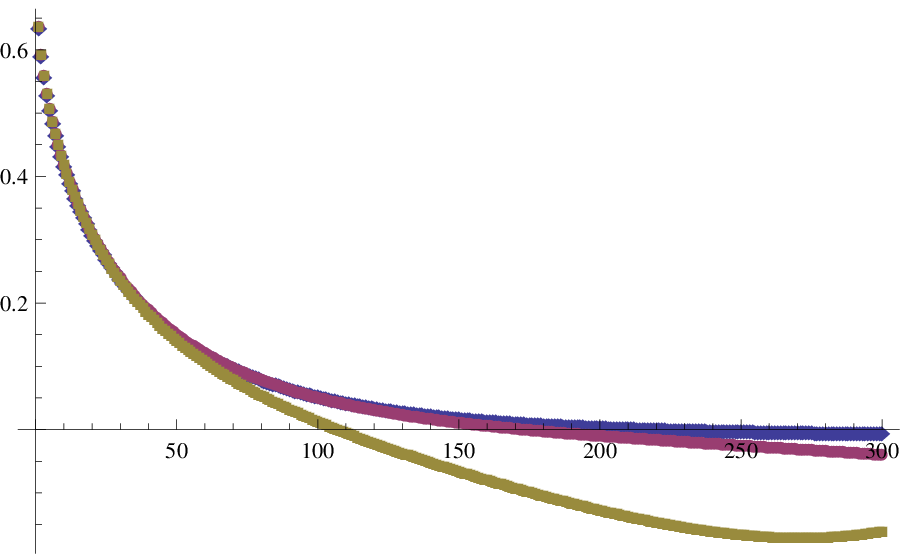}}
  \put(8,90){\fbox{$\lambda=-0.255$}}
  \put(32,51){\mbox{\small$L=2000$}}
  \put(60,55){\mbox{\small$L=10000$}}
  \put(60,63){\mbox{\small$L=40000$}}
  \put(47,91){\mbox{\small$n^{\mathcal{L}}(2000)=109$}}
  \put(45,86){\mbox{\small$n^{\mathcal{L}}(10000)=179$}}
  \put(45,81){\mbox{\small$n^{\mathcal{L}}(40000)=266$}}
  \put(82,50){\includegraphics[width=8.4cm,bb=0 0 288 171]{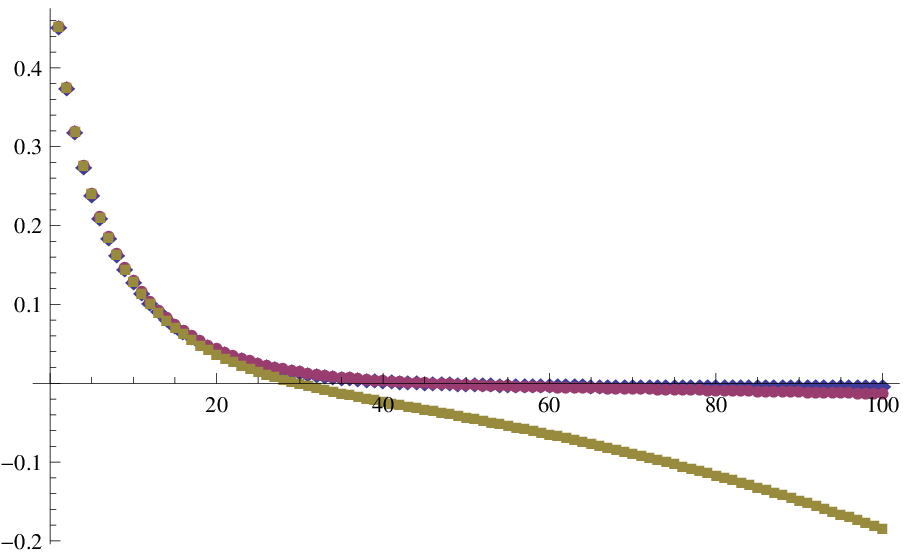}}
  \put(95,90){\fbox{$\lambda=-0.318$}}
  \put(120,55){\mbox{\small$L=2000$}}
  \put(140,59){\mbox{\small$L=10000$}}
  \put(140,67){\mbox{\small$L=40000$}}
  \put(132,91){\mbox{\small$n^{\mathcal{L}}(2000)=31$}}
  \put(130,86){\mbox{\small$n^{\mathcal{L}}(10000)=49$}}
  \put(130,81){\mbox{\small$n^{\mathcal{L}}(10000)=69$}}
  \put(0,0){\includegraphics[width=8.4cm,bb=0 0 288
    175]{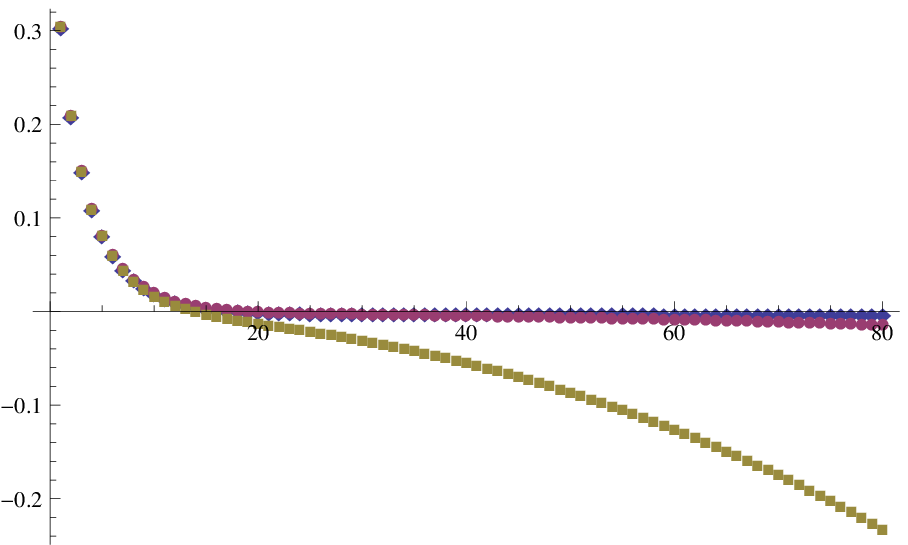}}
  \put(8,40){\fbox{$\lambda=-0.350$}}
  \put(36,8){\mbox{\small$L=2000$}}
  \put(55,14.5){\mbox{\small$L=10000$}}
  \put(55,21){\mbox{\small$L=40000$}}
  \put(47,41){\mbox{\small$n^{\mathcal{L}}(2000)=15$}}
  \put(45,36){\mbox{\small$n^{\mathcal{L}}(10000)=23$}}
  \put(45,31){\mbox{\small$n^{\mathcal{L}}(40000)=30$}}
  \put(82,0){\includegraphics[width=8.4cm,bb=0 0 288 171]{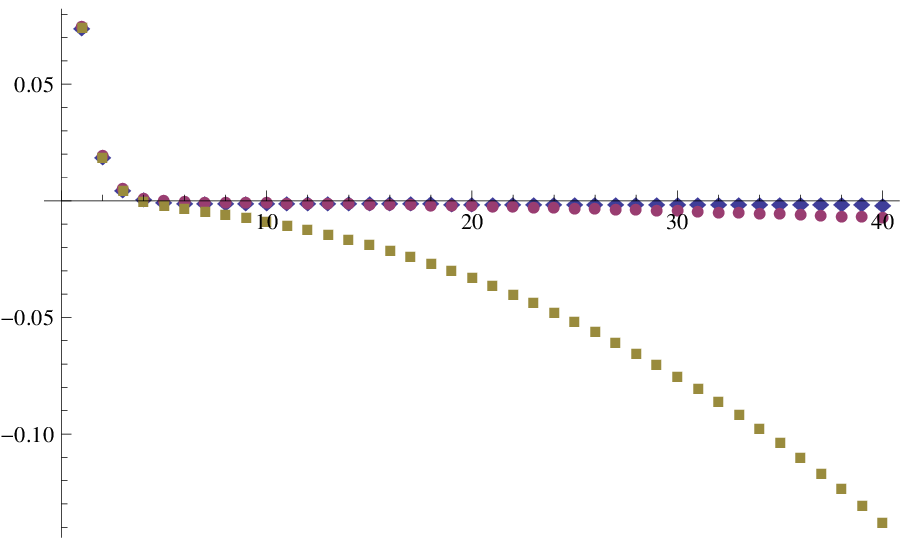}}
  \put(95,40){\fbox{$\lambda=-0.388$}}
  \put(140,14){\mbox{\small$L=2000$}}
  \put(135,23){\mbox{\small$L=10000$}}
  \put(135,31){\mbox{\small$L=40000$}}
  \put(92,11){\mbox{\small$n^{\mathcal{L}}(2000)=5$}}
  \put(90,6){\mbox{\small$n^{\mathcal{L}}(10000)=6$}}
  \put(90,1){\mbox{\small$n^{\mathcal{L}}(40000)=7$}}
\end{picture}
\caption{Plots of $\frac{(-1)^n}{(n-1)!}
(\log G_{0b})^{(n)}\big|_{b=0}$ as function of
  $n$ for various resolutions $L$. The boundary 2-point 
function $G_{0b}$ is logarithmically 
completely monotonic if these curves are positive for all $n$. The
discretisations fail this property at a critical index 
$n^{\mathcal{L}0}(L)$.
\label{Fig:Logcmon}}
\end{figure}

The strongest support for Conjecture~\ref{conj-Stieltjes} 
comes from the observation that the critical indices 
\begin{itemize} \itemsep 0pt\parskip 0pt
\item $n^{\mathcal{L}0}=\min\{n\;:~(-1)^n (\log G_{0b})^{(n)}\big|_{b=0} <0\}$,
\item $n^{\mathcal{L}}=\min\{n\;:~(-1)^n (\log G_{0b})^{(n)} <0\text{ for some
  }b\}$
  where logarithmically complete monotonicity of $G_{0b}$ fails,

\item $n^{\mathcal{C}}=\min\{n\;:~(-1)^n G_{0b}^{(n)} <0\text{ for some
  }b\}$
  where complete monotonicity of $G_{0b}$ fails,

\item $n_0^{\mathcal{S}}=\min\{n\;:~L_{n,t}(G_{0\bullet})<0\text{ for some
  }t\}$ where the Stieltjes  property of $G_{0b}$ fails, 

\item $n^{\mathcal{S}}=\min\{n\;:~L_{n,t}(G_{\bullet\bullet})<0\text{ for some
  }t\}$ where the Stieltjes  property of $G_{aa}$ fails,

\end{itemize}
satisfy for all tested values of $L$ and $\lambda$ the following 
relations:
\begin{align}
  n^{\mathcal{L}0}=n^{\mathcal{L}}\;,\quad
n^{\mathcal{C}} \gtrapprox n^{\mathcal{L}}\;,\quad
n_0^{\mathcal{S}}\geq n^{\mathcal{C}}+1\;.
\label{nLS}
\end{align}
Our results are given (together with $n^{\mathcal{S}}$ discussed
below) in table~\ref{tab:LS}.
\begin{table}[ht]
\parbox[b]{80mm}{$
\begin{array}{|c|r|r|r|c|c||c|}
\hline 
\lambda\pi  & L~~ & n^{\mathcal{L}0} & n^{\mathcal{L}}  &
n^{\mathcal{C}} & n^{\mathcal{S}}_0 & n^{\mathcal{S}} \\ \hline
-0.80  & 2000  & 109 & 109 &    &    & \\
-0.80  & 10000 & 179 & 179 &    &    & \\
-0.80  & 40000 & 266 & 266 &    &    & \\ \hline
-0.90  & 2000  & 58  & 58  &    &    &\\
-0.90  & 10000 & 95  & 95  &    &    & \\ 
-0.90  & 40000 & 137 & 137 &    &    & \\ \hline
-1.00  & 2000  & 31  & 31  & 35 & 37 &\\
-1.00  & 10000 & 49  & 49  & 55 &    & \\ 
-1.00  & 40000 & 69  & 69  & 75 &    & \\ \hline
-1.05  & 2000  & 22  & 22  & 25 & 26 &\geq 11\\
-1.05  & 10000 & 34  & 34  & 38 & 39 & \\ 
-1.05  & 40000 & 47  & 47  & 51 &    & \\ \hline
-1.10  & 2000  & 15  & 15  & 17 & 18 &\geq 11 \\ 
-1.10  & 10000 & 23  & 23  & 25 & 26 & \\ 
-1.10  & 40000 & 30  & 30  & 33 & 34 & \\ \hline
-1.15  & 2000  & 9   & 9   & 10 & 11 & 8 \\ 
-1.15  & 10000 & 14  & 14  & 15 & 16 & > 10  \\ 
-1.15  & 40000 & 18  & 18  & 20 & 21  &     \\ \hline
-1.20  & 2000  & 6   & 6   & 6  & 7  & 6 \\ \hline
-1.22  & 2000  & 5   & 5   & 5  & 6  & 5 \\ 
-1.22  & 10000 & 6   & 6   & 7  & 8  & 6 \\ 
-1.22  & 40000 & 7   & 7   & 9  & 10 &   \\ \hline
\end{array}
$}\quad
\parbox{80mm}{\begin{picture}(105,60)
  \put(0,-31){\includegraphics[width=7.5cm,bb=0 0 196 312]{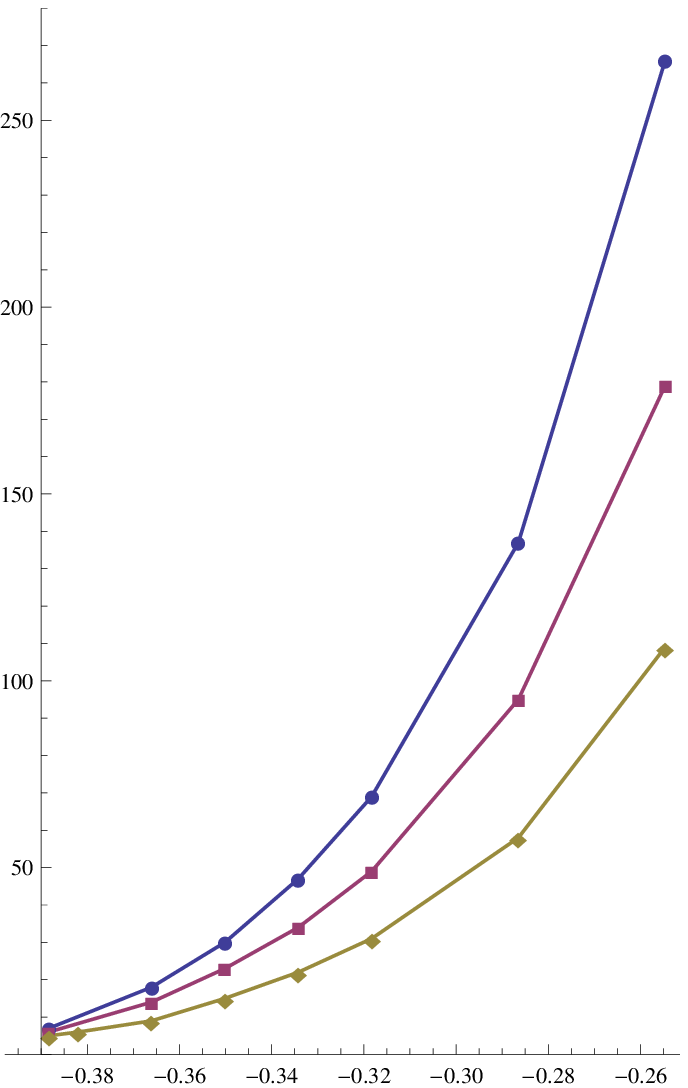}}
\put(49,-13){\mbox{\small$L=2000$}}
\put(58,25){\mbox{\small$L=10000$}}
\put(47,55){\mbox{\small$L=40000$}}
\end{picture}}
\caption{The critical indices where Stieltjes and complete
  monotonicity properties fail, as function of $\lambda$ and of the 
number $L$ of sample points. 
Because of discretisation errors we define $n^{\mathcal{S}}$ in a
coarse manner in (\ref{ns-new}), and due to noise we can only give
lower bounds in some cases. 
The figure shows 
$n^{\mathcal{L}}$ as function of $L$ and $\lambda$.
%
%
%
%
\label{tab:LS}}
\end{table}
The computation of $n^{\mathcal{L}0}$ is 
fast and therefore to perform to large values.
We can then look at $(-1)^n(\log G_{0b})^n$ for
$n\in\{n^{\mathcal{L}0},n^{\mathcal{L}0}-1\}$ and notice that 
the first wrong sign always arises for $b=0$. The computation of
$n^{\mathcal{C}}$ involves via the Bell polynomials $Y_{n,k}$ a sum
over all partitions of $n$. For $n\gtrapprox 80$ this cannot be done
anymore in reasonable time. For the same reason we can only test
$n^{\mathcal{S}}_0\lessapprox 40$. The relation $n^{\mathcal{C}}\geq
n^{\mathcal{L}}$ is clear by definition. But $n^{\mathcal{S}}_0$ and
$n^{\mathcal{C}}$ are, a priori, uncorrelated because a typical
completely monotonic function has no reason to be Stieltjes. The
observed relation $n^{\mathcal{S}}_0 \geq n^{\mathcal{C}}+1$ is
therefore extremely strong support for the claim that a 
completely monotonic solution of (\ref{master}) is automatically
Stieltjes for $\lambda_c<\lambda\leq 0$. Together with the 
observed dependence of $1+\mathcal{Y}_n=\frac{(-1)^n}{(n-1)!} (\log
G_{0b})^{(n)}\big|_{b=0}$ on $n$, shown for 
selected values of $\lambda$ and $L$ in fig.~\ref{Fig:Logcmon}, 
we have overwhelming support for the assertion that \emph{the exact
solution $G_{0b}$ of (\ref{master}) is  
a Stieltjes function for $\lambda_c<\lambda\leq
0$}. Near the critical coupling constant $\lambda_c$ the improvement 
at higher resolution $L$ slows down. In agreement with previous
considerations on the discontinuity of $\frac{d\mathcal{Y}}{d\lambda}$ 
we confirm that 
$\lambda=-\frac{1.22}{\pi}\approx -0.388$ is already very close to
$\lambda_c$, which we would define as the critical value where a 
finite $n^{\mathcal{L}}(\infty)$ remains. The curves in
fig.~\ref{Fig:Logcmon} suggest that for $0\leq b\leq 1$ one has a
power series representation
\begin{align}
G_{0b}=\exp\Big(\sum_{n=1}^\infty \frac{(-1)^n}{n} c_n
b^n\Big)\;,~ 
(c_n)_n \text{ positive monotonously decreasing null sequence}\;.
\raisetag{2ex}
\end{align}

\begin{figure}[t]
\begin{picture}(150,100)
  \put(0,55){\includegraphics[width=7.8cm,
bb=0 0 288 176]{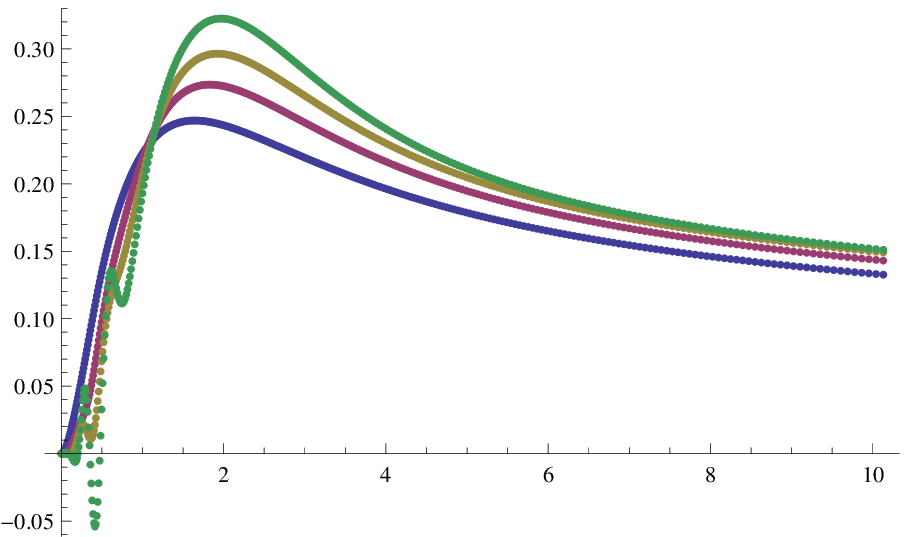}}
  \put(40,68){\fbox{$\lambda=-0.382$}}
  \put(47,92){\fbox{$n^{\mathcal{S}}=6$}}
 \put(30,78){\mbox{\small$L_3$}}
 \put(35,88){\mbox{\small$L_6$}}
  \put(82,55){\includegraphics[width=7.8cm,
bb=0 0 288 183]{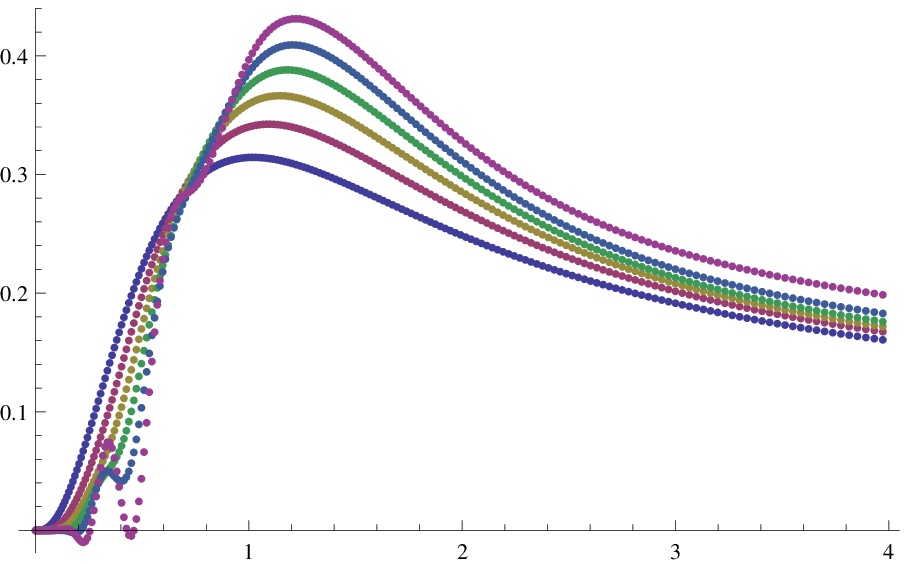}}
  \put(110,63){\fbox{$\lambda=-0.366$}}
 \put(110,80){\mbox{\small$L_4$}}
 \put(115,94){\mbox{\small$L_9$}}
 \put(136,92){\mbox{\small\fbox{$n^{\mathcal{S}}=8$}}}
\put(0,0){\includegraphics[width=7.8cm,
bb=0 0 288 176]{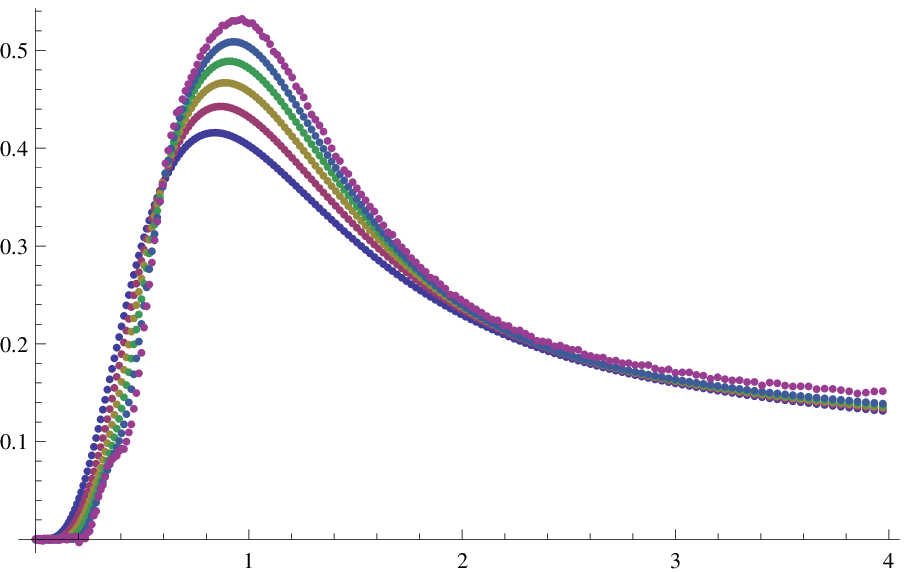}}
  \put(20,10){\fbox{$\lambda=-0.350$}}
  \put(44,38){\fbox{$n^{\mathcal{S}}=11$}}
 \put(21,24){\mbox{\small$L_6$}}
 \put(24,39){\mbox{\small$L_{11}$}}
\put(82,0){\includegraphics[width=7.8cm,
bb=0 0 288 176]{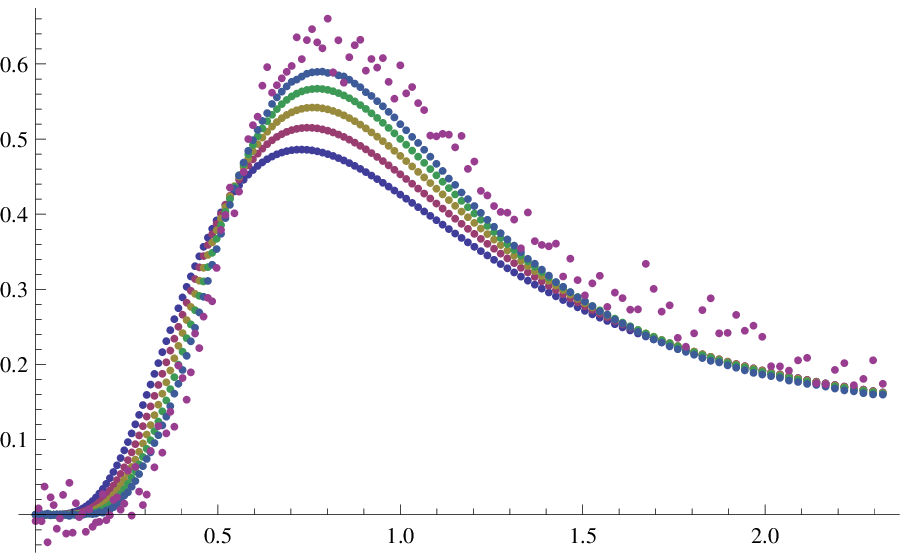}}
  \put(102,10){\fbox{$\lambda=-0.334$}}
  \put(134,38){\fbox{$n^{\mathcal{S}}=11$}}
 \put(112,23){\mbox{\small$L_7$}}
 \put(115,40){\mbox{\small$L_{12}$}}
\end{picture}
\caption{Widder's operations $L_{n,t}[G_{\bullet\bullet}]$ at
 $\lambda\pi \in \{{-}1.20,\;{-}1.15,\;{-}1.10,\;{-}1.05\}$ and
 $L=2000$ sample points. 
In order to define a Stieltjes
  function, $L_{n,t}$ has to be non-negative for all $n$ and $t$. 
For $n\geq 12$ there is too much noise to be conclusive. At 
$\lambda\pi \in \{{-}1.20,\;{-}1.15\}$ the curves turn negative,
but also these oscillations are possibly discretisation artifacts.
  \label{fig:Widder}}
\end{figure}

\begin{figure}[t]
\begin{picture}(150,48)
  \put(0,0){\includegraphics[width=7.8cm,bb=0 0 288
    175]{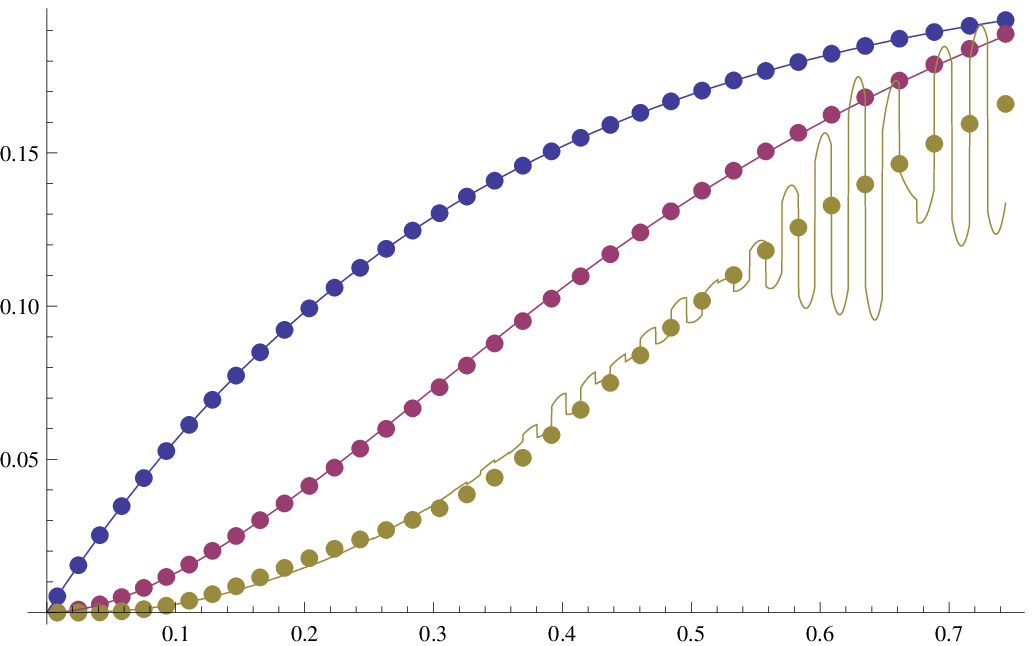}}
\put(24,32){\mbox{\small$L_2$}}
\put(42,32){\mbox{\small$L_3$}}
\put(50,18){\mbox{\small$L_4$}}
  \put(82,0){\includegraphics[width=7.8cm,bb=0 0 288
    171]{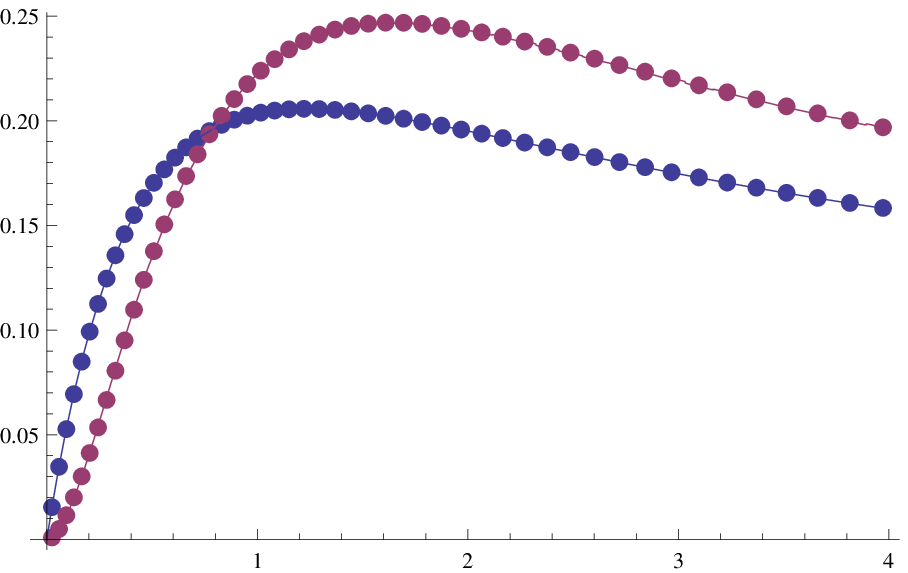}}
\put(127,43){\mbox{\small$L_3$}}
\put(113,30){\mbox{\small$L_2$}}
\end{picture}
\caption{Comparison of interpolation (solid line) and integral 
formulae (dots) for 
$L_{n,a}[G_{\bullet\bullet}]$ at $\lambda\pi=-1.20$. The integral formulae are
based on $\Lambda^2=10^7$ and $L=2000$ sample points.
\label{fig:Stieltjes-1.20}}
\end{figure}

\begin{figure}[t]
\begin{picture}(150,45)
  \put(0,0){\includegraphics[width=7.8cm,bb=0 0 288
    175]{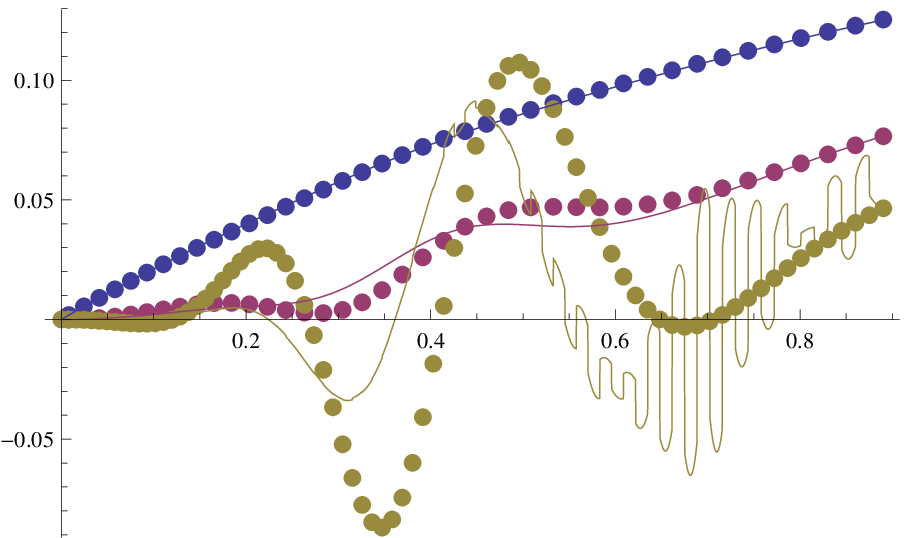}}
\put(60,42){\mbox{\small$L_2$}}
\put(60,32){\mbox{\small$L_3$}}
\put(38,7){\mbox{\small$L_4$}}
  \put(82,0){\includegraphics[width=7.8cm,bb=0 0 288
    171]{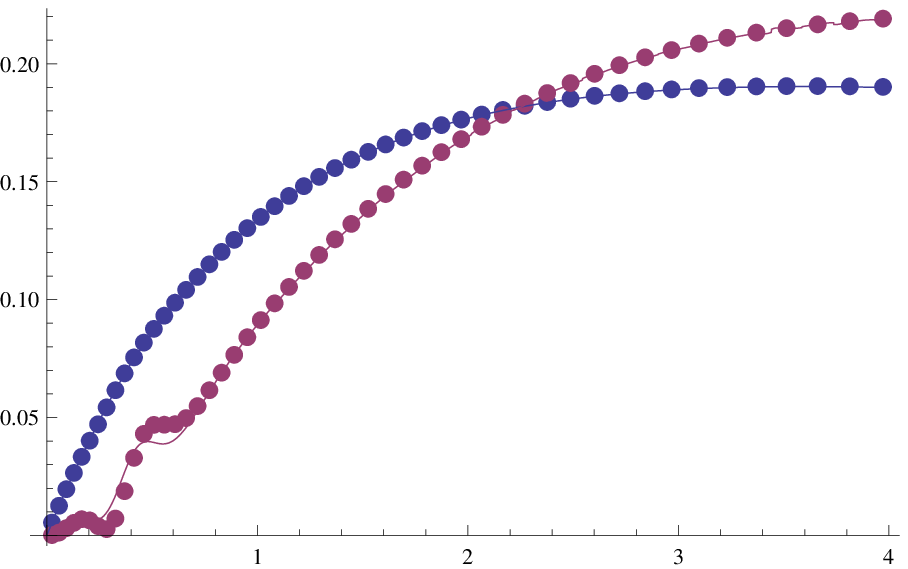}} 
\put(97,30){\mbox{\small$L_2$}}
\put(113,25){\mbox{\small$L_3$}}
\end{picture}
\caption{Comparison of interpolation (solid line) and integral 
formulae (dots) for 
$L_{n,a}[G_{\bullet\bullet}]$ at $\lambda\pi=-1.25$. The integral formulae are
based on $\Lambda^2=10^7$ and $L=2000$ sample points. There is a clear
discrepancy already in $L_{3,a}$ in the interval $a\in [0.15,0.7]$ 
which becomes dramatic in $L_4$. For larger $a$ the agreement
improves, subject to noise in the interpolation.
\label{fig:Stieltjes-1.25}}
\end{figure}

Our main interest is the diagonal 2-point function
$G_{aa}$. Due to much larger numerical errors we can only give
qualitative results. For instance, the value $n^{\mathcal{S}}_0$ at
$\lambda=-\frac{1}{\pi}$ and $L=2000$ results from 
$L_{37,x_2}(G_{0\bullet})=-1.36\times 10^{-57}$. If we apply such
strict criteria to $G_{aa}$ then $G_{aa}$ fails to be
logarithmically completely monotonic already at very low $n$. As we describe
below, there are clear hints that these wrong signs are due to noise
and systematic
discretisation errors. The values of $n^{\mathcal{S}}$ in
Table~\ref{tab:LS} do not reflect its strict definition but show the
critical index where the curve $L_{n,t}(G_{\bullet\bullet})$ becomes
``visibly'' negative for some $t$. Fig.~\ref{fig:Widder} 
shows these curves for typical values of $\lambda$ and $n$. We notice
that for $0\leq t<t_0(\lambda)$, the sequence $L_{n,t}[G_{\bullet\bullet}]$
converges to zero (which reflects a mass gap). 
Any small noise of the zero function produces values $<0$. Therefore 
we discard the interval $[0,t_0]$ from our definition of $n^{\mathcal{S}}$,
\begin{align}
n^{\mathcal{S}}:= \min\{n\,:~
L_{n,t}[G_{\bullet\bullet}] < 0\text{ for some } t\geq t_0(n):= 
\inf\{s\,:~ L_{s,n}[G_{\bullet\bullet}]>3\cdot 10^{-4}\} \}\;.
\label{ns-new}
\end{align}
In this way we require a certain amount of oscillation for a violation
of the Stieltjes property. For $n\geq 12$ and $L=2000$ the noise is so
large that we even violate the coarse condition. For $-1.22\leq
\lambda\pi \leq -1.15$ and $L=2000$ the plots of
$L_{n,t}(G_\bullet\bullet)$ become coarsely negative before the noise
sets in. We are convinced that also these visible oscillations 
are discretisation artifacts. 
In fig.~\ref{fig:Stieltjes-1.20} 
we show that for small $|\lambda|$ there is excellent
agreement between the interpolation formula and the integral formula of
$L_{n,t}[G_{\bullet\bullet}]$. For larger $|\lambda|$, as shown in
fig.~\ref{fig:Stieltjes-1.25}, 
\begin{figure}[h!]
\begin{picture}(150,94)
  \put(0,0){\includegraphics[width=15cm,bb=0 0 641 395]{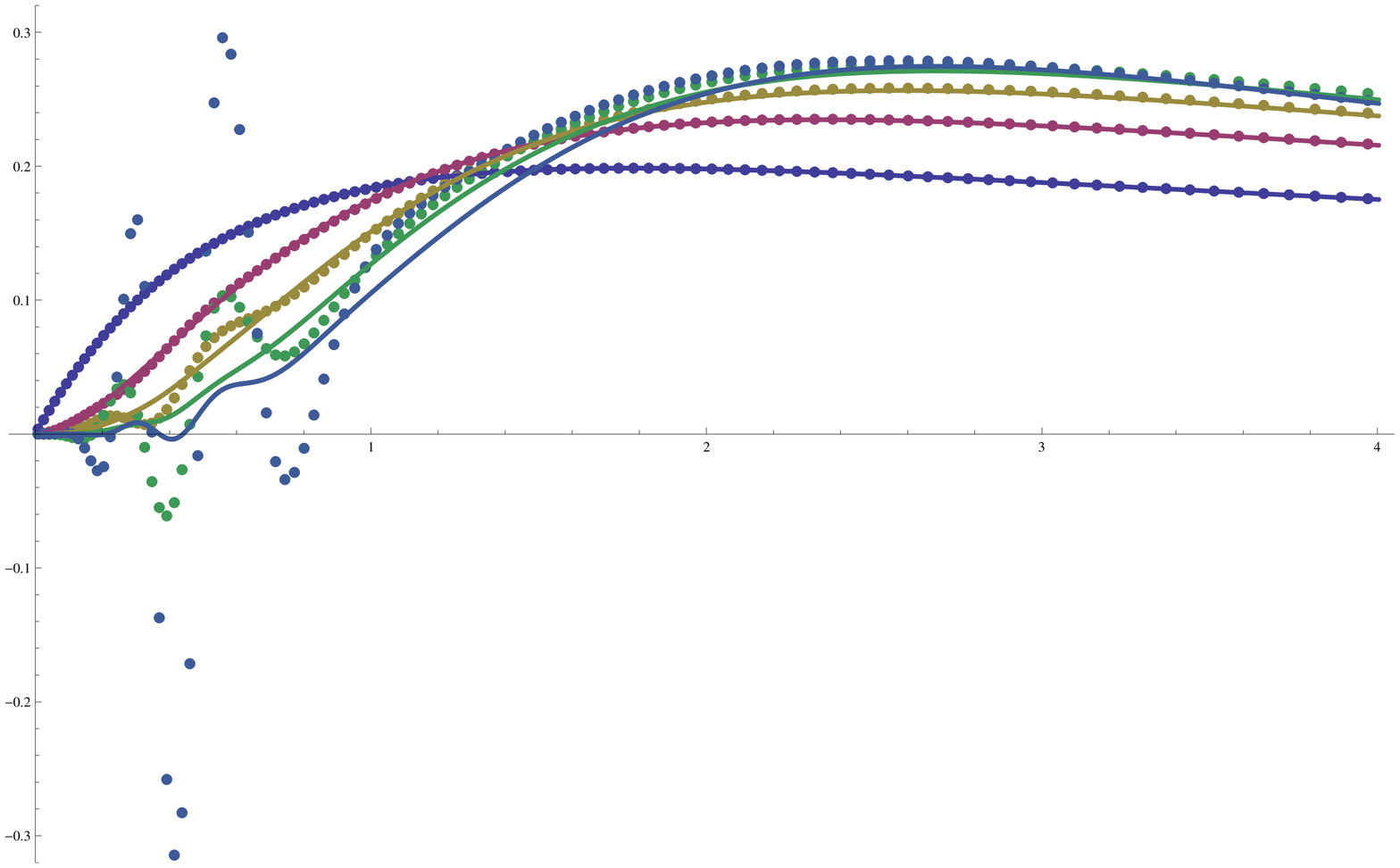}}
\put(130,68){\mbox{\small$L_2$}}
\put(130,87){\mbox{\small$L_5+L_6$}}
\put(43,61){\mbox{\small$L_6\,@10000$}}
\put(33,41){\mbox{\small$L_6@\,2000$}}
\put(19,35){\mbox{\small$L_5@\,2000$}}
\end{picture}
\caption{$L_{n,t}[G_{\bullet\bullet}]$ for 
$\lambda=-\frac{1.22}{\pi}$, $\Lambda^2=10^7$ and 
$\tilde{\Lambda}^2=15848$, but with $L=2000$ (squares) versus $L=10000$ sample
points (solid curves). 
\label{fig:Stieltjes-1.22}}
\end{figure}
we notice a severe discrepancy which by far exceeds our typical
reliability region of $5\%$. We think that for smaller $|\lambda|$
this discrepancy is still present in $L_{n,a}[G_{\bullet\bullet}]$ but
for larger $n$, leading to the oscillations noticed in
fig.~\ref{fig:Widder}. The reason is that both 
$\mathrm{sign}(\lambda)\mathcal{H}^{\tilde{\Lambda}}_a[\tau_b(\bullet)]$ and 
$\log \frac{\sin \tau_b(a)}{|\lambda| \pi a}$ have large derivatives 
but of opposite sign which almost compensate each other. Errors of 5\% 
in each of 
$\mathrm{sign}(\lambda)\mathcal{H}^{\tilde{\Lambda}}_a[\tau_b(\bullet)]$ and 
$\log \frac{\sin \tau_b(a)}{|\lambda| \pi a}$ can thus make their sum 
to $(\log G_{aa})^{(n)}$ unreliable.

\begin{figure}[t]
\begin{picture}(150,45)
\put(0,0){\includegraphics[width=7.8cm,
bb=0 0 288 176]{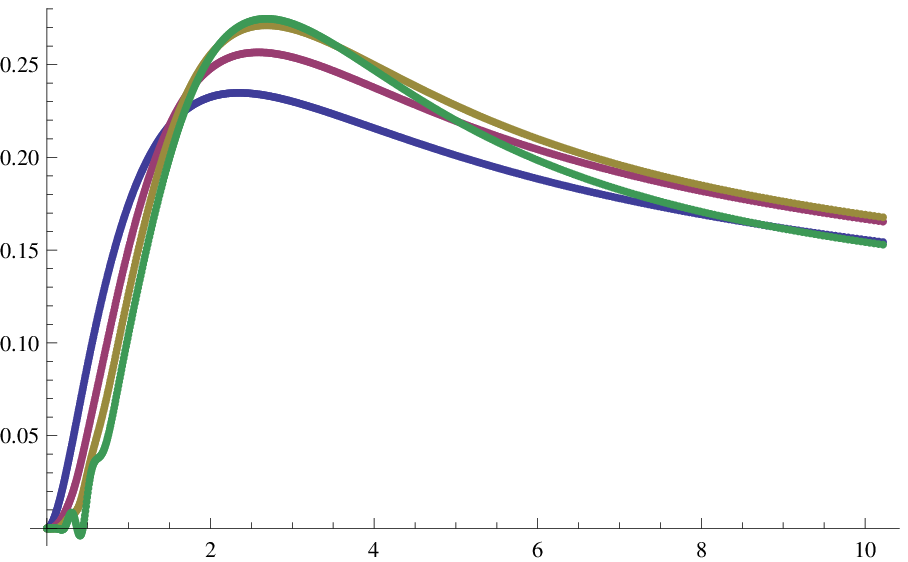}}
  \put(40,10){\fbox{$\lambda=-0.388$}}
  \put(54,38){\fbox{$n^{\mathcal{S}}=6$}}
 \put(20,33){\mbox{\small$L_3$}}
 \put(11,14){\mbox{\small$L_{6}$}}
\put(82,0){\includegraphics[width=7.8cm,
bb=0 0 288 176]{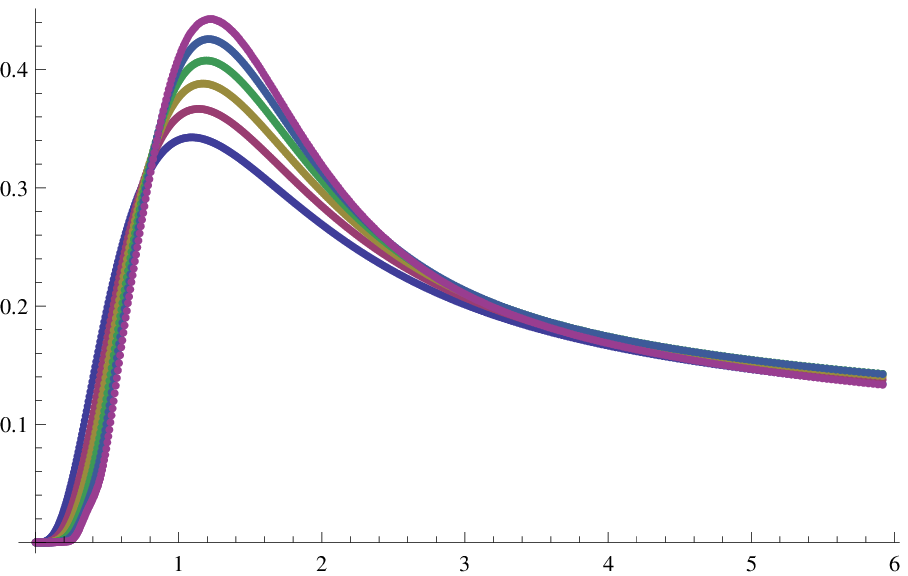}}
  \put(102,10){\fbox{$\lambda=-0.366$}}
  \put(134,38){\fbox{$n^{\mathcal{S}}> 10$}}
 \put(98,30){\mbox{\small$L_5$}}
 \put(102,42){\mbox{\small$L_{10}$}}
\end{picture}
\caption{Widder's operations $L_{n,t}[G_{\bullet\bullet}]$ at
 $\lambda\pi \in \{{-}1.22,\;{-}1.15\}$ and
 $L=10000$ sample points. These plots need much computation time so
 that $L_{n,t}$ is not yet available for $n>10$.  
\label{fig:Widder-10000}}
\end{figure}

At finer resolution $L$ the discretisation error should improve. This
is clearly visible in fig.~\ref{fig:Stieltjes-1.22} which compares
$L_{n,t}[G_{\bullet\bullet}]$ for iterations with the same values 
$\lambda=-\frac{1.22}{\pi}$, $\Lambda^2=10^7$ and 
$\tilde{\Lambda}^2=(\Lambda^2+1)^{\frac{3}{5}}-1\approx 15848$; but 
with $L=2000$ versus $L=10000$ sample points. 
Whereas the critical index $n^{\mathcal{S}}$ (coarsely defined by
(\ref{ns-new})) where the Stieltjes property is lost
increases only from $n^c=5$ at $L=2000$ to $n^c=6$ at $L=10000$, the
curves differ dramatically. Fig.~\ref{fig:Widder-10000} is the analogue 
of the first row in fig.~\ref{fig:Widder} for the 
resolution $L=10000$.
For $\lambda\pi=-1.15$ we have up to $n=10$ no hint of a visible
oscillation, whereas for $L=2000$ we had a failure already at
$n^{\mathcal{S}}=8$.  All this is overwhelming support for the
conjecture that for the exact solution of the master equation (i.e.\
$L\to\infty$) the critical index $n^{\mathcal{S}}$ diverges for
$\lambda_c<\lambda\leq 0$.

\section{Conclusions}

In summary, we are convinced that Conjecture~\ref{conj-Stieltjes} is
true. A proof is impossible by numerical methods, but the simulations
gave us a clear strategy how to proceed. One should first prove that
the fixed point equation~(\ref{master}) has for $\lambda<0$ a unique
\emph{stable} solution $G_{0b}$ inside $\exp(\mathcal{K}_\lambda)$, and 
that this solution is Stieltjes.  Our numerical results leave
no doubt that this is the case. The further steps, symmetry
$G_{ab}=G_{ba}$ and Stieltjes property of $G_{aa}$ should then make
use of the Stieltjes representation of $G_{0b}$.

Suppose all this succeeds and the Schwinger 2-point 
function $\mathcal{S}_c(\mu x_1,\mu x_2)$ defined in 
(\ref{Schwinger-final}) is reflection positive. 
Then one has to pass to the higher functions $G_{a_1\dots
  a_1|\dots |a_B\dots a_B}$. These are given by algebraic recursion
formulae \cite{Grosse:2012uv} if one of the cycles $a_i\dots a_i$
consists of $\geq 4$ indices, but solve their own linear singular
integral equations\footnote{The ($2{+}2$)-point function $G_{ab|cd}$
  involves an auxiliary function that solves the linear singular
  integral equation \cite[eq.~(A.20a)]{Grosse:2012uv}.  Contrary to
  the statement in \cite{Grosse:2012uv}, this equation is not of
  Carleman type, but the solution techniques of
  \cite{Carleman,Tricomi} allow to regularise this equation to an
  integral equation of Fredholm type which always has a unique
  solution for $|\lambda|$ small enough.} if all cycles have length
$2$. A representation of $G_{0b}$ as a Stieltjes transform will help
to control positivity of these solutions, but there is no guarantee
that this is enough. It might be necessary to have an explicit formula
for $G_{0b}$ in terms of known functions. We have looked for such a
formula in various directions; so far without success.

Suppose that all this leads to a proof of reflection positivity for
the family (\ref{Schwinger-final}) of Schwinger functions.  The
Osterwalder-Schrader theorem \cite{Osterwalder:1974tc} then
reconstructs Wightman functions of a relativistic quantum field theory
\cite{Streater:1964??}.  The final problem is then to decide between
triviality or non-triviality of the model. The Schwinger functions
(\ref{Schwinger-final}) do not permit momentum transfer, which in 4
dimensions is usually a sign of triviality. However, the model has two
features which might circumvent the triviality theorems: Absence of clustering 
and absence of a second gap ${]m^2,4m^2[}$ in the mass spectrum (deduced 
from an extrapolation of $L_{n,t}[G_{\bullet\bullet}]$ to $n=\infty$ in 
figs.~\ref{fig:Widder} and 
\ref{fig:Widder-10000}). On the other hand, absence of momentum transfer 
is a generic feature of any integrable model 
\cite{Moser:1975qp, Kulish:1975ba}.
One cannot expect the richness of two-dimensional integrable models: 
The Schwinger functions (\ref{Schwinger-final}) do not depend on 
$\langle p_i,p_j\rangle$ for $i\neq j$ so that the $S$-matrix cannot 
depend on rapidities. At best the model decomposes into different vacuum 
sectors (no clustering!), and in each sector the $S$-matrix
is a sector-dependent pure phase $S=e^{\mathrm{i}\alpha}$.
But even such a
simple $S$-matrix is outside the scope of any other four-dimensional
quantum field theory we know of. This provides enough motivation to
proceed.

\section*{Acknowledgements}

We would like to thank the Erwin-Schr\"odinger-Institute for
Mathematical Physics in Vienna and the Collaborative Research Centre
``Groups, Geometry and Actions'' (SFB 878) in M\"unster for financing
several mutual visits.

\begin{appendix}
\renewcommand{\theequation}{\thesection.\arabic{equation}}
\setcounter{equation}{0}
\makeatletter\@addtoreset{equation}{section}\makeatother

\section{Implementation in Mathematica$^{TM}$}

\label{sec:implementation}

\subsection{Main definitions}

We view $G_{0b}$ as linear interpolation between an increasing
sequence of sample points $x_k$ for $k=1,\dots L+1$, with $x_1=0$ and
$x_{L+1}=\Lambda^2$. We let $G_{0x_k}=:G(k)=\,$\texttt{lis[[k]]}.
This is a Lipschitz-continuous function so that the Hilbert transform
exists pointwise \cite{Okada-Elliott}.  For $s \in [x_{k},x_{k+1}]$ we
have $G_{0s}:= \frac{x_{k+1}-s}{x_{k+1}-x_{k}} G(k)+
\frac{s-x_{k}}{x_{k+1}-x_{k}} G(k+1)$. We are only interested in the
Hilbert transform at sample points $x_n$:
\begin{align}
&\pi \mathcal{H}^{\!\Lambda}_{x_n}[G_{0\bullet}]
\nonumber
\\
&= \lim_{\epsilon\to 0} \sum_{k=1}^{L}
\int_{x_k+\epsilon}^{x_{k+1}-\epsilon} \frac{ds}{s-x_n}
\Big(\frac{x_{k+1}-s}{x_{k+1}-x_{k}} G(k)+
\frac{s-x_{k}}{x_{k+1}-x_{k}} G(k+1)
\Big)
\nonumber
\\
&= G(L+1)-G(1)+\lim_{\epsilon\to 0} \sum_{k=1}^{L}
\int_{x_k+\epsilon}^{x_{k+1}-\epsilon} \frac{ds}{s-x_n}
\Big(\frac{x_{k+1}-x_n}{x_{k+1}-x_{k}} G(k)+
\frac{x_n-x_{k}}{x_{k+1}-x_{k}} G(k+1)
\Big)
\nonumber
\\
&= 
\Big(\sum_{k=1}^{n-2}+\sum_{k=n+1}^{L}\Big)
\frac{(x_n-x_k) G(k+1)- (x_n-x_{k+1}) G(k)}{x_{k+1}-x_k}
\log \Big( \frac{ x_n-x_{k+1}}{x_n-x_k}\Big) 
\nonumber
\\
&+ G(L+1)-G(1)+\left\{\begin{array}{cl}
G(1)\log \frac{x_2-x_1}{\epsilon} & 
\text{for } n =1
\\[1ex]
G(n) \log \frac{x_{n+1}-x_n}{x_n-x_{n-1}} & \text{for } n\neq 1,L+1
\\[1ex]
-G(L+1)\log \frac{x_{L+1}-x_L}{\epsilon} & 
\text{for } n =L+1
\end{array}\right. 
\label{Hilbert-discrete}
\end{align}
We assume the sample points $\{x_k\}$ given as list {\tt xi} of length
$L+1=${\tt len+1} with {\tt xi[[1]]=0} and {\tt xi[[len+1]]=co}$=\Lambda^2$. 
According to (\ref{Hilbert-discrete}) we implement 
the finite Hilbert transform of a function given as list {\tt
  lis} of length $\geq${\tt lng+1} as
\begin{calc}
\item Hilbert[lis_, xi_, n_, lng_] := (1/Pi)*( lis[[lng+1]]-lis[[1]] +
\\ \phantom{-}
  Sum[If[Or[n==k, n-1==k], 0,
\\ \phantom{---}
    ((xi[[n]] - xi[[k]])*lis[[k+1]] 
\\ \phantom{-------}
- (xi[[n]]-xi[[k+1]])*lis[[k]])/(xi[[k+1]]-xi[[k]])* 
\\ \phantom{---}
     Log[(xi[[n]]-xi[[k+1]])/(xi[[n]]-xi[[k]])]], 
\{k, 1, lng\}] +
\\ \phantom{-}
  If[n==1, lis[[1]]*infty, 
   If[n==lng+1, -lis[[lng+1]]*infty, 
\\ \phantom{-----}
    lis[[n]] Log[(xi[[n+1]]-xi[[n]])/(xi[[n]]-xi[[n-1]])]]])

\end{calc}

\noindent
We set both $\frac{x_{2}-x_1}{\epsilon}$ and
$\frac{x_{L+1}-x_L}{\epsilon}$ to the number {\tt infty}. 
We usually have {\tt lng=len}$\,=L$; only later for the Stieltjes
property we need another cut-off. The next
step is to implement the angle function (\ref{tau}). We assume
that $\{G_{0x_k}\}$ and $\{\mathcal{H}_{x_k}^{\!\Lambda}
[G_{0\bullet} ]\}$ are given as lists {\tt lis} and {\tt
  hilb} of length {\tt len+1}. The coupling constant is {\tt
  la}$=\lambda$. Then $\tau_{x_b}(x_a)$ given by (\ref{tau}) is
implemented as
\begin{calc}
\item 
CorrAT[x_] := If[x >= 0 , x, Pi+x]; 
\\ 
Tau[lis_, hilb_, xi_, a_, b_] := CorrAT[ArcTan[
\\ \phantom{---}
(Abs[la] Pi xi[[a]])/(xi[[b]] + 
\\ \phantom{-------}
(1 + la  Pi xi[[a]] hilb[[a]])/lis[[a]] )]]]

\end{calc}

\noindent
The function {\tt CorrAT} moves the branch of the $\arctan$ into the
interval $[0,\pi]$.

Always for $\lambda <0$ and under the assumption
$f_{\lambda,\Lambda^2}(b)=0$ also for $\lambda>0$, the equation
(\ref{G0b}) is a fixed point equation $G=\tilde{T}G$ for
the boundary 2-point function $G_{0b}$. Its solution gives the
full 2-point function $G_{ab}$ via (\ref{Gab}), always for $\lambda<0$ and
for $\lambda>0$ under the additional assumption
$C_{\lambda,\lambda^2}=f_{\lambda,\lambda^2}(b)=0$.
We thus implement the operator $(\tilde{T}[~])_{0x_b}$ and the full 2-point function
$G_{x_ax_b}$ as {\tt Gout} and {\tt Gfull}, respectively:
\begin{calc}

\item 
Gout[lis_, hilb_, xi_, b_] := (1/(1 + xi[[b]]))*Exp[-Sign[la]*
\\ \phantom{-}
Hilbert[Table[Tau[lis, hilb, xi, k, 1] - 
\\ \phantom{-----------}
Tau[lis, hilb, xi, k, b], \{k, 1, len+1\}], xi, 1, len]

\item 
Gfull[lis_, hilb_, xi_, a_, b_] := Exp[-Sign[la]*
\\ \phantom{-}
(Hilbert[Table[Tau[lis, hilb, xi, n, 1],\{n, 1, len+1\}], 
\\ \phantom{-----------} 
xi, 1, len] - 
\\ \phantom{--}
Hilbert[Table[Tau[lis, hilb, xi, n, b],\{n, 1, len+1\}], 
\\ \phantom{-----------} 
xi, a, len])]*
\\ \phantom{----}
If[a>1, Sin[Tau[lis, hilb, xi, a, b]]/(Abs[la] Pi xi[[a]]),
\\ \phantom{------------}
1/(1+xi[[b]])] //Quiet
\end{calc}

\noindent 
In principle we could spell out the Hilbert transform in (\ref{G0b})
as an integral to obtain the master equation (\ref{master}). Depending
on the numerical implementation of the integration there is then the
danger to violate the identity 
{\tt Gfull[lis, hilb, xi, 1, b]=
Gout[lis, hilb, xi, b]}. We therefore prefer (\ref{G0b}) to
(\ref{master}). The result of $G_{x_ax_b}$ for $x_a$ close to $\Lambda^2$ can
become smaller than the minimal positive machine number so that we 
turn off the corresponding error message via {\tt Quiet}.

We also need a few functions to control the convergence and the
quality of our discrete approximation. We define supremum norm,
Lipschitz seminorm, the absolute asymmetry $\sup_{a,b}
|G_{x_ax_b}-G_{x_bx_a}|$ and the relative asymmetry $\sup_{a,b}
\frac{|G_{x_ax_b}-G_{x_bx_a}|}{G_{x_ax_b}+G_{x_bx_a}}$:

\begin{calc}

\item 
 SupNorm[lis1_, lis2_] := 
\\  \phantom{-----}
Max[Table[Abs[lis1[[k]] - lis2[[k]]], \{k, 1, len+1\}]];
\\
LipNorm[lis1_, lis2_, xi_] := 
\\  \phantom{---}
Max[Table[Max[Table[
     Abs[((lis1[[n]] - lis2[[n]]) - 
\\  \phantom{------------}
(lis1[[k]] - lis2[[k]]))/ (xi[[n]] - xi[[k]])], 
\\  \phantom{-----}
\{k, n+1, len+1\}]], \{n, 1, len\}]];
\\
AbsAsm[lis_, hilb_, xi_, sx_, dx_, fx_, sy_, dy_, fy_] := 
\\ \phantom{--}
Max[Table[Max[Table[
\\
\phantom{-----}
Abs[Gfull[lis, hilb, xi, sx + n dx, sy + k dy] - 
\\
\phantom{----------}
       Gfull[lis, hilb, xi, sy + k dy, sx + n dx]], 
\\
\phantom{-----}
\{n, 0, Floor[Min[len+1-sx, fx-sx]/dx]\}]], 
\\
\phantom{-----}
\{k, 0, Floor[Min[len+1-sy, fy-sy]/dy]\}]];
\\
RelAsm[lis_, hilb_, xi_, sx_, dx_, fx_, sy_, dy_, fy_] := 
\\ \phantom{--}
Max[Table[Max[Table[ Abs[ 1-2/(1+
\\
\phantom{-----}
Gfull[lis, hilb, xi, sx + n dx, sy + k dy]/
\\
\phantom{----------}
       Gfull[lis, hilb, xi, sy + k dy, sx + n dx])], 
\\
\phantom{-----}
\{n, 0, Floor[Min[len+1-sx, fx-sx]/dx]\}]], 
\\
\phantom{-----}
\{k, 0, Floor[Min[len+1-sy, fy-sy]/dy]\}]];

\end{calc}

\noindent
In order to have tolerable computing time the asymmetries need 
to be evaluated for a subset 
$\{a_k=a_0+k \delta_1,~ b_n=b_0+n \delta_2\}$ of indices. We search
the region of maximal asymmetry by hand.

The 4-point function at vanishing arguments defines the effective
coupling constant $G_{0000}=-\lambda_{\mathit{eff}}$ which
according to \cite{Grosse:2012uv} is given by
\begin{align}
\lambda_\mathit{eff}
&= \frac{\lambda}{1+\mathcal{Y}_1} + \frac{\lambda^2\pi}{(1+\mathcal{Y}_1)^2}
\mathcal{H}^{\Lambda}_0\Big[
\frac{1-G_{0\bullet}}{G_{0\bullet}}  \Big(\frac{\sin \tau_0(\bullet)}{|\lambda| \pi
  \bullet}\Big)^2\Big]\;,
\label{lambdaeff}
\end{align}
where $\mathcal{Y}_\ell$ is defined in (\ref{calY-ell}).
We implement these functions as
\begin{calc}
\item 
calY[ell_, lis_, hilb_, xi_] := Sign[la]*Hilbert[ Table[
\\ \phantom{----}
Sin[ell Tau[lis, hilb, xi, n, 1]]*(If[n==1, 1,
\\ \phantom{-------}
        Sin[Tau[lis, hilb, xi, n, 1]]/(Abs[la] Pi xi[[n]])])^ell,
\\ \phantom{--}
    \{n, 1, len+1\} ], xi, 1, len];
\\
laeff[lis_, hilb_, xi_] := 
 la/(1 + calY[1, lis, hilb, xi]) + 
\\ \phantom{--}
la^2 Pi/(1 + calY[1, lis, hilb, xi])^2*
\\ \phantom{------}
Hilbert[
Table[((1 - lis[[n]])/lis[[n]])*
\\ \phantom{-----------}
If[Or[n==1, la==0], 1, 
\\ \phantom{--------------}
Sin[Tau[lis, hilb, xi, n, 1]]^2/(la Pi xi[[n]])^2],
\\ \phantom{------------}
     \{n, 1, len+1\}], xi, 1, len];
\end{calc}

\subsection{Iteration}

{}From a numerical simulation in an early version arXiv:1205.0465v1 of
\cite{Grosse:2012uv} we expect that $G_{b0} \approx
\frac{1}{(1+b)^\eta}$ shows a power-law behaviour. This suggests to
choose the sample points $x_k$ according to a geometric
progression:
\begin{calc}
\item co = 100; len = 1000; la = 1/Pi; infty = N[10^8];
\\
xs = Table[N[(1+co)^((n-1)/len) - 1], \{n, 1, len+1\}];
\label{co}
\end{calc}
\noindent
Coupling constant $\lambda=\,${\tt la}, cut-off {\tt co}$=\Lambda^2$
and the number {\tt len}$=L$ of sample points will be varied; but the
list of sample points will always be a geometric progression {\tt xs}.
We have also tried equidistant samples and finer resolutions near {\tt
  co}$=\Lambda^2$ to better deal with the singularity of the finite
Hilbert transform at the boundary $\Lambda^2$; all had worse quality
parameters than the geometric progression.

For definiteness of the result we start with the
constant function $G_{0b}^0=1$ (which would be an exact solution for
{\tt la<0} and {\tt co=}$\infty$, see \cite[Appendix A]{Grosse:2015fka}) and 
approximate $G_{0b}\mapsto (TG)_{0b}$ by the numerical
implementation {\tt lis[[\,.\,]]}$\mapsto${\tt
  Gout[lis,hilb,xs,\,.\,]} below in {\tt\bf In[\ref{For}]}:
\begin{calc}
\item 
gs[0] = Table[1., \{n, 1, len+1\}];
\\*
hs[0] = Table[Hilbert[gs[0], xs, n, len], \{n, 1, len+1\}];
\\*
For[i=1, i<=imax, i++,
\\ \phantom{--}
gs[i] = Table[Gout[gs[i-1], hs[i-1], xs, b], \{b, 1, len+1\}];
\\* \phantom{--}
 hs[i] = Table[Hilbert[gs[i], xs, n, len], \{n, 1, len+1\}];
\\ \phantom{--}
 Print[i, "  " ,
\\* \phantom{---}
  Interpolation[Table[{xs[[k]], gs[i][[k]]}, \{k, 1, len+1\}]][100], 
\\ \phantom{-----}
  "  " ,  gs[i][[-1]], "  ",
\\ \phantom{-----}
 SupNorm[gs[i], gs[i-1]], "  ",
LipNorm[gs[i], gs[i-1],xs]];
\\ \phantom{-}
If[i>=20, Break[]];
 ];
\\ 
gfull[i]=Table[Gfull[gs[i], hs[i], xs, k, k], \{k, 1, len+1\}];
\label{For}

\end{calc}

\noindent
We set {\tt imax} to a sufficiently large number but actually stop
here at {\tt i=20}. During the iteration we print out several
parameters to control the quality. We notice that both supremum norm
and Lipschitz seminorm improve (for $\lambda=\frac{1}{\pi}$) by a
factor $>3$ in the step from {\tt i} to {\tt i+1}. This is strong
support for norm convergence of the iteration. We also list the
approximation of $G_{0b}$ for $b=100$ (kept fixed when varying
$\Lambda^2$) and $b=\Lambda^2$. The first value is to check the
pointwise convergence of $G_{0b}$ as $\Lambda\to \infty$.  The second
value affects the absolute asymmetry if $\Lambda$ is chosen too small.
The asymmetry is tested with the function {\tt AbsAsm} for various
ranges of parameters. We plot the functions $G_{0b}$ and $G_{aa}$ in
double logarithmic coordinates:
\begin{calc}
\item  
ListPlot[
\\ \phantom{-}
\{Table[{Log[1+xs[[k]]], Log[gfull[20][[k]]]\}, \{k, 1, 
    len-40\}], 
\\ \phantom{---}
Table[\{Log[1+xs[[k]]], Log[gs[20][[k]]]}, \{k, 1, len\}]\}, 
\\ \phantom{----} 
AxesOrigin -> \{0, 0\}, PlotStyle->PointSize[Tiny]]
\end{calc}

\noindent
These functions are decreasing and approximately linear (see
fig.~\ref{fig:G}). The diagonal 
function $G_{aa}$ shows boundary artifacts which we cut off by {\tt
  len-40}. We fit $\log G_{0,\exp (x)-1}$ to a line $A+Bx$ and  
$\log G_{\exp (x)-1,\exp(x)-1}$ to a line $C+Dx$:
\begin{calc}
\item  
\{Fit[Table[\{Log[1+xs[[k]]], Log[gs[20][[k]]]\}, \{k, 1, len\}], 
\\* \phantom{------}
\{1, x\}, x], 
\\* \phantom{-}
Fit[Table[\{Log[1+xs[[k]]], Log[gfull[20][[k]]]\}, \{k, 1, len-40\}], 
\\*
\phantom{------}
\{1, x\}, x]\}
\end{calc}

For $\lambda>0$ the general theory leads to undetermined parameters
$C_{\lambda,\Lambda^2}$ and $f_{\lambda,\Lambda^2}(b)$ in the formula
(\ref{Gab}) for the 2-point function. In a first step we assume
$f_{\lambda,\Lambda^2}(b)=0$ so that the fixed point equation
(\ref{master}) is unchanged. Under this assumption,
$C_{\lambda,\Lambda^2}$ is computable from (\ref{ClL}) which we implement
as 
\begin{calc}
\item  
ClL[lis_, hilb_, xi_, a_] := (1 - xi[[a]]/co)*(Exp[
\\* \phantom{-}
 Hilbert[Table[Tau[lis, hilb, xi, n, a], \{n, 1, len+1\}], 
\\* \phantom{--------}
xi, 1, len] - 
\\ \phantom{-}
Hilbert[Table[Tau[lis, hilb, xi, n, 1], \{n, 1, len+1\}], 
\\* \phantom{--------}
xi, a, len]]*
\\ \phantom{-----}
      Sqrt[(la Pi xi[[a]]/(1 + xi[[a]]))^2 + 
\\* \phantom{-----}
((1 +  la Pi xi[[a]] hilb[[a]])/((1+xi[[a]]) lis[[a]]))^2]  
- \\* \phantom{--} 1 )/xi[[a]]

\label{calc:ClL}
\end{calc}
\noindent
Fig.~\ref{fig:ClL} shows typical results.

\subsection{The Stieltjes property}

For a first impression we implement Widder's operators $L_{n,t}$ 
defined in (\ref{Widder}) for $n\geq 1$ via an interpolation formula
\begin{calc}
\item WidderInterpolation[xi_, lis_, n_, t_] := (
\\ \phantom{--}
(-x)^(n-1)/If[n>=2, n!(n-2)!, 1]*
\\ \phantom{-----}
   D[Interpolation[
      Table[\{xi[[k]], xi[[k]]^n lis[[k]]\}, 
\\ \phantom{---------------------}
\{k, 1, len+1\}],       InterpolationOrder->2n][x], 
\\ \phantom{--------}
\{x, 2n-1\}]) /.x->t
\label{WidderI}
\end{calc}

\noindent
The discrete list of $G_{aa}$ is interpolated by a polynomial of
degree $2n$. Clearly, this is only reliable for small $n$.
We have given typical results in \cite[Fig.~3]{Grosse:2014nza}.

The implementation of the integral formula for $L_{n,t}$ starts with 
the formula (\ref{dlogG0b-b}) for the derivatives $(\log G_{0b})^{(\ell)}$: 
\begin{calc}
\item 
DLogG0[ell_, lis_, hilb_, xi_, b_] := (-1)^ell (ell-1)! *
\\ \phantom{---}
(1/(1 + xi[[b]])^ell + Sign[la]* Hilbert[ Table[
\\ \phantom{-------}
Sin[ell Tau[lis, hilb, xi, k, b]]*(If[k>1,  
\\ \phantom{-----------}
Sin[Tau[lis, hilb, xi, k, b]]/(Abs[la] Pi xi[[k]]), 
\\ \phantom{----------------}
    1/(1 + xi[[b]])])^ell, 
\\ \phantom{--------}
\{k, 1, len+1\}], xi, 1, len])
\end{calc}
       
\noindent
We arrange them in a table $\mathtt{dlogg[i]}=\{(\log
G_{0b})^{(\ell)}\}_{\ell b}$ of the following type
\begin{calc}
\item 
dlogg[i] = 
  Table[DLogG0[n, gs[i], hs[i], xs, b], \\ \phantom{------------------}
\{n, 1, 11\}, \{b, 1, len+1\}];
\label{dlogg}
\end{calc}

\noindent
Here {\tt [i]} refers to the value reached in  
{\tt\bf In[\ref{For}]}, and the length $11$ can vary, of course.
We compute the derivatives $(G_{0b})^{(n)}$ via (\ref{DGab}):
\begin{calc}
\item 
DG0[n_, lis_, dlogg_, b_] := lis[b]*If[n==0, 1, Sum[
\\ \phantom{---} 
   BellY[n, k, Table[dlogg[[m]][[b]], 
\{m, 1, n-k+1\}]], 
\{k, 1, n\}]]
\end{calc}

\noindent
We arrange them in a table $\mathtt{dg[i]}=\{
(G_{0b})^{(n)}\}_{n b}$ of the type
\begin{calc}
\item 
dg[i] = Table[DG0[n, gs[i], dlogg[i], b], \\
\phantom{------------------}
 \{n, 1, 11\}, \{b, 1, len+1\}];
\label{dg}
\end{calc}

\noindent
and implement the functions $F^\Lambda_{n,k}(a)$ of (\ref{Diff-Hilbert-F}) and 
$C^n_0(a)$ of (\ref{Cbna}) and (\ref{Gamma-sa}) as
\begin{calc}
\item 
FSum[n_, k_, xi_, lng_, a_] := (1/(1-xi[[a]]/xi[lng+1]))* (1+
\\* \phantom{------} 
Sum[(Binomial[n-k-1,p]/ Binomial[n-1,p])*
\\* \phantom{-----------} 
(-xi[[a]]/(xi[[lng+1]]-xi[a]))^p, \{p, 1, n-k-1\}]);

\item DCotTau0[n_, lis_, dg_, xi_, a_] := 
 If[a == len+1, \\ \phantom{--} 
  InterpolatingPolynomial[
   Table[{xi[[j]], 
\\ \phantom{-----} 
DCotTau0[n, lis, dg, xi, j]}, \{j, len-2, len\}], 
   xi[[len+1]]],
\\ \phantom{--} 
  (n!/lis[[a]])* (1 + la Pi xi[[a]] Hilbert[lis, xi, a, len])*
\\ \phantom{-----} 
    Sum[((-1)^l l!/n!) BellY[n, l, 
       Table[(-xi[[a]])^kappa *
\\ \phantom{---------} 
dg[[kappa]][[a]]/lis[[a]], \{kappa, 1,  n-l+1\}]], 
\\ \phantom{-------} 
\{l, 1, n\}] +
\\ \phantom{--} 
   (n!/lis[[a]])*Sum[(1 + 
\\ \phantom{-------} 
(la xi[[a]]/k)*Sum[ If[l==0, lis[[len+1]], 
\\ \phantom{--------------} 
          (-xi[[len+1]])^l dg[[l]][[len+1]]/ l!]*
\\ \phantom{----------------} 
FSum[k, l, xi, len, a], \{l, 0, k-1\}] + 
\\ \phantom{-------} 
          la Pi xi[[a]]* Hilbert[
            Table[(-xi[[c]])^k *
\\ \phantom{---------------} dg[[k]][[c]]/k!, \{c, 1, len+1\}], 
            xi, a, len])*
\\ \phantom{----} 
      If[n==k, 1, Sum[((-1)^l l!/(n-k)!)*
\\ \phantom{---------------} 
BellY[n-k, l, 
          Table[(-xi[[a]])^kappa *
\\ \phantom{-----------------} dg[[kappa]][[a]]/lis[[a]], \{kappa, 1,
             n-k-l+1\}]], 
\\ \phantom{--------------} 
\{l, 1, n-k\}]], 
\\ \phantom{-----} \{k, 1, n\}]]
\end{calc}

\noindent
The variable length {\tt lng} in {\tt FSum} is necessary for a
subsequent step. To avoid ``$0^0$'' we have to separately 
implement the case $p=0$ in (\ref{Diff-Hilbert-F}).
Since $C^n_0(\Lambda^2)=\,${\tt DCotTau0[n, gs[i],
  dg[i], xs, len+1]} is undefined, we extrapolate
it via the quadratic function through its values at
{\tt xs[[len-2]]}, {\tt xs[[len-1]]} and 
{\tt xs[[len]]}.
We arrange these functions in a table
$\mathtt{dcottau[i]}=\{C^n_0(a)\}_{n a}$
implemented as 
\begin{calc}
\item dcottau[i] = 
  Table[DCotTau0[n, gs[i], dg[i], xs, a], \\ \phantom{--------------} 
\{n, 1, 11\}, \{a, 1, len+1\}];
\label{dcottau}
\end{calc}

The next step consists in implementing the functions 
$A^{(n,\ell)}$ defined in (\ref{Anl}) and $L^{(n,\ell)}$ defined in (\ref{Lnl}):

\begin{calc}
\item
NegXDADaDb[n_, ell_, lis_, hilb_, dg_, dcottau_, xi_, a_, b_] := 
\\ \phantom{-} 
If[n+ell==0, Tau[lis, hilb, xi, a, b],
\\ \phantom{--} 
  If[ell==0,
   Sum[((-1)^k (k-1)!/n!)*
\\ \phantom{----} 
Sin[k Tau[lis, hilb, xi, a, b]] *
 BellY[n, k, Table[
\\ \phantom{-------} 
(kappa!*xi[[b]] + dcottau[[kappa]][[a]])*
\\ \phantom{---------} 
        If[a==1, 1/(1+xi[[b]]), 
\\ \phantom{-------------} 
         Sin[Tau[lis, hilb, xi, a, b]]/(Abs[la] Pi xi[[a]])], 
\\ \phantom{---------------} 
\{kappa, 1, n-k+1\}]], 
\\ \phantom{-----} 
\{k, 1, n\}],
\\ \phantom{--} 
   Sum[Binomial[n-m+ell-1, ell-1] ((-1)^k (ell+k-1)!/(m! ell!))*
\\ \phantom{-----} 
Sin[(ell+k) Tau[lis, hilb, xi, a, b]] *
\\ \phantom{-----} 
 (xi[[b]]* If[a==1, 1/(1+xi[[b]]), 
\\ \phantom{--------}  
   Sin[Tau[lis, hilb, xi, a, b]]/(Abs[la] Pi xi[[a]])])^ell*
\\ \phantom{-----} 
     BellY[m, k, Table[(kappa!*xi[[b]] + dcottau[[kappa]][[a]])*
\\ \phantom{---------} 
        If[a==1, 1/(1+xi[[b]]), 
\\ \phantom{------------} 
         Sin[Tau[lis, hilb, xi, a, b]]/(Abs[la] Pi xi[[a]])], 
\\ \phantom{---------} 
\{kappa, 1, m-k+1\}]], 
\\ \phantom{----} 
\{m, 0, n\}, \{k, 0, m\}] ]]

\item NegXDLDaDb[n_, ell_, lis_, hilb_, dg_, dcottau_, xi_, a_, b_] := 
\\ \phantom{-} 
 If[n+ell==0, Log[If[a==1, 1/(1+xi[[b]]), 
\\ \phantom{----} 
    Sin[Tau[lis, hilb, xi, a, b]]/(Abs[la] Pi xi[[a]])]],
\\ \phantom{--} 
  If[ell==0, 1/n +  Sum[((-1)^k (k-1)!/n!)*
\\ \phantom{----} 
Cos[k Tau[lis, hilb, xi, a, b]] *
 BellY[n, k, Table[
\\ \phantom{------} 
(kappa! xi[[b]] + dcottau[[kappa]][[a]])*
\\ \phantom{---------} 
        If[a==1, 1/(1+xi[[b]]), 
\\ \phantom{-------------} 
         Sin[Tau[lis, hilb, xi, a, b]]/(Abs[la] Pi xi[[a]])], 
\\ \phantom{---------------} 
\{kappa, 1, n-k+1\}]], 
\\ \phantom{-----} 
\{k, 1, n\}],
\\ \phantom{--} 
   Sum[Binomial[n-m+ell-1, ell-1] ((-1)^k (ell+k-1)!/(m! ell!))*
\\ \phantom{-----} 
Cos[(ell+k) Tau[lis, hilb, xi, a, b]] *
\\ \phantom{-----} 
 (xi[[b]]* If[a==1, 1/(1+xi[[b]]), 
\\ \phantom{--------}  
   Sin[Tau[lis, hilb, xi, a, b]]/(Abs[la] Pi xi[[a]])])^ell*
\\ \phantom{-----} 
     BellY[m, k, Table[(kappa!*xi[[b]] + dcottau[[kappa]][[a]])*
\\ \phantom{---------} 
        If[a==1, 1/(1+xi[[b]]), 
\\ \phantom{------------} 
         Sin[Tau[lis, hilb, xi, a, b]]/(Abs[la] Pi xi[[a]])], 
\\ \phantom{---------} 
\{kappa, 1, m-k+1\}]], 
\\ \phantom{----} 
\{m, 0, n\}, \{k, 0, m\}] ]]

\end{calc}

According to (\ref{dlogGab-ab}), the derivatives 
$\frac{(-a)^n(-b)^\ell}{n!\ell!} 
\frac{\partial^{n+\ell}(\log G_{ab})}{\partial a^n\partial  b^\ell}$
are a sum of $L^{(n,\ell)}(a,b)$ defined before and the more
complicated remainder 
$\frac{(-a)^n(-b)^\ell}{n!\ell!} 
\frac{\partial^{n+\ell}(\mathrm{sign}(\lambda)\mathcal{H}_a^{\tilde{\Lambda}}
[\tau_b(\bullet)])}{\partial a^n\partial  b^\ell}$. The latter
function is for $n+\ell>0$
and $a<\tilde{\Lambda}^2={\tt lng}$ implemented as
\begin{calc}
\item
NegXDHTauDaDb[n_,ell_,lis_,hilb_,dg_,dcottau_,xi_,lng_,a_,b_] := 
\\
\phantom{--}
If[n==0, Sign[la] Hilbert[ Table[
\\
\phantom{--------}
NegXDADaDb[0, ell, lis, hilb, dg, dcottau, xi, bu, 
         b], 
\\
\phantom{-------}
\{bu, 1, lng+1\}], xi, a, lng], 
\\
\phantom{---}
    la Pi xi[[a]] Hilbert[  Table[If[bu==1,  If[n<=1, 
\\
\phantom{------}
(-1)^n If[ell==0, 1, xi[[b]]^ell]/(1 + xi[[b]])^(ell+1), 0], 
\\
\phantom{------}
NegXDADaDb[n,ell,lis,hilb,dg,dcottau,xi,bu,b]/
\\
\phantom{--------------------}(Abs[la] Pi xi[[bu]])], 
\\
\phantom{----}
\{bu, 1, lng+1\}], xi, a, lng] -
\\
\phantom{---}
     (Sign[la]/(Pi n))*(-xi[[a]]/(xi[[lng+1]] - xi[[a]]))^n * 
\\
\phantom{----------}
NegXDADaDb[0, ell, lis, hilb, dg, dcottau, xi, a, b] +
\\
\phantom{---}
If[n<=1, 0, (Sign[la]/(Pi *n(n-1)))*(xs[[a]]/xs[[lng+1]]) *
\\
\phantom{--------}
Sum[k*NegXDADaDb[k,ell,lis,hilb,dg,dcottau,xi,lng+1,b]*
\\
\phantom{----------}
    FSum[n-1, k-1, xi, lng, a] , \{k, 1, n-1\}]]]
\end{calc}  

\noindent 
The 6$^{\text{th}}$ line {\tt (-1)\^{ }n If[\dots ]} uses 
the limit (\ref{limAnl}).

We intercept the cases $n+\ell=0$ and 
$a<\tilde{\Lambda}^2$ to obtain the following implementation of 
$\frac{(-a)^n(-b)^\ell}{n!\ell!} 
\frac{\partial^{n+\ell}(\log G_{ab})}{\partial a^n\partial  b^\ell}$:
\begin{calc}
\item NegXDLogGDaDb[n_,ell_,lis_,hilb_,dg_,dcottau_,xi_,lng_,a_,b_] := 
\\
\phantom{---}
If[a==lng+1, 
  InterpolatingPolynomial[
   Table[\{xi[[j]], 
\\*
\phantom{--------}
     NegXDLogGDaDb[n,ell,lis,hilb,dg,dcottau,xi,lng,j,b]\}, 
\\*
\phantom{-----}
\{j, lng-2, lng\}], xi[[lng+1]]],
\\
\phantom{---}
  If[n+ell==0, Log[Gfull[lis, hilb, xi, a, b]],
\\
\phantom{---}
   NegXDHTauDaDb[n, ell, lis, hilb, dg, dcottau, xi, lng, a, b] +
\\
\phantom{---}
    NegXDLDaDb[n, ell, lis, hilb, dg, dcottau, xi, a, b]
   ]]
 \end{calc}

\noindent
It remains to sum these contributions to $(-a)^n(\log G_{aa})^{(n)}$
according to (\ref{dlogGaa-a}): 
\begin{calc}
\item 
NegXDLogGfull[n_, lis_, hilb_, dg_, dcottau_, xi_, lng_, a_] := 
\\
\phantom{---}
  n!*Sum[NegXDLogGDaDb[n-ell,ell,lis,hilb,dg,dcottau,xi,lng,a,a], 
\\
\phantom{------}
\{ell, 0, n\}]
   \end{calc}

\noindent
We collect these values in a table and use Fa\`a di Bruno to obtain 
$(-a)^n(G_{aa})^{(n)}$ according to (\ref{aGaa-n}):
\begin{calc}
\item negxdloggfull[i]=Table[
\\
\phantom{--}
NegXDLogGfull[n, gs[i], hs[i], dg[i], dcottau[i],xs,1200,a], 
\\
\phantom{------}
\{n, 1, 11\}, \{a, 1, 200\}];
\\
negxdgfull[i] = Table[
\\
\phantom{--}
DG0[n, gfull[i], negxdloggfull[i], a], \{n, 1, 11\}, 
\{a, 1,  200\}];%
\label{negxdgfull}
 \end{calc}

\noindent
The sizes {\tt n=1\dots 11} and {\tt a=1\dots 200} can be adapted, of
course, but require the tables {\tt dg[i]} defined in {\bf
  In[\ref{dg}]} and {\tt dcottau[i]} defined in {\bf
  In[\ref{dcottau}]} of length not shorter than {\tt n}.
Also the secondary cutoff $\tilde{\Lambda}^2=x_{1201}$ can be adapted.
It remains to define the integral formula for Widder's operators
$L_{n,t}[G_{\bullet\bullet}]$ according to 
(\ref{Widder-K})
and to visualise the results:
\begin{calc}

\item WidderL[n_, t_, tab_] := 
 Sum[((-1)^(n-l) *
\\
\phantom{--}
Binomial[2n-1, l] Binomial[n, l] l!/ If[n<=1, 1, (n-2)! n!])*
\\
\phantom{------}
tab[[2n-l-1]][[t]], \{l, 0, n\}]

\item ListPlot[\{
\\
\phantom{---}
Table[\{xs[[k]], WidderL[2, k, negxdgfull[i]]\}, \{k, 1,  200\}], 
\\
\phantom{---}
Table[\{xs[[k]], WidderL[3, k, negxdgfull[i]]\}, \{k, 1,  200\}], 
\\
\phantom{---}
Table[\{xs[[k]], WidderL[4, k, negxdgfull[i]]\}, \{k, 1,  200\}], 
\\
\phantom{---}
Table[\{xs[[k]], WidderL[5, k, negxdgfull[i]]\}, \{k, 1,  200\}], 
\\
\phantom{---}
Table[\{xs[[k]], WidderL[6, k, negxdgfull[i]]\}, \{k, 1,  200\}]\}]
\end{calc}

\noindent
Typical results are shown in figs.~\ref{fig:Widder} and 
\ref{fig:Stieltjes-1.22}. Figs.~\ref{fig:Stieltjes-1.20} 
and \ref{fig:Stieltjes-1.25} compare {\tt WidderL} with 
{\tt WidderInterpolation} defined in {\bf In[\ref{WidderI}]}.  
Note that {\tt WidderL[s,\dots]} requires
lengths {\tt n=2s-1} or bigger in  {\bf In[\ref{dlogg}]}, 
{\bf In[\ref{dg}]}, {\bf
  In[\ref{dcottau}]} and {\bf In[\ref{negxdgfull}]}.  For comparison
(table~\ref{tab:LS} and fig.~\ref{Fig:Logcmon}) 
we can evaluate $L_{n,t}[G_{0\bullet}]$ starting from {\tt dlogg[i]}
computed in {\bf In[\ref{dlogg}]} as follows:
\begin{calc}
\item 
negxdg0[i] = Table[DG0[n, gs[i], Table[
\\
\phantom{--------} 
(-xs[[a]])^k dlogg[i][[k]][[a]],\{k, 1, Length[dlogg[i]]\}], 
\\
\phantom{----} 
\{n, 1, 11\}, \{a, 1, 200\}];
\\
Table[\{xs[[k]], WidderL[6, k, negxdg0[i]]\}, \{k, 1,  200\}]
\end{calc}

\section{The derivatives $L^{(n,\ell)}(a,b)$ and 
$(A^{(n,\ell)}(a,b))/(|\lambda|\pi a)$ at $a=0$}

\label{app:limit}

Vanishing of Widder's operators  $L_{n,t}[G_{\bullet\bullet}]$ at
$t=0$ requires $L^{(n,\ell)}(0,b)=0$ for the functions (\ref{Lnl}).
With 
$\tau_b(0)=0$, $\lim_{a\to 0} \frac{\sin(\tau_b(a))}{|\lambda|\pi
  a}=\frac{1}{1+b}$  and
$\lim_{a\to 0} \frac{C^k_b(a)\sin(\tau_b(a))}{|\lambda|\pi a}=k!$
as well as $Y_{m,k}(1!,2!,\dots,(m{-}k{+}1)!)=
\frac{m!(m-1)!}{k!(k-1)!(m-k)!}$ for $m\geq 1$ \cite{Bell}, we have:
\begin{align}
&L^{(n,\ell)}(0,b)\big|_{\ell,n\geq 1}
\nonumber
\\
& = \Big(\frac{b}{1+b}\Big)^\ell 
\sum_{m=0}^{n} \sum_{k=0}^m 
\binom{n{-}m{+}\ell{-}1}{\ell-1} 
\frac{(-1)^{k}(\ell+k-1)!}{m!\ell!} 
Y_{m,k}\Big(1!,2!,\dots,(m-k+1)!\Big)
\nonumber
\\
& = \Big(\frac{b}{1+b}\Big)^\ell \Big\{
\frac{(n{+}\ell{-}1)! }{n!\ell!} 
-
\sum_{m=1}^{n} 
\binom{n{-}m{+}\ell{-}1}{\ell-1} 
\sum_{k=0}^{m-1} 
\frac{(-1)^{k} 1!(\ell+k)!(m-1)!}{(k+1)!\ell!(m-1-k)!k!}\Big\}
\nonumber
\\
&= \Big(\frac{b}{1+b}\Big)^\ell \Big\{
\frac{(n{+}\ell{-}1)! }{n!\ell!} 
-
\sum_{m=1}^{n} 
\binom{n{-}m{+}\ell{-}1}{\ell-1} 
\;{}_2F_1\Big( \di{\ell+1,1-m}{2}\Big|1\Big)
\Big\}\;.
\label{Lnl-0}
\end{align}
Now we use the recursion formula \cite[\S 9.137.7]{Gradsteyn:1994??} to obtain
\begin{align*}
{}_2F_1\Big( \di{\ell+1,1-m}{2}\Big|1\Big)
&= \bigg(\prod_{p=2}^m \Big(1-\frac{\ell+1}{p}\Big)\bigg)
{}_2F_1\Big( \di{\ell+1,m-m}{1+m}\Big|1\Big)
= \frac{(-1)^{m-1} (\ell-1)!}{(\ell-m)!m!}\;.
\end{align*}
The remaining $m$-summation in (\ref{Lnl-0}), including the $m=0$ case 
$\frac{(n{+}\ell{-}1)! }{n!\ell!}$, yields
\begin{align*}
L^{(n,\ell)}(0,b)\big|_{\ell,n\geq 1}
= \Big(\frac{b}{1+b}\Big)^\ell 
\frac{(n{+}\ell{-}1)! }{n!\ell!} \;
{}_2F_1\Big(\di{-n,-\ell}{1-n-\ell}\Big|1\Big) =0\;,
\end{align*}
using \cite[\S 9.137.7]{Gradsteyn:1994??} again.
The proof of $L^{(n,0)}(0,b)=0$ is much simpler.

For the numerical implementation we have to control the function under
the Hilbert transform in (\ref{dlogGab-ab}) at $\bullet=0$. The same
considerations as before yield
\begin{align}
&\lim_{a\to 0} \frac{A^{(n,\ell)}(a,b)}{|\lambda|\pi a}
\Big|_{\ell,n\geq 1}
\nonumber
\\
& = \Big(\frac{b}{1+b}\Big)^\ell 
\sum_{m=0}^{n} \sum_{k=0}^m 
\binom{n{-}m{+}\ell{-}1}{\ell-1} 
\frac{(-1)^{k}(\ell+k)!}{m!\ell!} 
Y_{m,k}\Big(1!,2!,\dots,(m-k+1)!\Big)
\nonumber
\\
&= \Big(\frac{b}{1+b}\Big)^\ell 
\Big(\binom{n{+}\ell{-}1}{\ell-1} 
- (\ell+1) \sum_{m=1}^{n} 
\binom{n{-}m{+}\ell{-}1}{\ell-1} 
\sum_{k=0}^{m-1} \frac{(-1)^{k}(\ell+1+k)!1!
(m-1)!}{(\ell+1)! (k+1)!(m-1-k)! k!}\Big)
\nonumber
\\
&= 
 \Big(\frac{b}{1+b}\Big)^\ell 
\Big(\binom{n{+}\ell{-}1}{\ell-1} 
- (\ell+1) \sum_{m=1}^{n} 
\binom{n{-}m{+}\ell{-}1}{\ell-1} 
{}_2F_1\Big( \di{\ell+2,1-m}{2}\Big|1\Big)
\Big)
\nonumber
\\
&= 
 \Big(\frac{b}{1+b}\Big)^\ell 
\sum_{m=0}^{n} 
\binom{n{-}m{+}\ell{-}1}{\ell-1} 
\frac{(-1)^{m} (\ell+1)!}{(\ell+1-m)!m!}
\nonumber
\\
&=  \Big(\frac{b}{1+b}\Big)^\ell 
\frac{(n{+}\ell{-}1)!}{n!(\ell-1)!} 
\sum_{m=0}^{n} 
\frac{(-1)^{m} (\ell+1)! (n{+}\ell{-}1-m)!n!}{(\ell+1-m)!
  (n{+}\ell{-}1)! (n-m)! m!}
\nonumber
\\
&= \left\{ \begin{array}{cl}
\displaystyle 
 \Big(\frac{b}{1+b}\Big)
\sum_{m=0}^{n} 
\frac{(-1)^{m} 2!}{(2-m)!m!}
&\text{ for } \ell=1\;,
\\
\displaystyle
 \Big(\frac{b}{1+b}\Big)^\ell 
\frac{(n{+}\ell{-}1)!}{n!(\ell-1)!} 
{}_2F_1\Big(\di{-(\ell+1),-n}{1-(n+\ell)}\Big| 1\Big)
&\text{ for } \ell>1\;.
\end{array}\right.
\end{align}
For $n=2$ we have
${}_2F_1\big(\di{-(\ell+1),-2}{-(\ell+1)}\big| 1\big)
=0$ and thus from the recursion
 \cite[\S 9.137.7]{Gradsteyn:1994??} 
${}_2F_1\big(\di{-(\ell+1),-n}{1-(n+\ell)}\big| 1\big)=0$ for all
$n\geq 2$. The cases $n=1$ and $n=0$ can easily be discussed so that in
summary we obtain
\begin{align}
\lim_{a\to 0} \frac{A^{(n,\ell)}(a,b)}{|\lambda|\pi a}
= \Big(\frac{b}{1+b}\Big)^\ell \cdot \left\{
\begin{array}{cl}
(-1)^n & \text{ for } n\in \{0,1\}\;, \\
0 & \text{ for } n \geq 2\;.
\end{array}\right.
\label{limAnl}
\end{align}
Repeating these arguments for (\ref{Anl-0}) we find that 
(\ref{limAnl}) also holds for $\ell=0$.

\end{appendix}

\end{document}